\documentclass[a4paper,fleqn,usenatbib]{mnras}

\usepackage{newtxtext,newtxmath}
\usepackage[T1]{fontenc}
\usepackage{ae,aecompl}
\usepackage{graphicx}	% Including figure files
\usepackage{amsmath}	% Advanced maths commands
\usepackage{natbib}
\usepackage{xcolor}
\usepackage{braket}

\title[Discriminating between Dark Matter and MOND Modified Inertia]{A Method for Discriminating Between Dark Matter Models and MOND Modified Inertia  via Galactic Rotation Curves}

\author[J. Petersen and M. Frandsen]{
Jonas Petersen$^{1}$\thanks{E-mail: petersen@cp3.sdu.dk}
and Mads T. Frandsen$^{1}$\thanks{E-mail: frandsen@cp3.sdu.dk}
\\
$^{1}$ Centre for Cosmology and Particle Physics Phenomenology, University of Southern Denmark, Campusvej 55, DK-5230 Odense M, Denmark}

\date{Accepted 2019 September 12. Received 2019 September 12; in original form 2019 June 25}

\pubyear{2019}

\begin{document}
\maketitle

\begin{abstract}
Dark Matter (DM) and Modified Newtonian Dynamics (MOND) models of rotationally supported galaxies lead to curves with different geometries in $(g_{N},g_{tot})$-space ($g2$-space). Here $g_{tot}$ is the total acceleration and $g_{N}$ is the acceleration as obtained from the baryonic matter via Newtonian dynamics. 
In MOND modified inertia (MI) models the curves in $g2$-space are closed with zero area and so curve segments at radii $r\geq r_{N}$ (large radii) and $r< r_{N}$ (small radii) coincide, where $r_{N}$ is the radius where $g_N$ is greatest. In DM models with cored density profiles where $g_{tot}$ is also zero at the galactic centre, the curves are again closed, but the area of the closed curves are in general non-zero because the curve segments at radii $r\geq r_{N}$ and $r<r_{N}$ do not coincide. 
Finally in DM models with cuspy density profiles such as the NFW profile where $g_{tot}$ is formally non-zero at the galactic origin the curves are open, and again the curve segments at radii $r\geq r_{N}$ and $r< r_{N}$ do not coincide.\newline 
We develop a test of whether data at small and large radii coincide and investigate rotation curves from the SPARC database in order to discriminate between the above geometries. Due to loosely quantified systematic uncertainties we do not underline the result of the test, but instead conclude that the test illustrates the relevance of this type of analysis and demonstrate the ability to discriminate between the considered DM and MI models in this way.
\end{abstract}


\section{Introduction}
There is a significant amount of astrophysical evidence for missing gravity on scales ranging from galactic to cosmological. The evidence includes the measured rotation curves of galaxies~\citep{Rubin:1970zza,Bosma,Rubin:1980zd} which appear to be well described by a modification of Newtonian dynamics (MOND) for accelerations below a characteristic acceleration scale close to the value $a_0\sim c H_0$, where $c$ is the speed of light and $H_0$ is the value of the Hubble constant today~\citep{Milgrom:1983ca}. In particular MOND provides an explanation of the Tully-Fisher relation~\citep{Tully:1977fu} and successfully predicted the Baryonic Tully-Fisher relation~~\citep{McGaugh:2000sr}. On larger scales the lensing of galaxy clusters \citep{Mellier}, observations of cluster mergers~\citep{Clowe:2006eq} and large scale structure surveys \citep{Dodelson:2006zt} all appear consistent with the DM hypothesis while it is not obvious how to explain this body of observations within the MOND hypothesis. \newline

\noindent MOND models can arise from a modification of gravity (MOND modified gravity, MG) or a modification of inertia (MOND modified inertia, MI)~\citep{Milgrom:1983ca}. In a recent analysis~\citep{McGaugh:2016leg} it was found that the rotation curve data from the galaxies in the SPARC database~\citep{Lelli:2016zqa} follow an analytical relation (called the radial acceleration relation (RAR)) closely related to MI. Data from all galaxies in the SPARC database are pooled together in the analysis of \citet{McGaugh:2016leg}. The pooling af data wash out radial information from each galaxy. In their analysis \Citet{Petersen:2020vks} have shown that the radial information hold the potential to discriminate between MI and MG. In this article we will analyze the radial information in an attempt to discriminate between MI and DM. It will be shown how dark matter (DM) and MI lead to curves with different geometries in $(g_{N},g_{tot})$-space ($g2$-space) (section $1$). Here $g_{tot}$ is the total centripetal acceleration and $g_{N}$ is the centripetal acceleration as obtained from the baryonic matter via Newtonian dynamics -- both for rotationally supported galaxies. The differences in geometry are also apparent from e.g. \citet{Salucci:2016vxb}, but in the present study we explore how the differences may be used to test MI (sections $2-4$).\newline

\noindent Previous analyses of rotation curve data from the SPARC database~\citep{Lelli:2016zqa} in $g2$-space has found it to be both consistent with MI \citep{McGaugh:2016leg,Lelli:2017vgz} and with DM~ \citep{DiCintio:2015eeq,Salucci:2016vxb,Keller:2016gmw,Ludlow:2016qzh}. See also \citet{Desmond:2016azy} for a detailed statistical analysis of the SPARC data. 
This is a priori not at odds with MI and DM yielding different geometries in $g2$-space since the differences manifests themselves in each galaxy mainly at small radii $r \lesssim r_{N}$, where $r_{N}$ is the radius of the maximum Newtonian baryonic acceleration $g_{N}$. The SPARC database is by far dominated by points at large radii ($r \gtrsim r_{bar}$ and so analyses that do not retain radial information (global analyses) will mostly test the behavior at large radii - at which the theoretical predictions of dark matter and MI are similar, namely $g_{tot}\sim \sqrt{g_N}$ -- dictated by the baryonic Tully-Fisher relation~\citep{Tully:1977fu}. Hence, with data from the SPARC database, the global analyses are not optimal for distinguishing between DM and MI. A better approach is to investigate the behavior of galaxies at small radii relative to large radii -- that is the radial information from each galaxy is needed in order to distinguish between different models.\newline 
In this manuscript, we use $tot$ and $N$ subscripts to denote theoretical predictions for quantities with subscript $obs$ and $bar$, respectively.

\section{Rotation curves in \lowercase{g2}-space}
\begin{figure}
	\centering
	\includegraphics[width=0.4\textwidth]{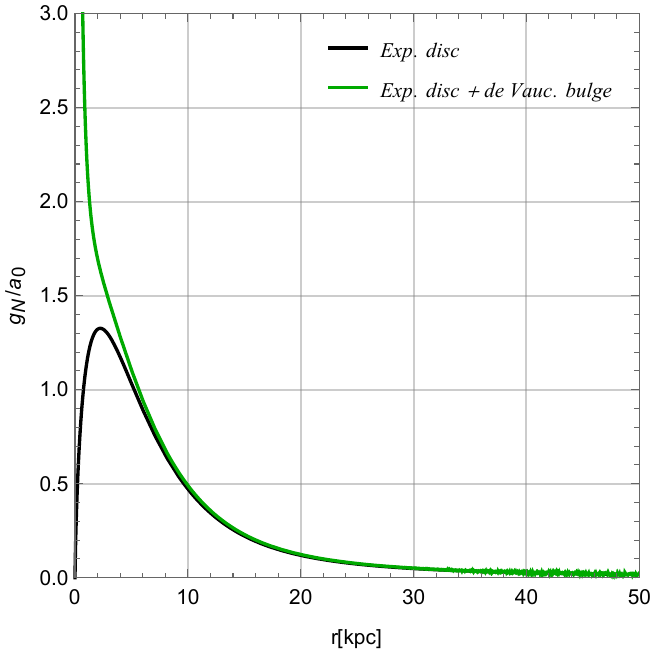}
	\caption
	{$g_{N}$ in the case where the baryonic content is modeled as an exponential disk (black) and an exponential disk as well as a de Vaucouleurs bulge (green). Here, we use $r_b=0.50\,\text{kpc}$, $\kappa=7.67$, $\tilde{\Sigma}_0=2.83\times10^8\frac{\rm M_\odot}{\text{kpc}^2}$, $r_d=3.00\,\text{kpc}$, $\Sigma_0=6.36\times 10^8\frac{\rm M_\odot}{\text{kpc}^2}$ and $a_0=1.2 \cdot 10^{-10} \frac{m}{s^2}$.}
	\label{fig:g(r)}
\end{figure}

The centripetal baryonic acceleration in a galaxy assuming Newtonian gravity, $g_{N}$, is given by
\begin{equation}
g_{N}(r)= \frac{\sum_{i\in \{\rm gas, \rm disk, \rm bulge\}}v_i(r)^2}{r} , 
\label{eq1}
\end{equation}
where $\{\rm gas, \rm disk, \rm bulge\}$ refer to the different baryonic components of the galaxies. As a first approximation for analyzing the baryonic centripetal acceleration the baryonic content in a rotationally supported galaxy can be approximated as consisting of an exponential disk~\citep{Patterson,freeman,Sparke,binney} and a spherical de Vaucouleurs bulge~\citep{1953MNRAS.113..134D} with surface mass densities
\begin{equation}
\Sigma_d(r)=\Sigma_0e^{-\frac{r}{r_d}}, \quad \Sigma_b(r)=\tilde{\Sigma}_0e^{-\kappa\big((\frac{r}{r_b})^{1/4}-1\big)},
\label{18}
\end{equation}
where $\kappa=7.67$ \citep{binney}. The centripetal accelerations corresponding to equation \eqref{18} are shown in figure \ref{fig:g(r)}.
As is evident from it; the de Vaucouleurs bulge dominates the baryonic acceleration at small radii, making it grow down to very small radii. Only around $20\%$ of galaxies in the SPARC database have a bulge, hence the majority of galaxies investigated are expected to be well described by an exponential disk. It is however emphasized that the data analysis does not rely on any assumptions on behalf of the mass distribution -- these are only used to illustrate the theoretical expectations that are tested.

\subsection{MOND Modified Inertia and the RAR}
\begin{figure}
	\centering
	\includegraphics[width=0.4\textwidth]{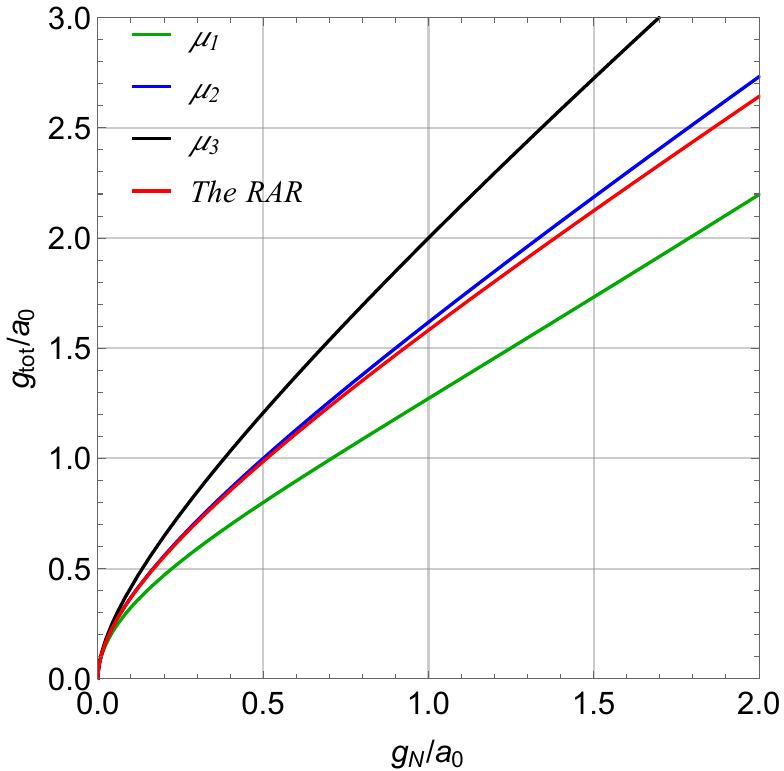}
	\caption{The $g2$-space geometries of MOND modified inertia from using the interpolation functions of equation \eqref{eq4} alongside the $g2$-space geometry of equation \eqref{mcgaugh}.}
	\label{fig:interoplation3}
\end{figure}

In MI the total acceleration, $g_{tot}^{(MI)} $, on a test mass is related to the Newtonian one, $g_{N}$,  via
\begin{equation}
g_{N}(g_{tot}^{(MI)})=\mu(x)g_{tot}^{(MI)} 
\label{eq3}
\end{equation}
or equivalently
\begin{equation}
g_{tot }^{(MI)}(g_{N})=\nu(y)g_{N} 
\end{equation}
where $x\equiv \frac{g_{tot}^{(MI)}}{a_0}$, $y\equiv \frac{g_{N}}{a_0}$ and $\nu(y)=\mu(I^{-1}(y))^{-1}$ with  $I(x)=\mu(x)x$ and $a_0= 1.2 \cdot 10^{-10} \frac{m}{s^2}$ the characteristic acceleration scale of MOND~\citep{Begeman1991}. The function $\mu(x)$ smoothly interpolates between $\mu(x)=x$ for $x\ll 1$ and $\mu(x)=1$ for $x\gg 1$. Commonly applied, experimentally motivated, interpolations include~\citep{Famaey:2005fd,Lelli:2017vgz}
\begin{equation}
\begin{split}
&\mu_1(x)=\frac{x}{\sqrt{1+x^2}}\\
&\mu_2(x)=\frac{x}{1+x}\\
&\mu_3(x)=\frac{\sqrt{1+4x}-1}{\sqrt{1+4x}+1}.\\
\end{split}
\label{eq4}
\end{equation}
The RAR is given by~\citep{McGaugh:2016leg}
\begin{equation}
g_{tot}^{(RAR)}(g_{N})=\frac{g_{N}}{1-e^{-\sqrt{\frac{g_{N}}{g^{\dagger}}}}},
\label{mcgaugh}
\end{equation}
with $g^\dagger\simeq 1.2\cdot 10^{-10}\frac{m}{s^2}$ obtained from globally fitting to the SPARC database. Although the RAR is motivated by the fit to data performed in~\citep{McGaugh:2016leg}, it is classified as an interpolation function of MI~\citep{Milgrom:2007br,McGaugh:2008nc}. Figure \ref{fig:interoplation3} show the $g2$-space geometries obtained from equations \eqref{eq4} and \eqref{mcgaugh}.

\subsection{Dark Matter}
In DM models the total centripetal acceleration $g_{tot}^{(DM)}(r)=g_{N}(r)+g_{halo}(r)$ is a sum of $g_{N}(r)$ and the acceleration from the DM halo $g_{halo}(r)$. 
Examples of DM density profiles are the Navarro-Frenk-White \citep[NFW,][]{Navarro:1995iw} and pseudo-isothermal profiles 
\begin{equation}
\rho_{NFW}(r)=\frac{\rho_s}{\frac{r}{r_s}(1+\frac{r}{r_s})^2} , \quad \rho_{iso}(r)=\frac{\rho_0}{1+(\frac{r}{r_0})^2} 
\end{equation}
with characteristic scale lengths $r_s,r_0$ and central densities $\rho_s,\rho_0$, respectively. The Navarro-Frenk-White profile $\rho_{NFW}(r)$ is motivated at large radii by fits to simulations of cold collisionless DM~\citep{Navarro:1995iw}. The isothermal DM density profile $\rho_{iso}(r)$ is motivated (at small radii) by models with sizeable DM self interactions~\citep{Kamada:2016euw,Creasey:2016jaq}. They recently proposed that the diversity of galactic rotation curves~\citep{Navarro:2008kc} can be accommodated in models of self interacting DM where the DM density follows the quasi-isothermal profile below a characteristic radius proportional to the self-interaction cross-section and reduces to the NFW profile at large radii. 
\begin{figure}
	\centering
		\includegraphics[width=0.4\textwidth]{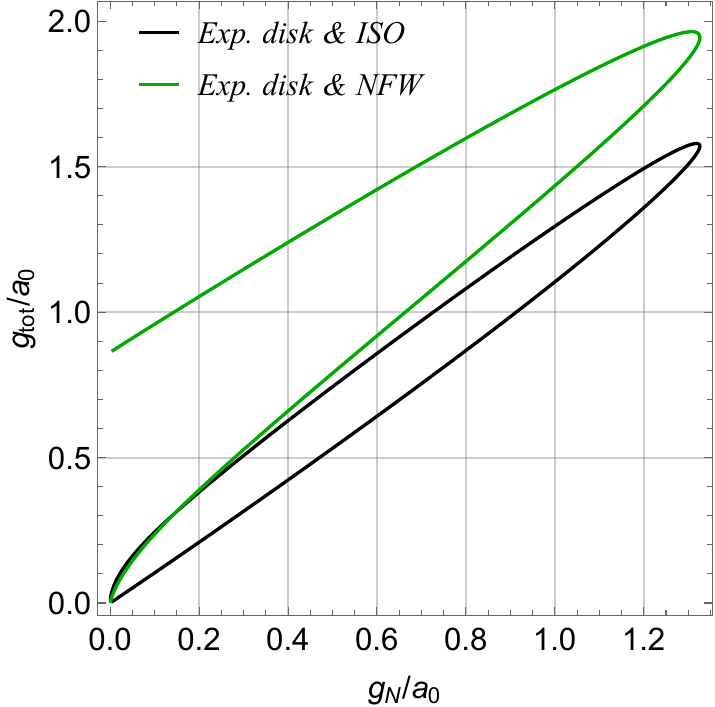}
		\includegraphics[width=0.4\textwidth]{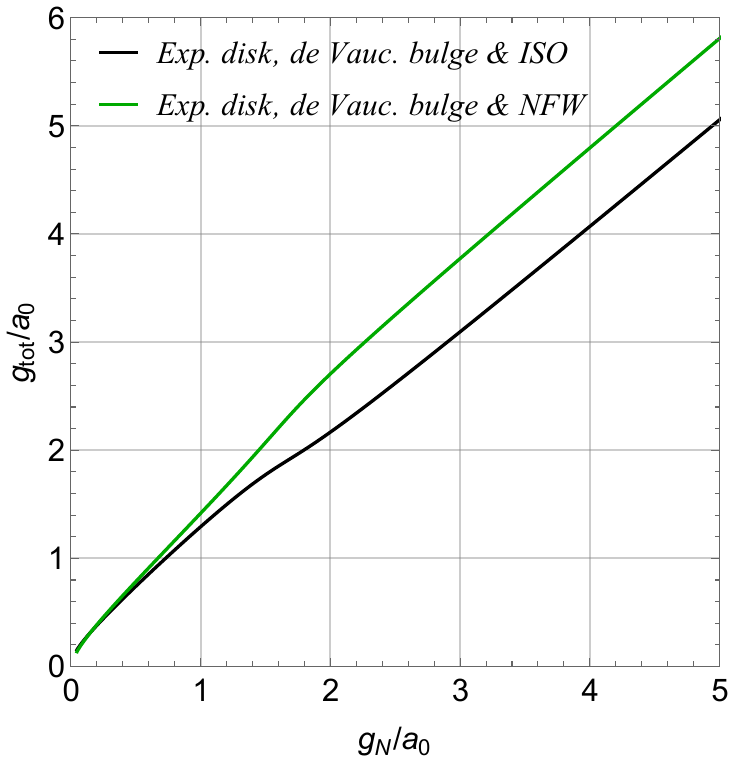}
	\caption
	{The $g2$-space geometries for the NFW (green) and pseudo-isothermal (black) density profiles. Top: the geometries using an exponential disk as baryonic matter. (bottom) shows the geometries using an exponential disk and a de Vaucouleurs bulge as baryonic matter. For the figures $r_s=8\,\text{kpc}$, $\rho_{0}^{NFW}=10^{-21}\frac{kg}{m^3}$, $\rho_{0}^{ISO}=2\cdot10^{-21}\frac{kg}{m^3}$, $r_0=3\, \text{kpc}$ alongside the baryonic data used in figure \ref{fig:g(r)}.}
	\label{fig:gtot}
\end{figure}
The geometry of each considered dark matter model in $g2$-space depend on the baryonic content of the considered galaxy. In presence of a de Vaucouleurs bulge the geometry will be monotonically increasing for radii realistic for sampling whereas in the absence of a de Vaucouleurs bulge this will not be the case as the curve segments above and below $r_N$ will not coincide (see figure \ref{fig:gtot}). 
Hence, when the peak in $g_{bar}$ can be sampled (i.e. in the absence of de Vaucouleurs-like bulges), MI predicts that data at large and small radii coincide whereas the considered DM profiles do not. This is the core idea of this article and what will be tested in the following using data from the SPARC database.

\section{Data}
The SPARC database consists of rotation curve data from $175$ rotationally supported galaxies \citep{Lelli:2016zqa,Lelli:2017vgz}. The database contain observed rotational velocities ($v_{obs}$), the corresponding uncertainties ($\delta v_{obs}$), distance ($D$) and inclination ($\alpha$) measurements for each galaxy and the rotational velocities due to the baryonic components ($v_{gas}\equiv v_g,v_{bul}\equiv v_b$ and $v_{disk}\equiv v_d$). In line with \citep{McGaugh:2016leg,Lelli:2017vgz} a $10\%$ relative uncertainty on the mass of the gas, $m^g_i$, ($\delta m^g_i$) is utilized.\newline\newline
Following \citet{Lelli:2017vgz} $22$ galaxies are dropped from the analysis based on the inclination angle and further quality criteria defined in \citep{Lelli:2017vgz}. $1$ additional galaxy is dropped because of large negative values of $v_g$ at the innermost radii, leaving $152$ galaxies making up $3143$ data points in $g2$-space. \newline\newline

\noindent The baryonic velocity ($v_{bar}$) for the $k$'th data point in the $i$'th galaxy is computed via 
\begin{equation}
v_{bar}(r_{k,i})=\sqrt{|v_{g}(r_{k,i})|v_{g}(r_{k,i})+\tilde{\Upsilon}^{d}_iv_{d}^2(r_{k,i})+\tilde{\Upsilon}^{b}_iv_{b}^2(r_{k,i})},
\label{v_bar}
\end{equation}
where $\tilde{\Upsilon}^{d}_i$ and $\tilde{\Upsilon}^{b}_i$ are the unitless\footnote{By definition they must be since $\tilde{\Upsilon}^{d}_iv_{d}^2(r_{k,i})$ must have units of "velocity squared".} mass to light ratios of the $i$'th galaxy, the radial measurement corresponding to the individual measurements of $v_g,v_d$ and $v_b$ is used to label the individual data points and we have imposed the physically motivated restriction that $v_d,v_b>0$. \newline

\section{Method}
\begin{figure}
	\centering
	\includegraphics[width=0.4\textwidth]{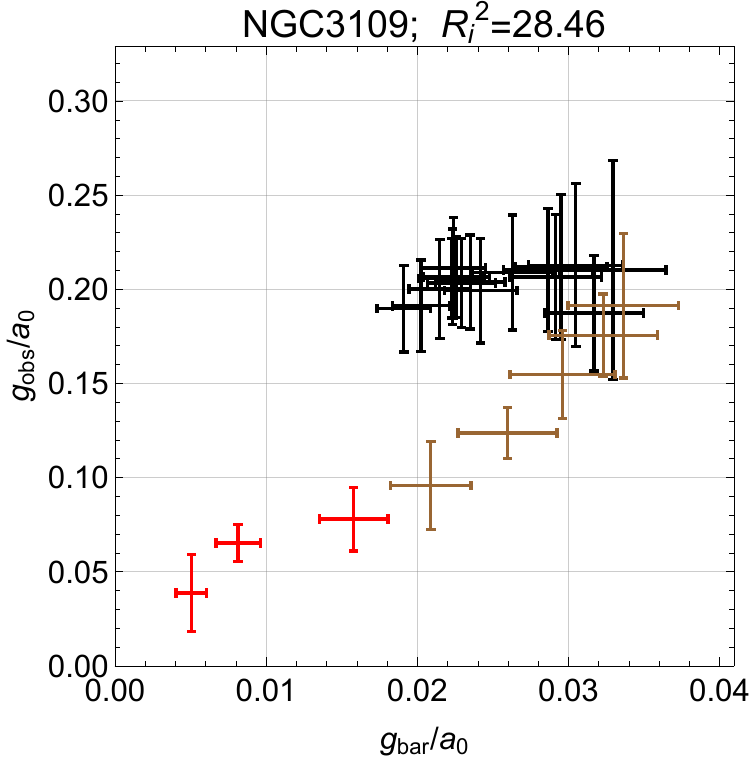}
	\includegraphics[width=0.4\textwidth]{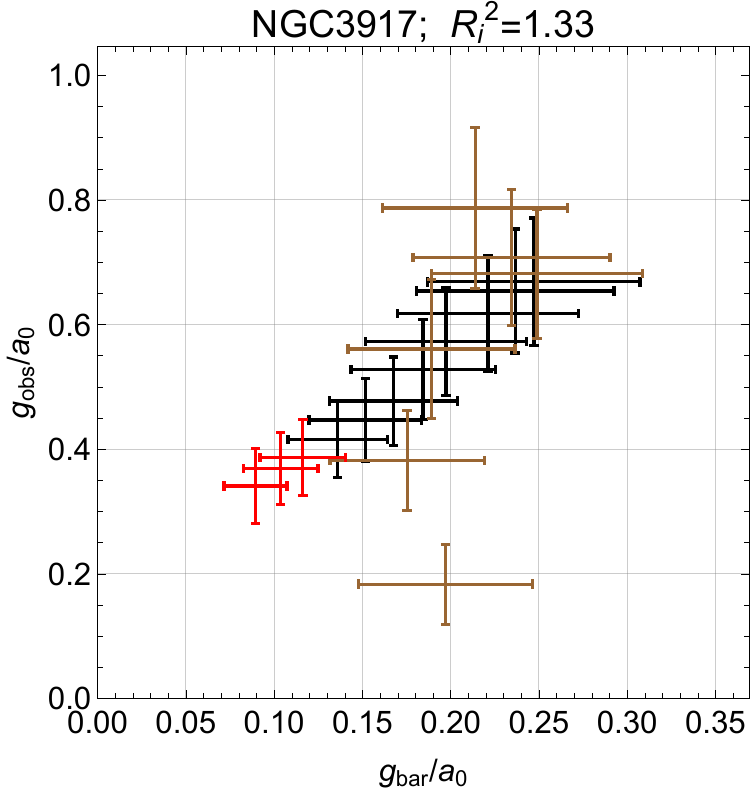}
	\caption{$g2$-space plots that illustrate the set definitions and discarded points. Note the errors bars include systematic uncertainties that propagate non-trivially to $\bar{g}^{obs}$. Top: A $g2$-space plot of NGC 3109. Black denotes points in $\tilde{G}_1$ (large radii), brown denotes points in $\tilde{G}_2$ (small radii) whereas red denotes points discarded. The red and brown points form $G_2$ whereas $\tilde{G}_1=G_1$ in this case. Bottom: A $g2$-space plot of NGC 3917. Black denote points in $\tilde{G}_1$, brown denote points in $\tilde{G}_2$ whereas red denote points discarded. The red and black points form $G_1$ whereas $\tilde{G}_2=G_2$ in this case. }
	\label{fig:p2}
\end{figure}

To test whether data at small and large radii coincide, the mean value of $g_{obs}$ for points with $r\leq r_{bar}$ is subtracted from the mean value of $g_{obs}$ for points with $r\geq r_{bar}$. Ideally, the number of points at $r\leq r_{bar}$ and $r\geq r_{bar}$ should be equal and have the same $g_{bar}$ values. However, as this is not a feasible experimental requirement, we relax it and instead discard points (red points in figure \ref{fig:p2}) symmetrically around $r_{bar}$ such that points at $r\leq r_{bar}$ (brown points in figure \ref{fig:p2}) and $r\geq r_{bar}$ (black points in figure \ref{fig:p2}) approximately refer to the same interval\footnote{This is done so the test performed is as independent of expected value of MI (equation \eqref{bias}) as possible.} of $g_{bar}$ (see figure \ref{fig:p2} or appendix \ref{app:galaxies}). We do not require an equal number of points, however, we do require that there be at least two points at both $r\leq r_{bar}$ and $r\geq r_{bar}$ in order for the average to make sense. This approximation mean that the subtraction of averages in $g_{obs}$ is not expected to be exactly zero for MI. We estimate the expected value for MI by using the RAR. 
\newline To formalize the above detailed procedure, let points in $g2$-space for $r\geq r_{bar}$ belong to the set $G_1$ and points for $r\leq r_{bar}$ belong to the set $G_2$ ($r_{bar}$ appear in both sets). Define the reference radius, $r_{ref}$, as 
\begin{equation}
r_{ref}\equiv \arg\max_r[\arg\min_{r\in G_1}(g_{bar}),\arg\min_{r\in G_2}(g_{bar})].
\end{equation}
Then let $g_{cut}\equiv g_{bar}(r_{ref})-\delta g_{bar}(r_{ref})$, with $\delta g_{bar}(r_{ref})$ being the uncertainty of $g_{bar}(r_{ref})$, denote a cutoff such that points in $g2$-space with $g_{bar}<g_{cut}$ are discarded. After discarding points $G_1\rightarrow \tilde{G}_1$ and $G_2\rightarrow \tilde{G}_2$. Figure \ref{fig:p2} illustrates the above procedure for two different galaxies (UGC 12506 and NGC 3109).
Whether data at small and large radii coincide for a given galaxy is measured by
\begin{equation}
\bar{g}^{obs}\equiv \frac{\sum_{j\in \tilde{G}_1}g_{obs}(r_{j})}{N_{\tilde{G}_1}}-\frac{\sum_{j\in \tilde{G}_2}g_{obs}(r_{j})}{N_{\tilde{G}_2}},
\label{e2}
\end{equation}
with $N_{\tilde{G}_1}$ and $N_{\tilde{G}_2}$ denoting the amount of points in $\tilde{G}_1$ and $\tilde{G}_2$, respectively. If the amount of points in $\tilde{G}_1$ and $\tilde{G}_2$ were the same and had the same $g_{bar}$, the expected value of MI would be zero. This is however not the case and so the expected value of MI is estimated as
\begin{equation}
\bar{g}^{tot}_{MI}\approx \bar{g}^{RAR},
\label{approx}
\end{equation}
with
\begin{equation}
\bar{g}^{RAR}\equiv \frac{\sum_{j\in \tilde{G}_1}g_{tot}^{(RAR)}(g_{bar}(r_{j}))}{N_{\tilde{G}_1}}-\frac{\sum_{j\in \tilde{G}_2}g_{tot}^{(RAR)}(g_{bar}(r_{j}))}{N_{\tilde{G}_2}}
\label{bias}
\end{equation}
Then, the $\chi^2$ can be approximated by
\begin{equation}
\chi^2\approx \sum_{i\in galaxies}R_i^2,
\label{chi2}
\end{equation}
with
\begin{equation}
R_i\equiv \frac{\bar{g}^{obs}_i-\bar{g}^{RAR}_i}{\sqrt{(\delta \bar{g}^{obs}_i)^2+(\delta \bar{g}^{RAR}_i)^2}}
\label{residuals}
\end{equation}
being the residuals and $i$ running over different galaxies. $\delta \bar{g}^{obs}_i$ and $\delta \bar{g}^{RAR}_i$ are derived in appendix \ref{sec:errors}. The approximation in equation \eqref{chi2} consists of estimating the expected value of MI by the RAR (equation \eqref{approx}).

\section{Results}
\label{sec:res}
\begin{figure}
	\centering
	\includegraphics[width=0.4\textwidth]{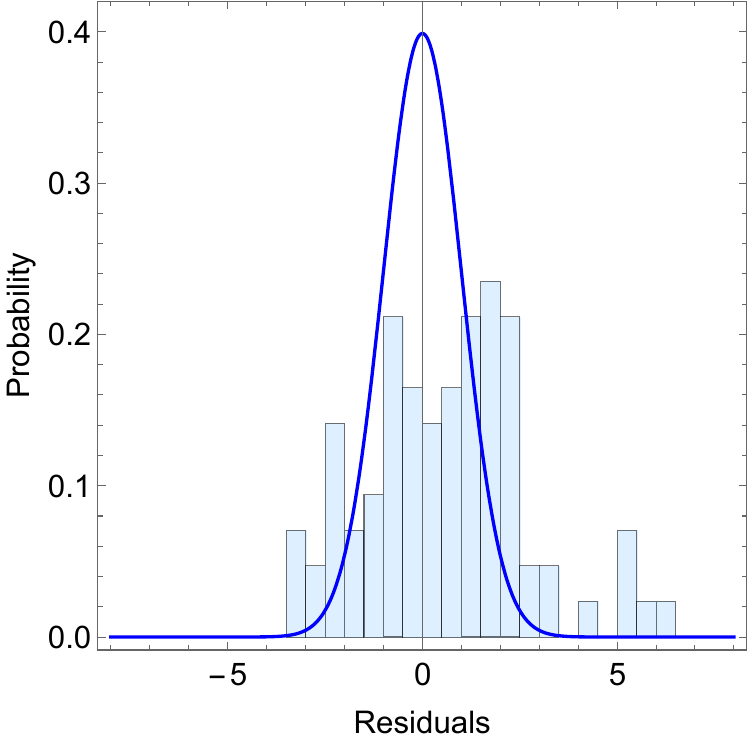}
	\caption{A histogram distribution of the residuals (equation \eqref{residuals}) alongside a standard normal distribution.}
	\label{fig:p1}
\end{figure}

In line with \citet{Lelli:2016zqa}, we take $\tilde{\Upsilon}^{d}_i=0.5$ and $\tilde{\Upsilon}^{b}_i=0.7$ for numerical computations, both with a $25 \%$ relative uncertainty.\newline 
Requiring that there be $2$ data points at $r\leq r_{bar}$ and $r\geq r_{bar}$ leaves $85/152$ galaxies. Figure \ref{fig:p1} shows a histogram of the residuals from these galaxies alongside a standard normal distribution. Several things can be noted from the figure; i) the residuals appear roughly normally distributed but with parameters that deviate significantly from a standard normal and ii) the histogram hints at the existence of subpopulations of galaxies with residuals above/below $0$. The residuals correspond to $\chi^2\approx 408$ with $85$ degrees of freedom, meaning that coinciding data at small and large radii can be rejected with $\gtrsim 8 \sigma$ confidence -- with the "$\gtrsim$" stemming from the approximation in equation \eqref{chi2}. 

\section{Summary and Discussion}
In this article the $g2$-space geometries of MOND modified inertia (MI) including the RAR (equation \eqref{mcgaugh}) and NFW and isothermal (ISO) dark matter (DM) density profiles have been reviewed and discussed. It has been shown how MI leads to a $g2$-space geometry that is covered twice with radius (the MI $g2$-space values for large and small radii coincide) whereas both ISO and NFW DM lead to $g2$-space geometries that are covered only once (the ISO and NFW DM $g2$-space values for large and small radii do not coincide). Using this knowledge a test of whether data at small and large radii coincide is developed and applied in order to discriminate between MI on one side and NFW and ISO dark matter on the other. \newline

\noindent Using data from the SPARC database, and under the approximations and assumptions made (e.g. regarding mass to light ratios being constant and having no radial dependence as well as estimating the expected value for MI in the described way), the test disfavors MI with $\gtrsim 8 \sigma$, where "$\gtrsim$" account for the approximation in using the RAR to estimate the expected value of MI in general. The expectation value of MI used here increases the $\chi^2$ by $9\%$ from $\chi^2\approx375$ to $\chi^2\approx408$ as specified in section \ref{sec:res}. To facilitate discussion we define $\chi^2_{ref}\equiv408$ as a reference value. In order for MI to be disfavored by less than $3\sigma$, $\chi^2_{ref}$ must be lowered by $\ga69\%$ with fixed degrees of freedom. Hence, it would likely require a combination of 1) a different definition of $r_{bar}$, 2) different values for the mass to light ratios, possibly including radial dependence, 3) removal of statistical outliers and 4) other effects.\newline 
Regarding point 1: There is an unaccounted bias in that the bins are defined by $r_{bar}$, a point which is affected by uncertainty. The impact of this bias can be roughly estimated by moving $r_{bar}$ one point toward smaller (larger) radii. Doing so yields a $52\%\, (-13\%)$ reduction in $\chi^2$ relative to $\chi^2_{ref}$, however with the accompanying degrees of freedom $44\,(79)$. The reduction in degrees of freedom (galaxies) are due to the criterion of at least two points at small radii. Considering the reduced chi square changes from $\frac{\chi^2_{ref}}{85}=4.8$ to $\frac{\chi^2_{ref}(1-0.52)}{44}=4.4$ after the shift to smaller radii, it can be concluded that although the shift in $r_{bar}$ greatly affect the $\chi^2$ it only slightly reduce the statistical disfavoring of MI.\newline
Regarding point 2: The dependence on the mass to light ratios may significantly affect $g_{bar}$ and consequently both the expected value of MI and the specification of $r_{bar}$. The fixed mass to light ratios used here ($\tilde{\Upsilon}^{d}_i=0.5$ and $\tilde{\Upsilon}^{b}_i=0.7$) are not optimal for all galaxies. This is clear from the rotation curve plots in appendix \ref{app:galaxies}, where for some galaxies $v_{bar}>v_{obs}$ for limited radial range (e.g. as is the case for CAMB and NGC4217). To roughly gauge the effect of radically different mass to light ratios, we take $\tilde{\Upsilon}^d_i=\tilde{\Upsilon}^b_i=0.2$ (corresponding to galaxies dominated by DM at all radii~\citep{Lelli:2017vgz}), yielding an $8\%$ increase in the $\chi^2$ relative to $\chi^2_{ref}$ (the result become slightly more significant). This is however just an example and one can imagine changing individual mass to light ratios and uncertainties potentially having a great impact on the result. \newline
Regarding point 3: Figure \ref{fig:p1} shows the distribution of residuals. Of the galaxies with $|R_i|>4$ 
UGC11455, NGC4217, NGC3109, ESO563-G021 and D631-7 have quality flag 1 whereas CAMB have quality flag 2 (listed in the SPARC database). This indicates that the disfavoring of MI is not rooted in bad data (figure \ref{fig:p1tlde} in appendix \ref{app:hist} show residuals after discarding all galaxies with quality flag larger than 1). That being said, as mentioned in point 2, the fixed mass to light ratios used are too large in case of NGC4217 and CAMB and possibly also affect the innermost radii of UGC11455 and ESO563-G021. It can also be argued that the high residual value in case of ESO563-G021 is at least partially rooted in a peculiar selection of points (e.g. see figure \ref{fig:p3} in appendix \ref{app:galaxies}). Hence, although data are of good quality, for several of the significant outliers, there are valid arguments that could at least partially explain the high residuals. To be conservative and test the dependence on the outliers, we proceed to discard all galaxies with $|R_i|>4$. Doing so yield a $42\%$ drop in the $\chi^2$ relative to $\chi^2_{ref}$, with an associated drop in degrees of freedom from $85$ to $79$.  
\newline
Regarding point 4: Another factor that potentially could affect the conclusion is that we take the uncertainties from the SPARC database at face value and propagate these to the observables constructed in this article. Significant uncertainty in the uncertainties may therefore have a potentially significant impact on the conclusion. Quantifying this systematic uncertainty is beyond the scope of this paper. Instead we emphasize that all quantitative results in this paper rely on the reliability of the reported uncertainties of data from the SPARC database.\newline
Combining points 1-3 by first changing the mass to light ratio, then shifting $r_{bar}$ one point toward smaller radii and lastly discarding galaxies with $|R_i|>4$, yield a $66\%$ reduction in $\chi^2$ relative to $\chi^2_{ref}$, however, with an associated reduction in degrees of freedom from $85$ to $47$. Hence, even in this case MI is still disfavored by $\approx 6.5\sigma$. That being said, points 1-3 are roughly estimated (and point 4 not quantified) and so instead of underlining the numerical result, we conclude that our test illustrates the relevance of analyses that include radial information and demonstrates the ability to discriminate between the considered DM and MI models. Lastly, we emphasize that MI models do not constitute all MOND models. MOND modified gravity (MG) models (another class of MOND models) lead to $g2$-space geometries similar to those of ISO DM and may still be consistent with our results. From figure \ref{fig:p1} it is clear that there is an overweight of outliers with $R_i>3$ relative to $R_i<-3$. This indicate that $g_{obs}$ for small radii may be systematically lower than $g_{obs}$ at large radii. This tendency is in line with what is expected from both ISO DM and MG~\citep{banik2019,Eriksen:2019xgl}.
\newline\newline 

{\bf Acknowledgments:}
We would like to thank Indranil Banik for helpful suggestions and detailed feedback on draft. We acknowledge partial funding from The Council For Independent Research, grant number DFF 6108-00623. The CP3-Origins center is partially funded by the Danish National Research Foundation, grant number DNRF90.\newline\newline

\appendix
\section{Uncertainties related to the residuals}
\label{sec:errors}
In this appendix the uncertainties related to the residuals $R_i$ are derived. In order to do so the notation is expanded with a galaxy index, $i$, throughout. The uncertainty of the measure of coinciding data at small and large radii can be written
\begin{equation}
\delta \bar{g}^{obs}_i=|\bar{g}^{obs}_i|\sqrt{\sum_{k\in \text{gal}_i}\bigg(C_{k,i}^{(1)}\frac{\delta v_{obs}(r_{k,i})}{v_{obs}(r_{k,i})}\bigg)^2+\bigg(\frac{\delta D_i}{D_i}\bigg)^2+\bigg(\frac{2\delta \alpha_i}{\tan(\alpha_i)}\bigg)^2}.
\end{equation}
with
\begin{equation}
C_{k,i}^{(1)}\equiv \frac{2}{\bar{g}^{obs}_i}\bigg(\frac{1}{N_{\tilde{G}_1}}\sum_{j\in \tilde{G}_1}g_{obs}(r_{j,i})\delta_{j,k}-\frac{1}{N_{\tilde{G}_2}}\sum_{j\in \tilde{G}_2}g_{obs}(r_{j,i})\delta_{j,k}\bigg).
\end{equation}
The uncertainty of the (approximated) expected value of MI estimate come from propagating the uncertainty in $g_{bar}$
\begin{equation}
\delta \bar{g}^{RAR}_i=\sqrt{\bigg(C^{(2)}_{k,i}\frac{\delta m_i^{g}}{m_i^{g}}\bigg)^2+\bigg(C^{(3)}_{k,i}\frac{\delta \tilde{\Upsilon}_i^{d}}{\tilde{\Upsilon}_i^{d}}\bigg)^2+\bigg(C^{(4)}_{k,i}\frac{\delta \tilde{\Upsilon}_i^{b}}{\tilde{\Upsilon}_i^{b}}\bigg)^2}
\end{equation}
with
\begin{equation}
\begin{split}
C_{k,i}^{(2)}\equiv& \frac{1}{N_{\tilde{G}_1}}\sum_{j\in \tilde{G}_1}\frac{\partial g_{tot}^{(RAR)}(g_{bar}(r_{j,i}),g^\dagger)}{\partial g_{bar}(r_{j,i})}\frac{|v_{g}(r_{j,i})|v_{g}(r_{j,i})}{r_{j,i}}\\
&-\frac{1}{N_{\tilde{G}_2}}\sum_{j\in \tilde{G}_2}\frac{\partial g_{tot}^{(RAR)}(g_{bar}(r_{j,i}),g^\dagger)}{\partial g_{bar}(r_{j,i})}\frac{|v_{g}(r_{j,i})|v_{g}(r_{j,i})}{r_{j,i}},
\end{split}
\end{equation}
\begin{equation}
\begin{split}
C_{k,i}^{(3)}\equiv& \frac{1}{N_{\tilde{G}_1}}\sum_{j\in \tilde{G}_1}\frac{\partial g_{tot}^{(RAR)}(g_{bar}(r_{j,i}),g^\dagger)}{\partial g_{bar}(r_{j,i})}\tilde{\Upsilon}_i^d\frac{v_{d}^2(r_{j,i})}{r_{j,i}}\\
&-\frac{1}{N_{\tilde{G}_2}}\sum_{j\in \tilde{G}_2}\frac{\partial g_{tot}^{(RAR)}(g_{bar}(r_{j,i}),g^\dagger)}{\partial g_{bar}(r_{j,i})}\tilde{\Upsilon}_i^d\frac{v_{d}^2(r_{j,i})}{r_{j,i}},
\end{split}
\end{equation}
and
\begin{equation}
\begin{split}
C_{k,i}^{(3)}\equiv& \frac{1}{N_{\tilde{G}_1}}\sum_{j\in \tilde{G}_1}\frac{\partial g_{tot}^{(RAR)}(g_{bar}(r_{j,i}),g^\dagger)}{\partial g_{bar}(r_{j,i})}\tilde{\Upsilon}_i^b\frac{v_{b}^2(r_{j,i})}{r_{j,i}}\\
&-\frac{1}{N_{\tilde{G}_2}}\sum_{j\in \tilde{G}_2}\frac{\partial g_{tot}^{(RAR)}(g_{bar}(r_{j,i}),g^\dagger)}{\partial g_{bar}(r_{j,i})}\tilde{\Upsilon}_i^b\frac{v_{b}^2(r_{j,i})}{r_{j,i}}.
\end{split}
\end{equation}

\section{Residual histograms}
\label{app:hist}
\begin{figure}
	\centering
	\includegraphics[width=0.4\textwidth]{p1a}
	\includegraphics[width=0.4\textwidth]{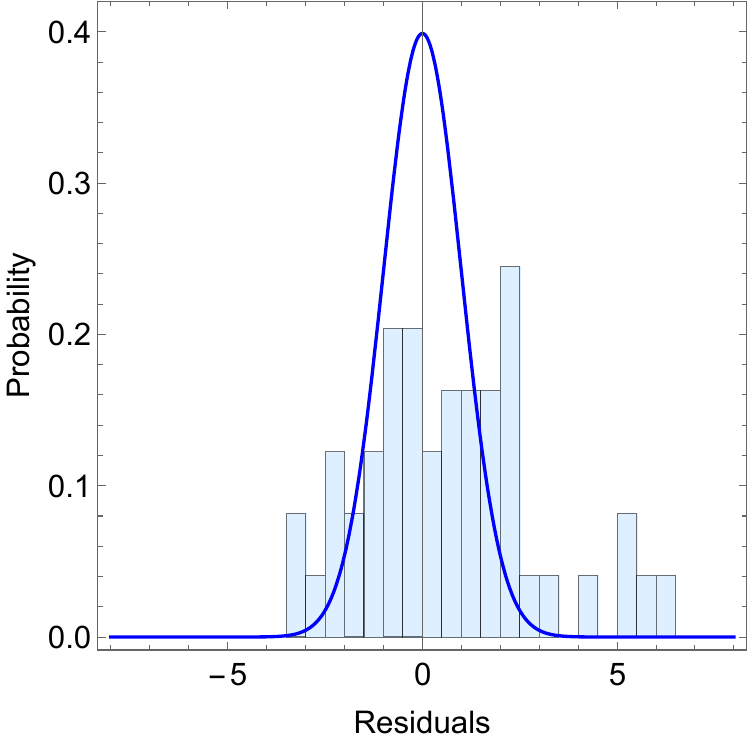}
	\caption{Top: A histogram distribution of the residuals (equation \eqref{residuals}) alongside a standard normal distribution. Bottom: A histogram corresponding to the top panel, but with galaxies with quality flag larger than 1 removed.}
	\label{fig:p1tlde}
\end{figure}
\clearpage
\section{Galaxies}
\label{app:galaxies}
\begin{figure*}
	\centering
	\includegraphics[width=0.2\textwidth]{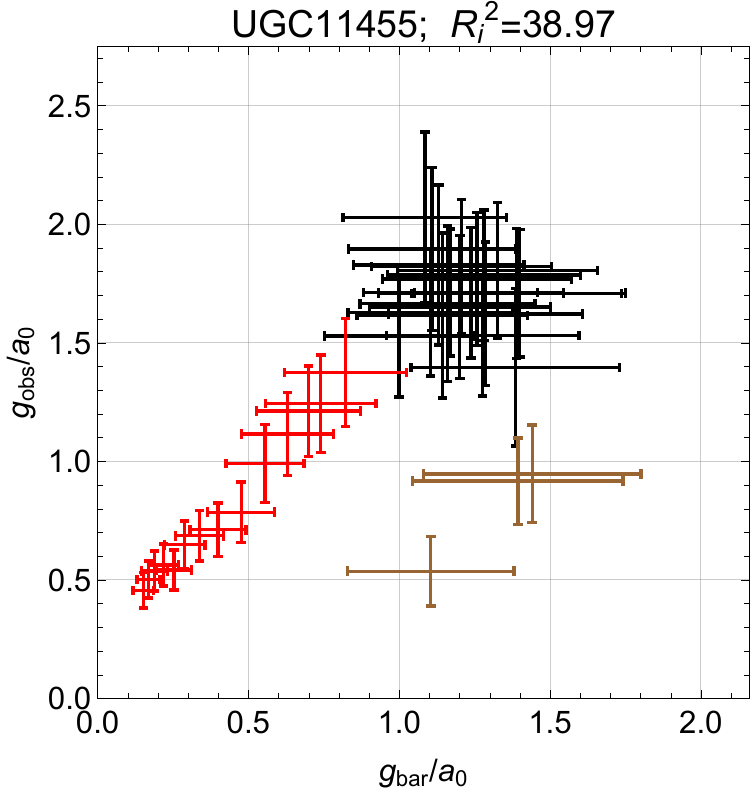}
	\includegraphics[width=0.2\textwidth]{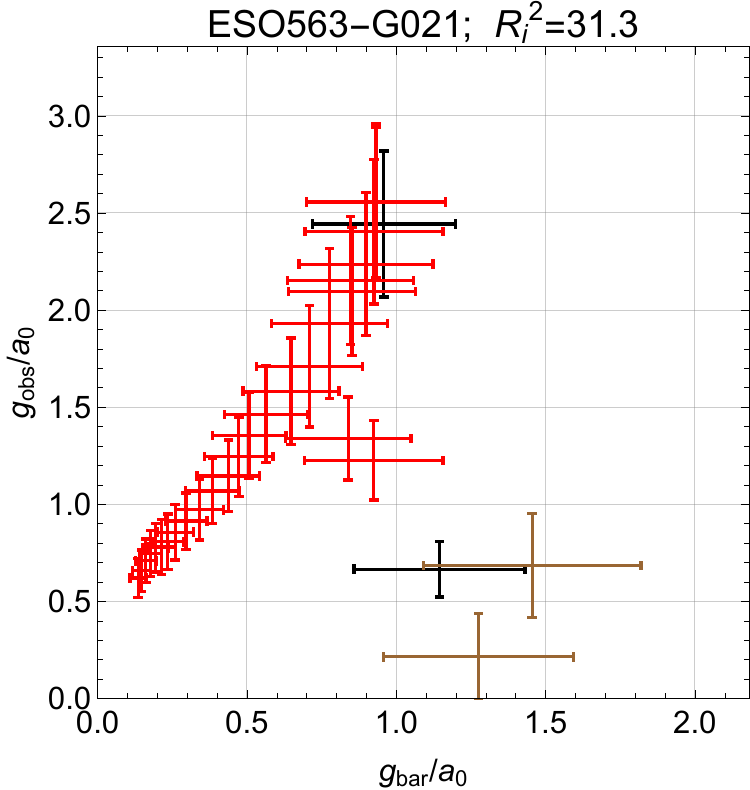}
	\includegraphics[width=0.2\textwidth]{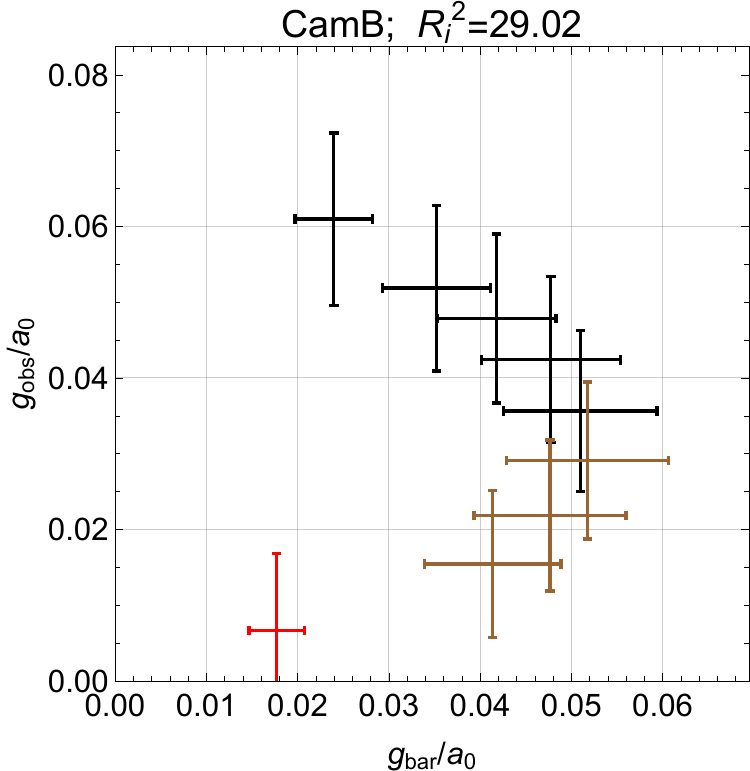}
	\includegraphics[width=0.2\textwidth]{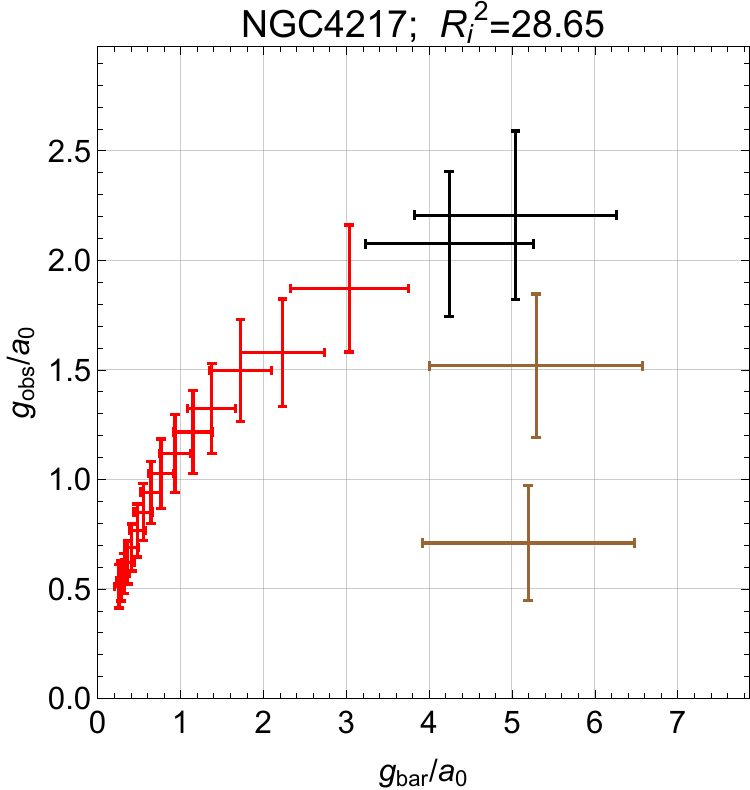}
	\includegraphics[width=0.2\textwidth]{p5}
	\includegraphics[width=0.2\textwidth]{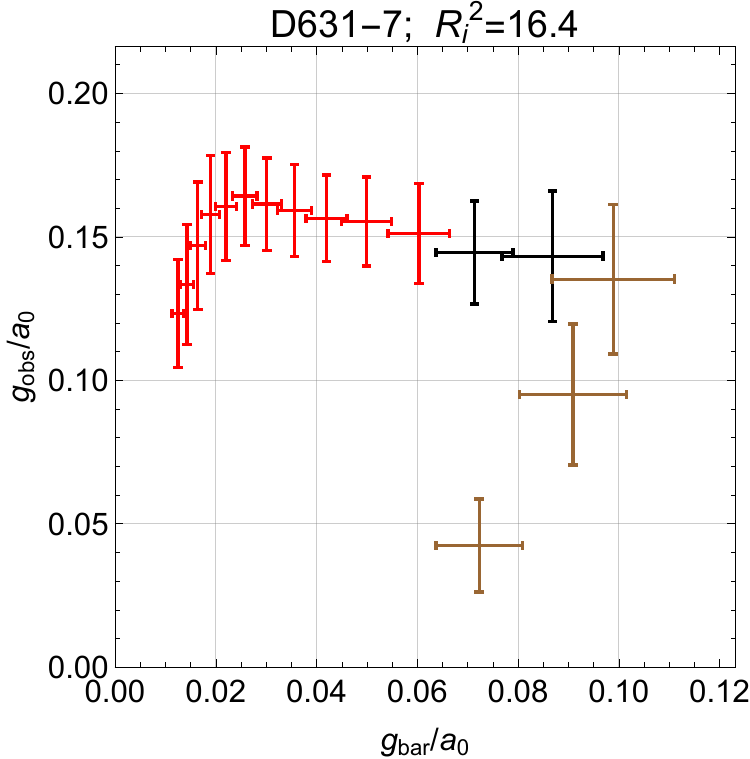}
	\includegraphics[width=0.2\textwidth]{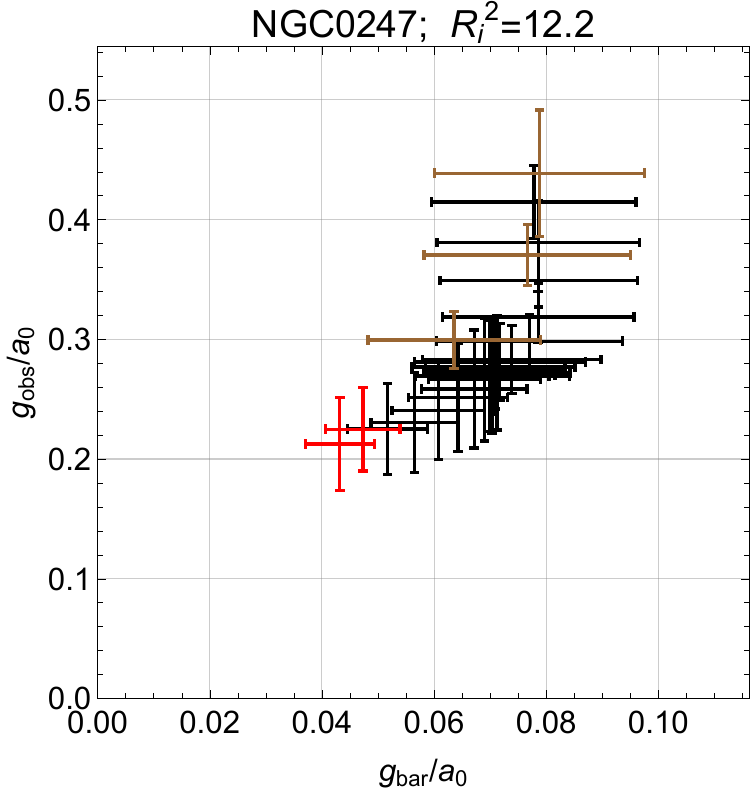}	
	\includegraphics[width=0.2\textwidth]{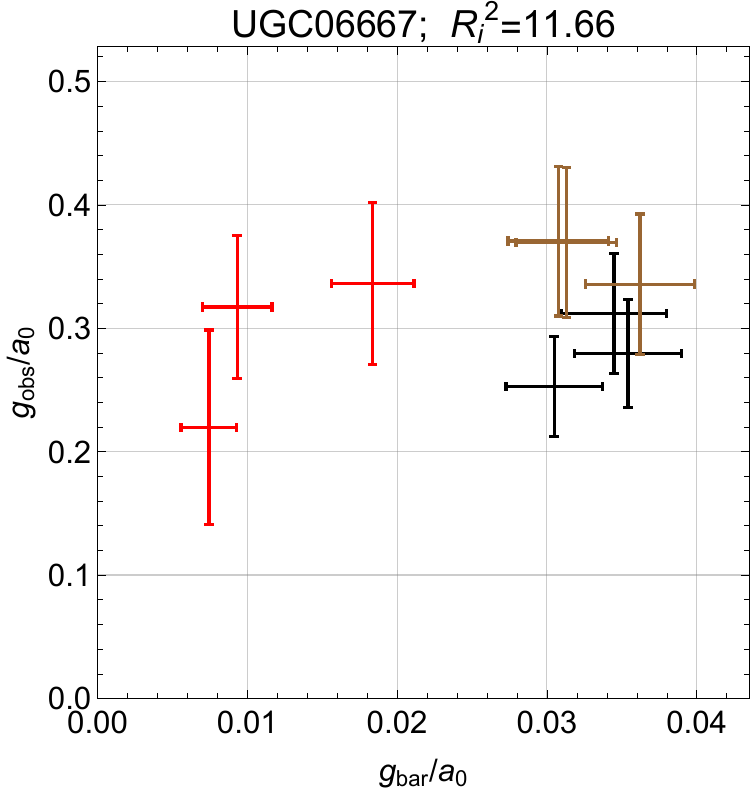}
	\includegraphics[width=0.2\textwidth]{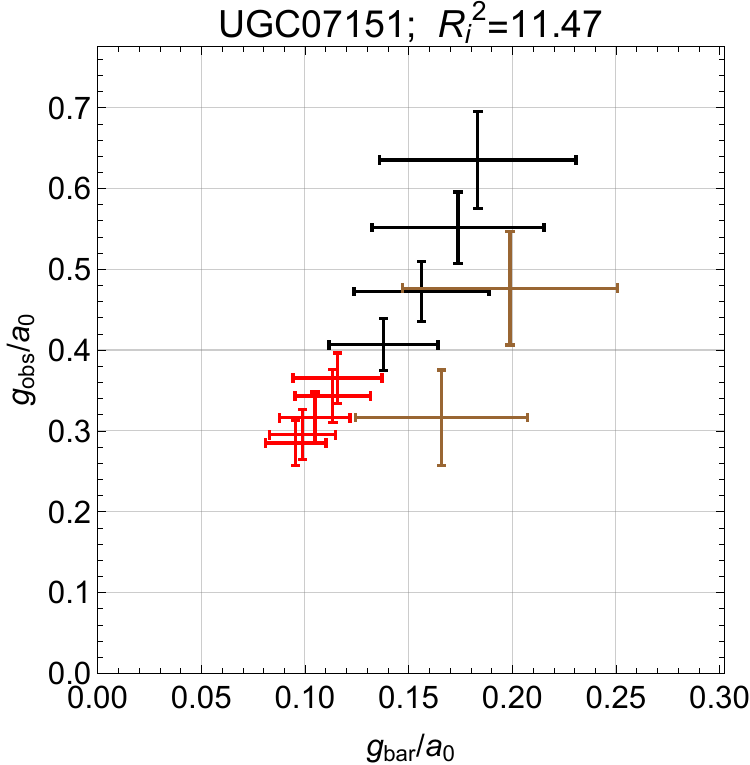}
	\includegraphics[width=0.2\textwidth]{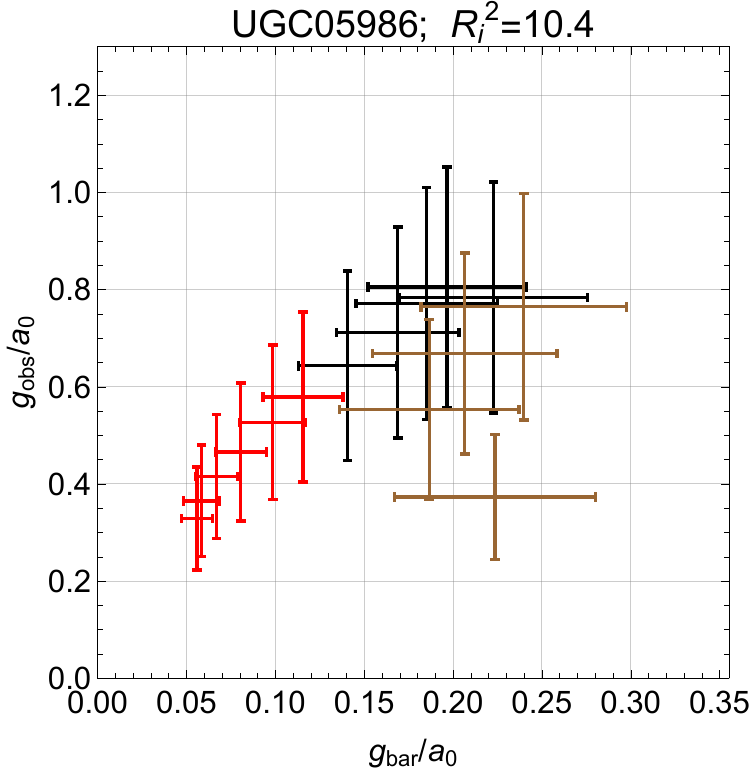}
	\includegraphics[width=0.2\textwidth]{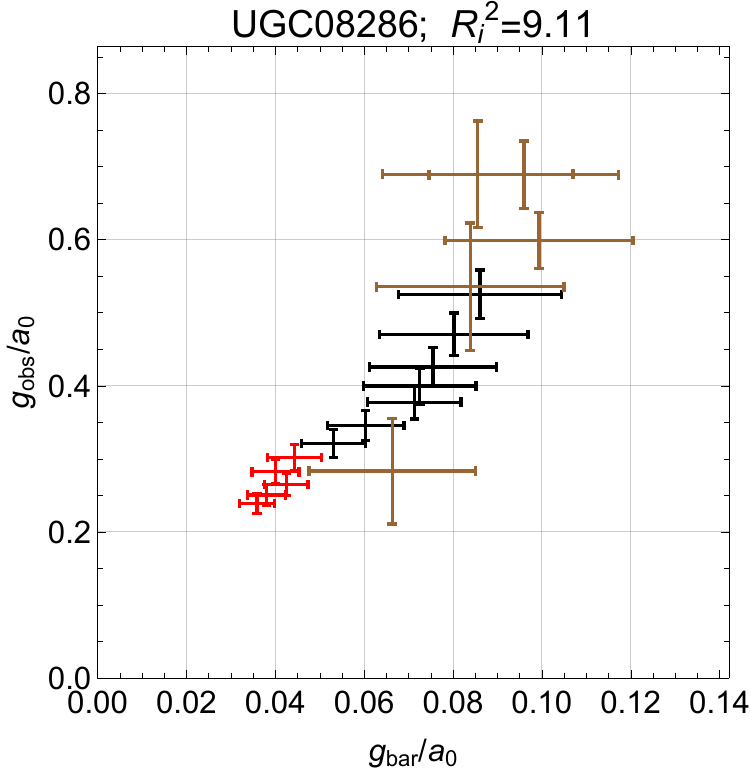}
	\includegraphics[width=0.2\textwidth]{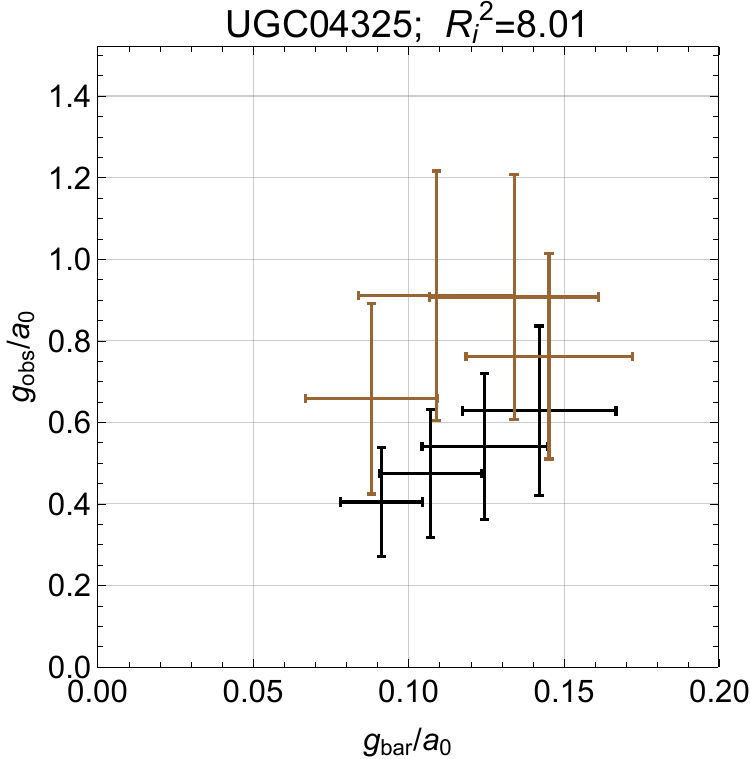}
	\includegraphics[width=0.2\textwidth]{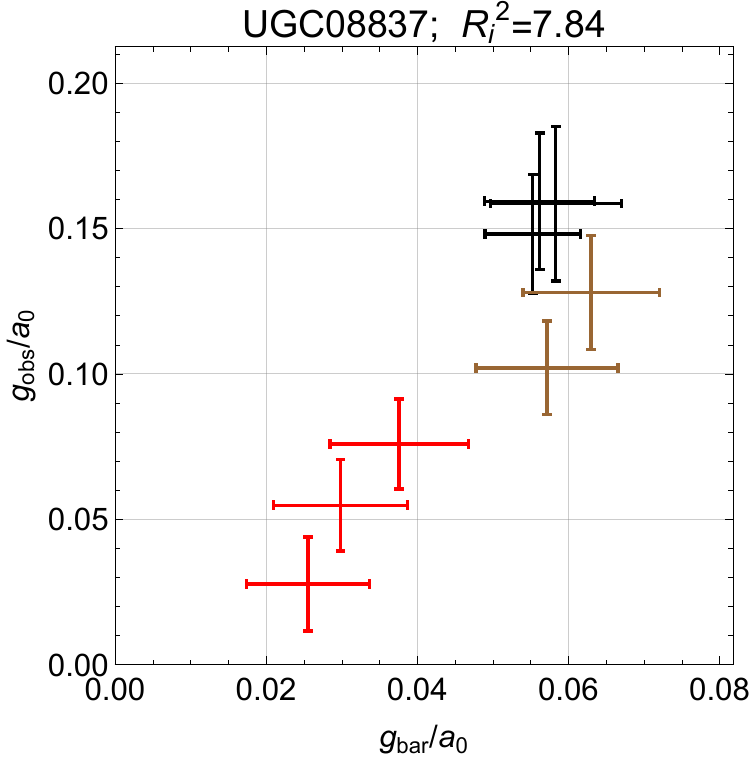}
	\includegraphics[width=0.2\textwidth]{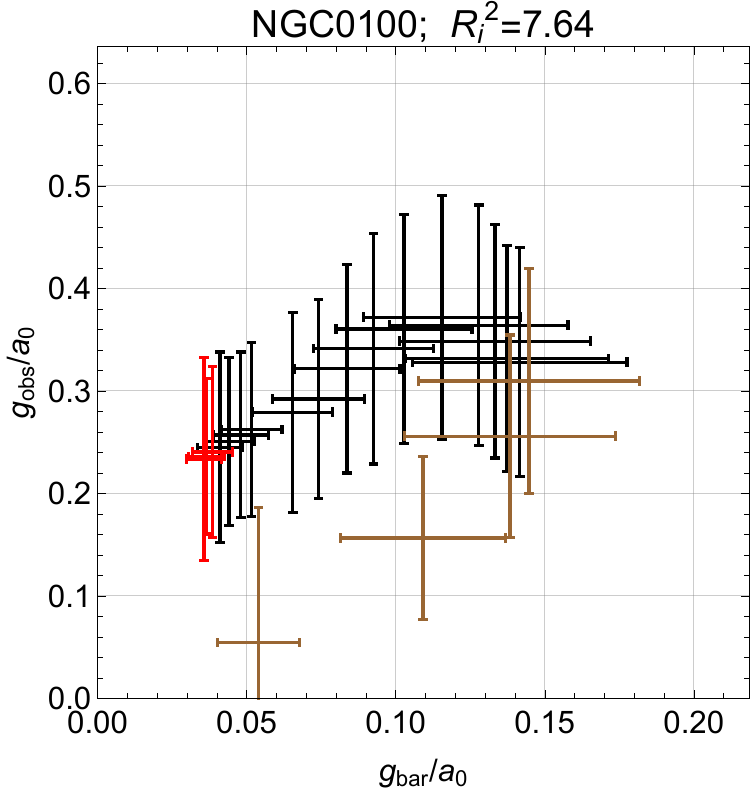}
	\includegraphics[width=0.2\textwidth]{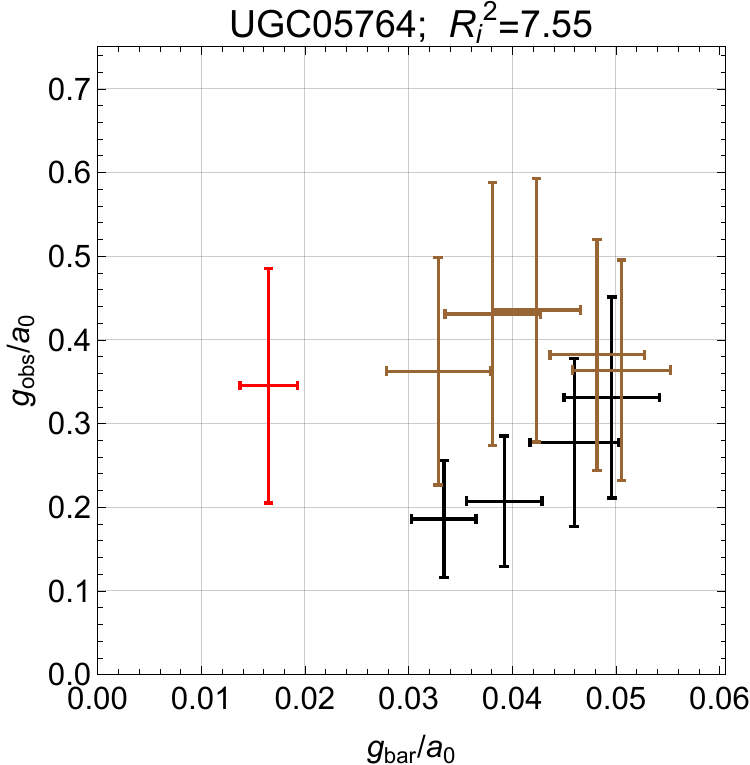}
	\includegraphics[width=0.2\textwidth]{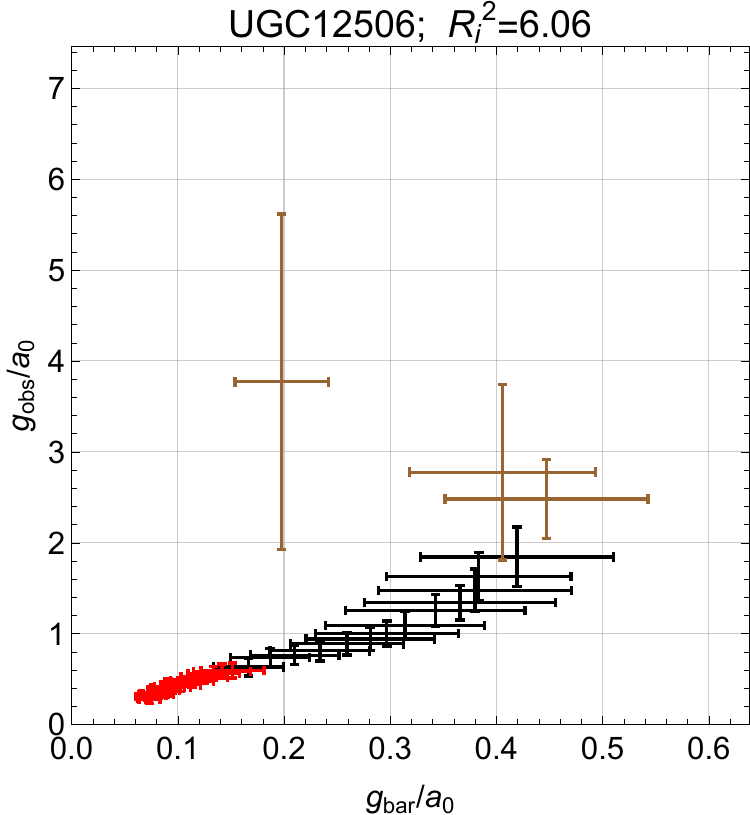}
	\includegraphics[width=0.2\textwidth]{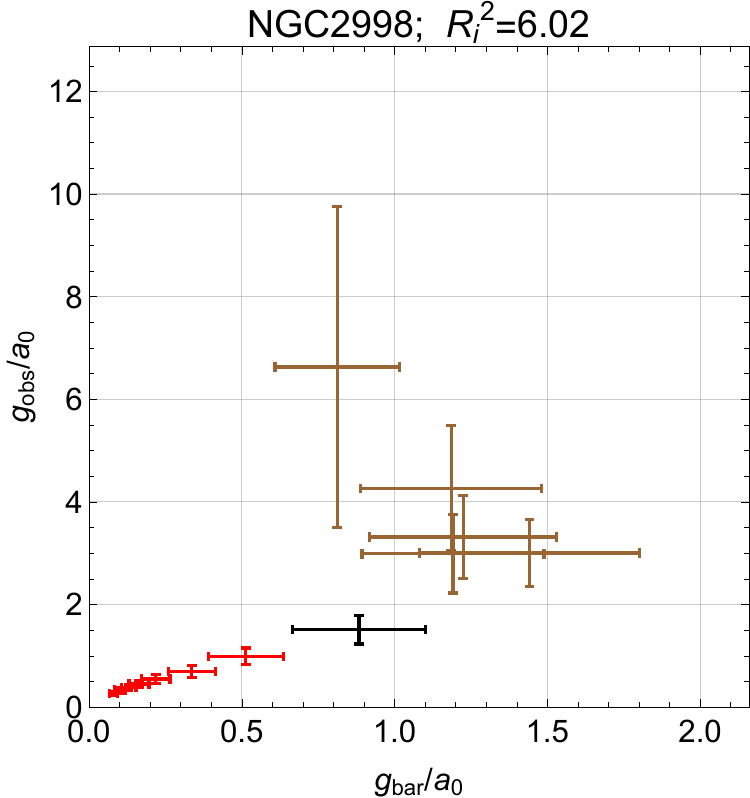}
	\includegraphics[width=0.2\textwidth]{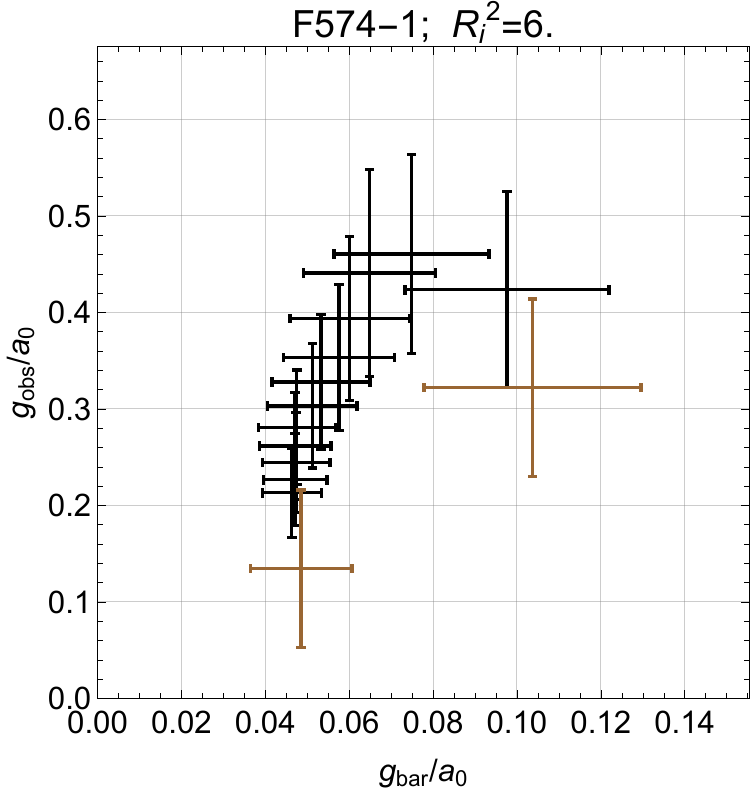}
	\includegraphics[width=0.2\textwidth]{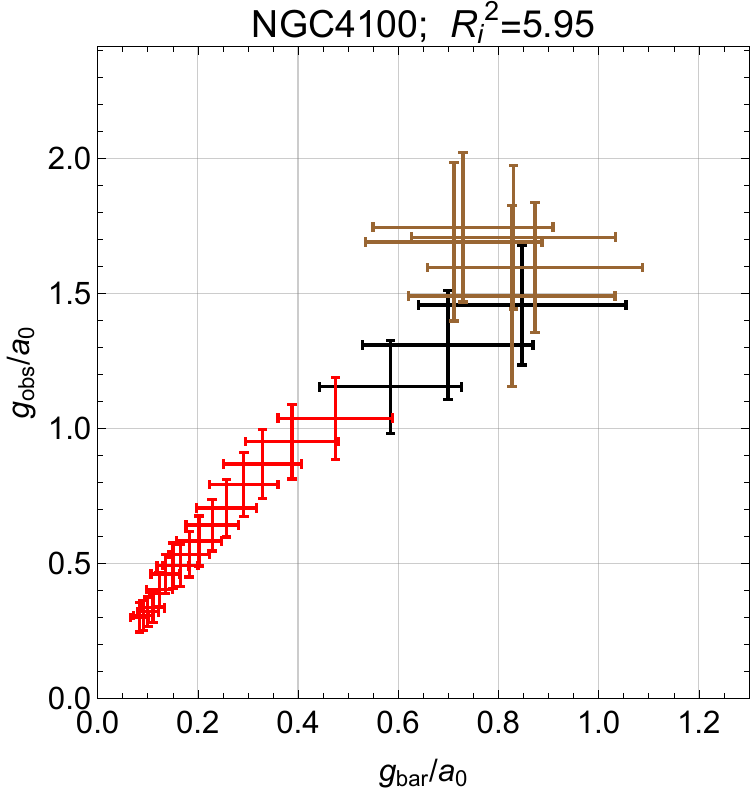}
	\includegraphics[width=0.2\textwidth]{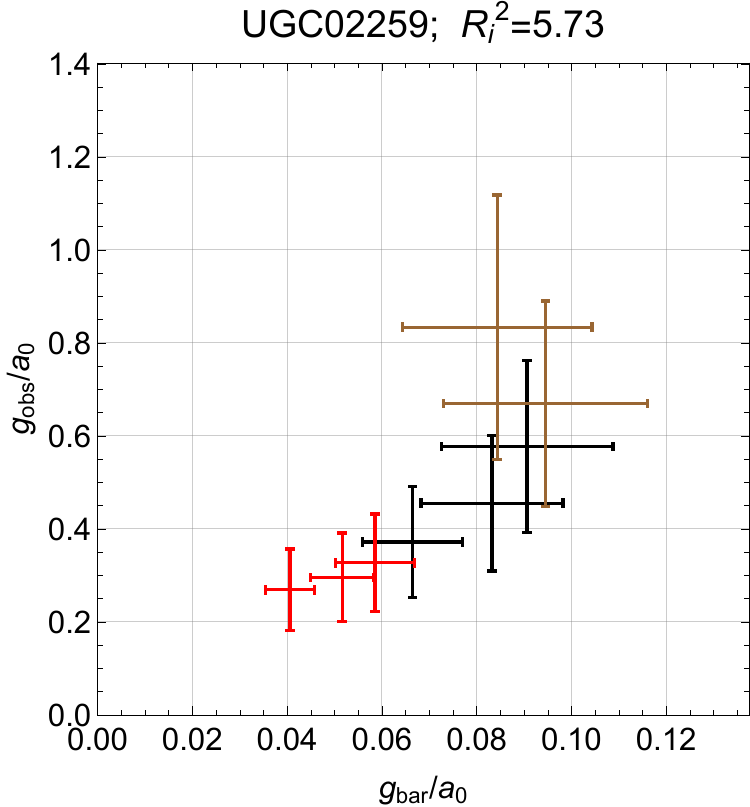}
	\includegraphics[width=0.2\textwidth]{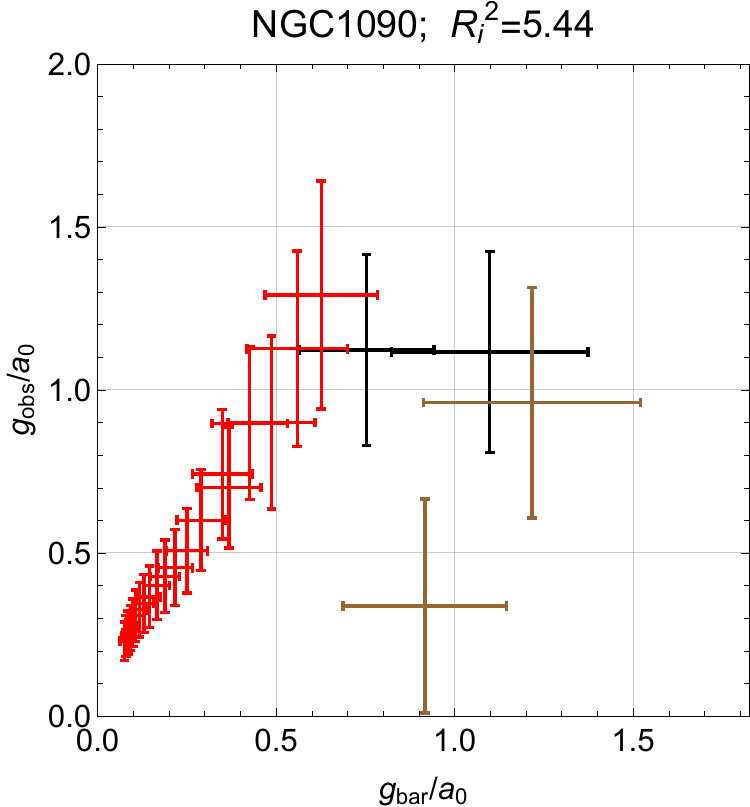}
	\includegraphics[width=0.2\textwidth]{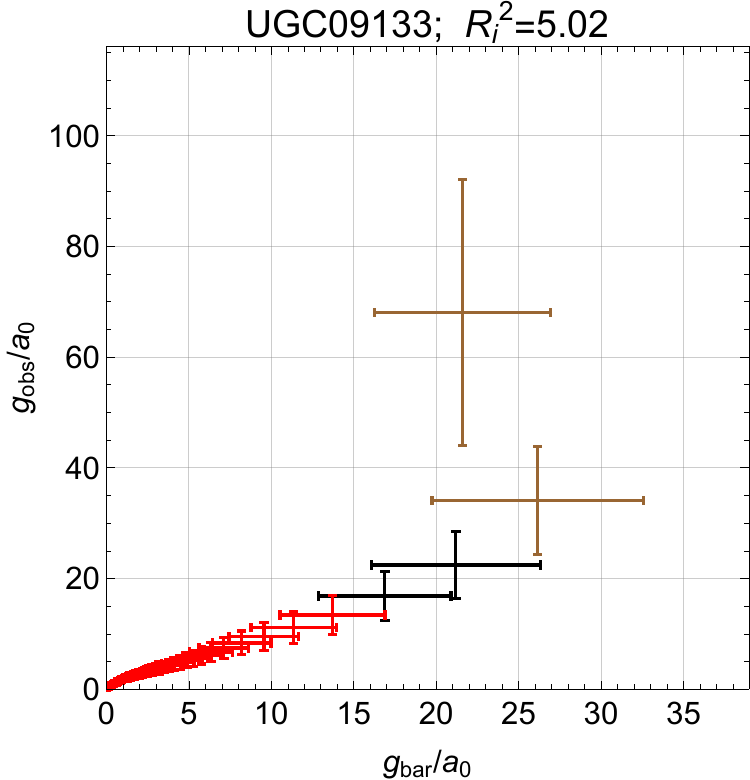}
	\includegraphics[width=0.2\textwidth]{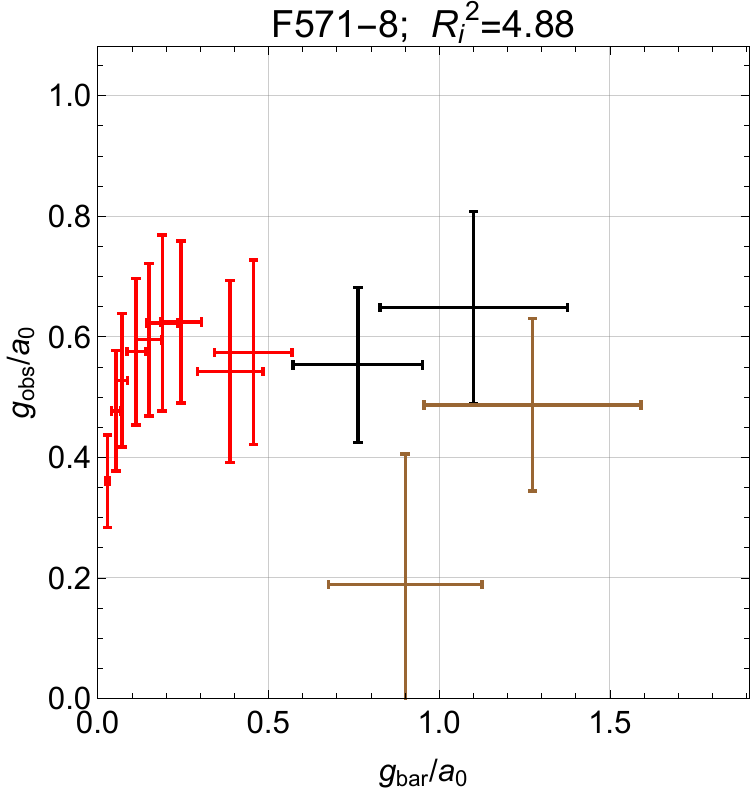}
	\includegraphics[width=0.2\textwidth]{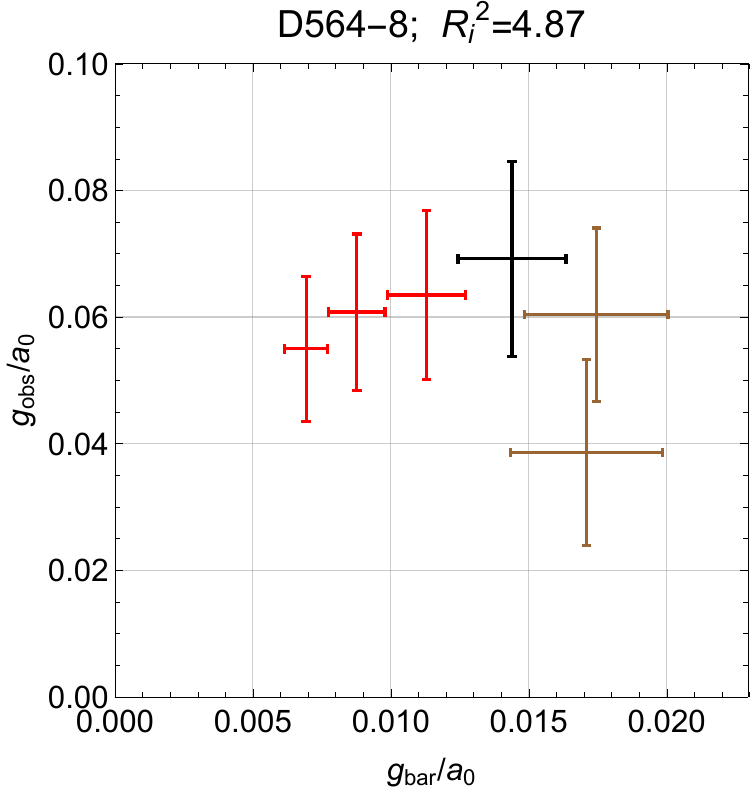}
	\caption{$g2$-space plots corresponding to those of figure \ref{fig:p2}.}
	\label{fig:p3}
\end{figure*}

\begin{figure*}
	\centering
	\includegraphics[width=0.2\textwidth]{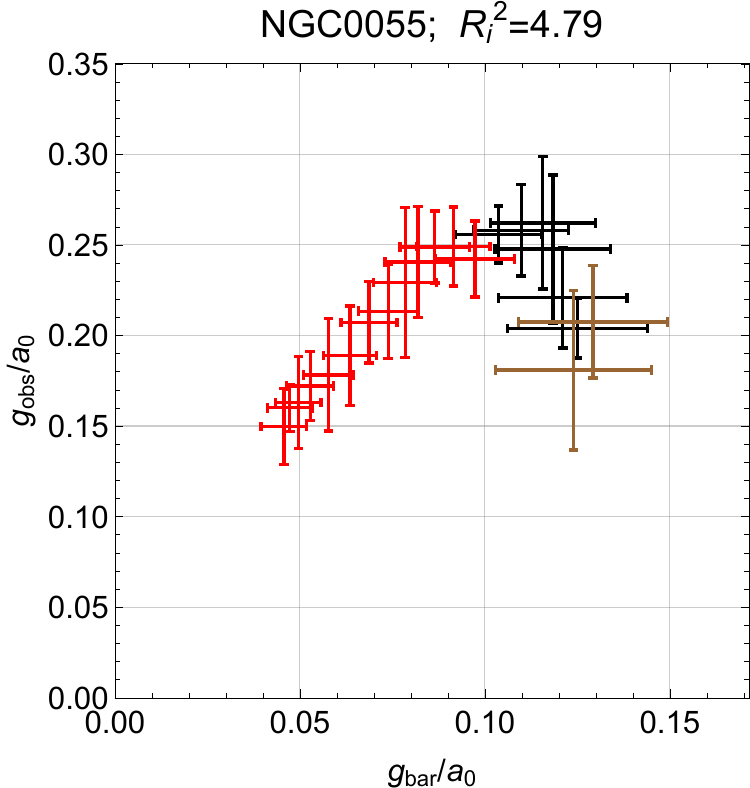}
	\includegraphics[width=0.2\textwidth]{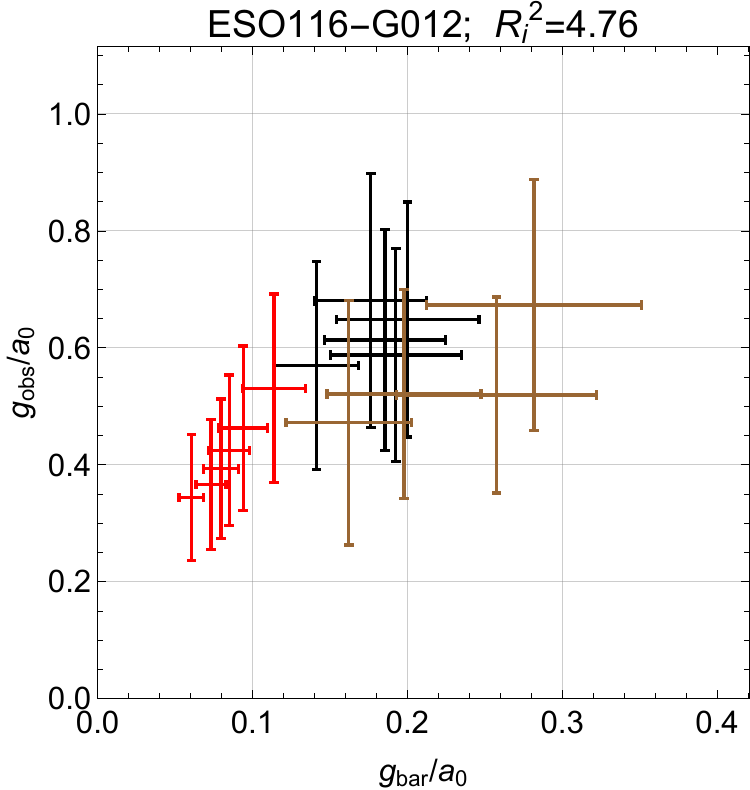}
	\includegraphics[width=0.2\textwidth]{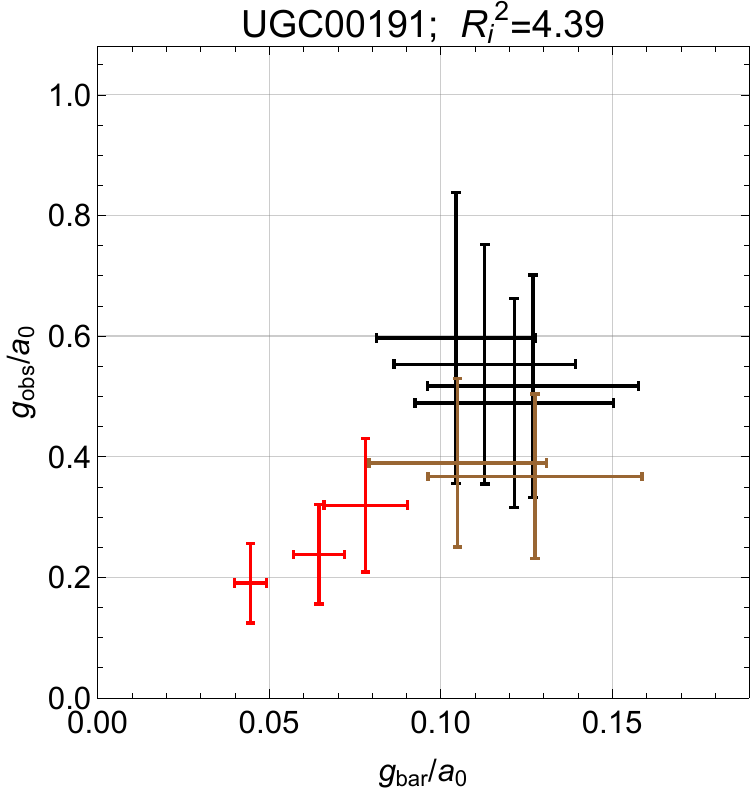}
	\includegraphics[width=0.2\textwidth]{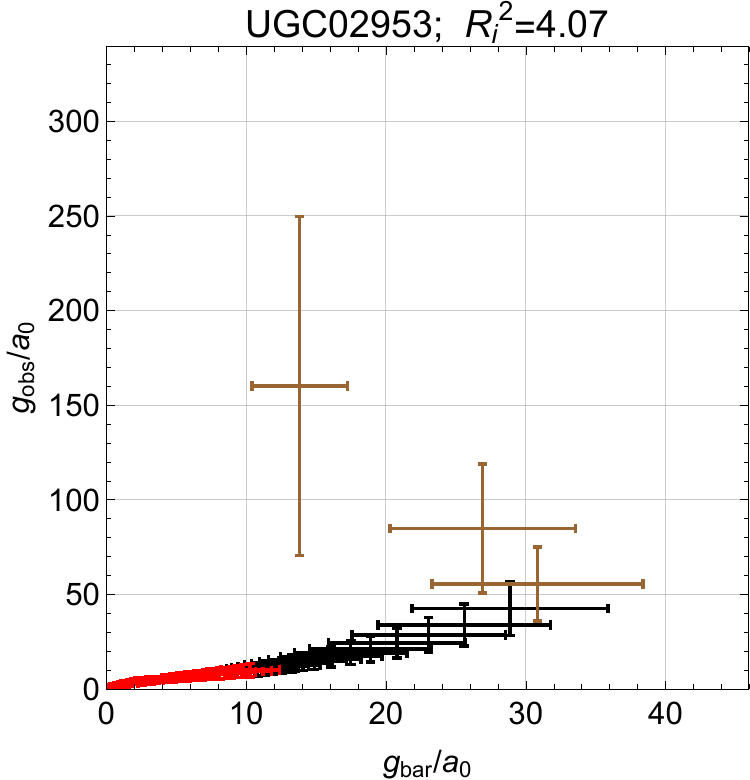}
	\includegraphics[width=0.2\textwidth]{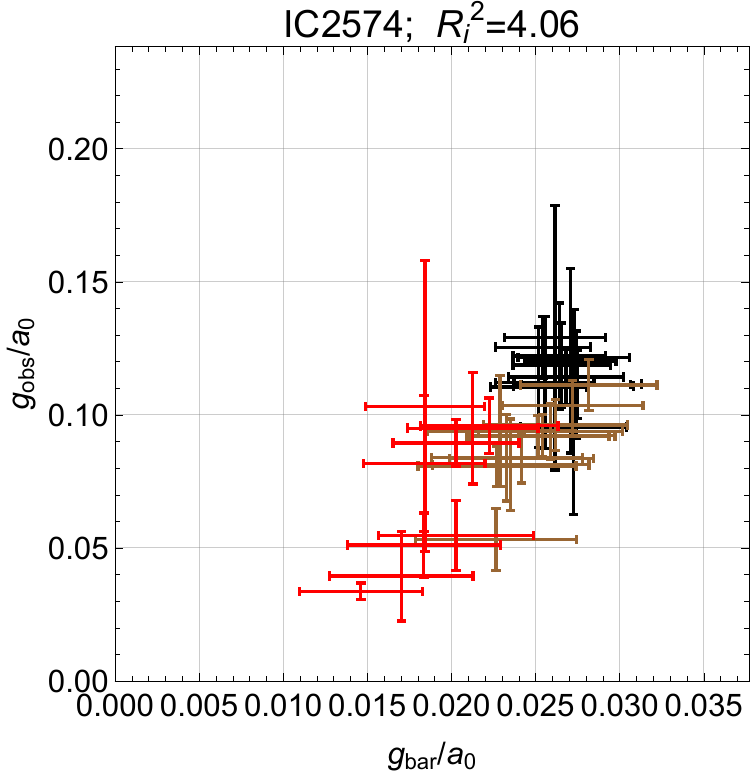}
	\includegraphics[width=0.2\textwidth]{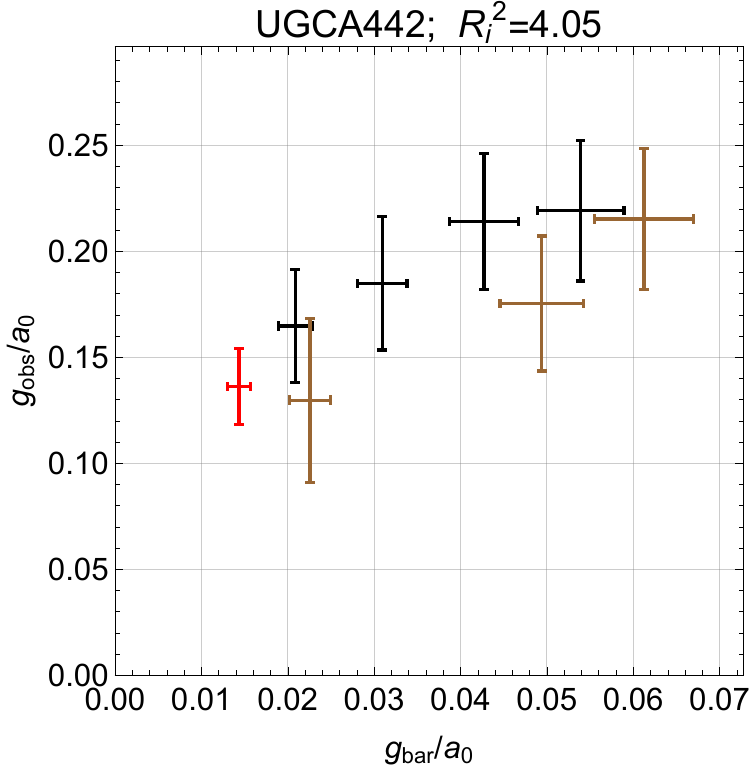}
	\includegraphics[width=0.2\textwidth]{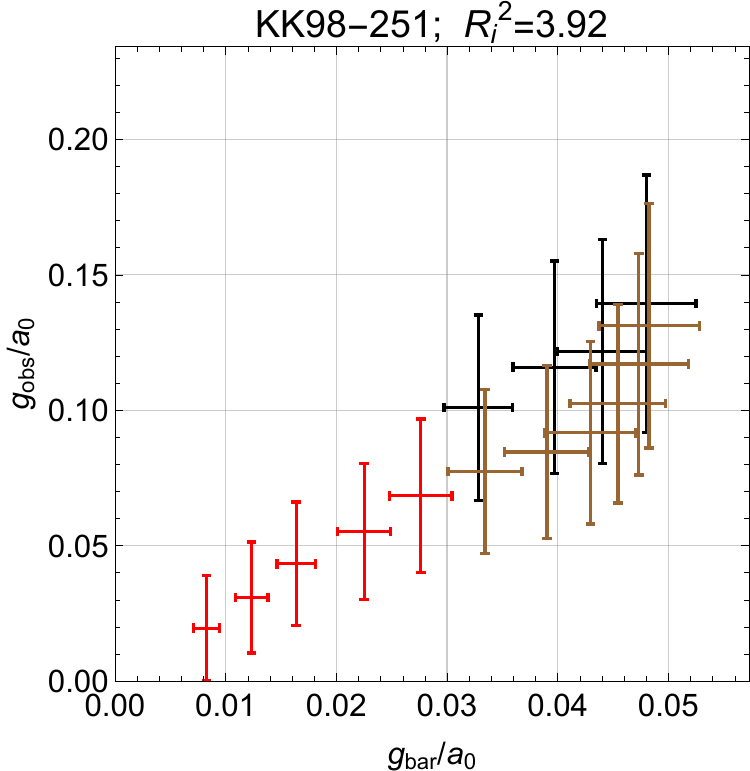}
	\includegraphics[width=0.2\textwidth]{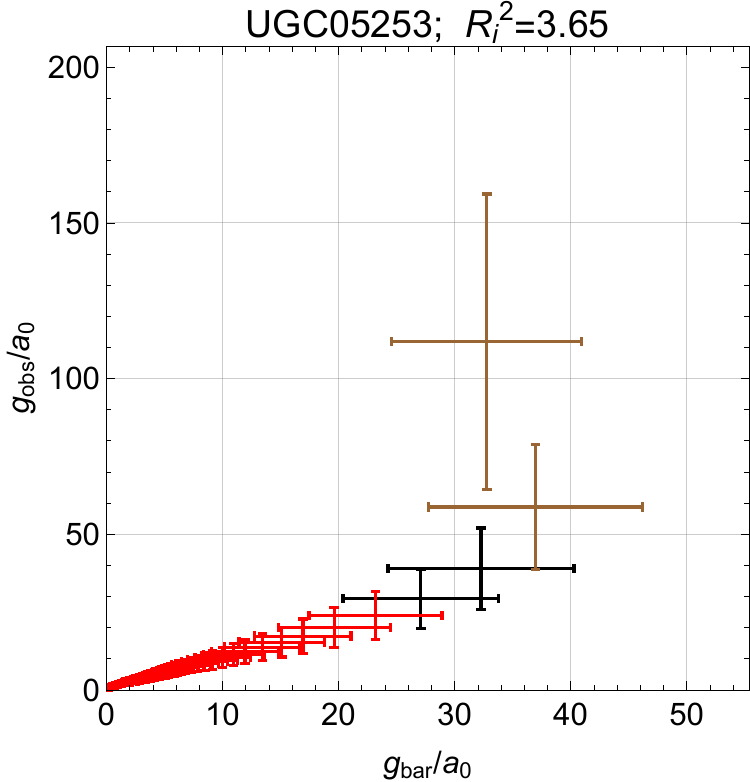}
	\includegraphics[width=0.2\textwidth]{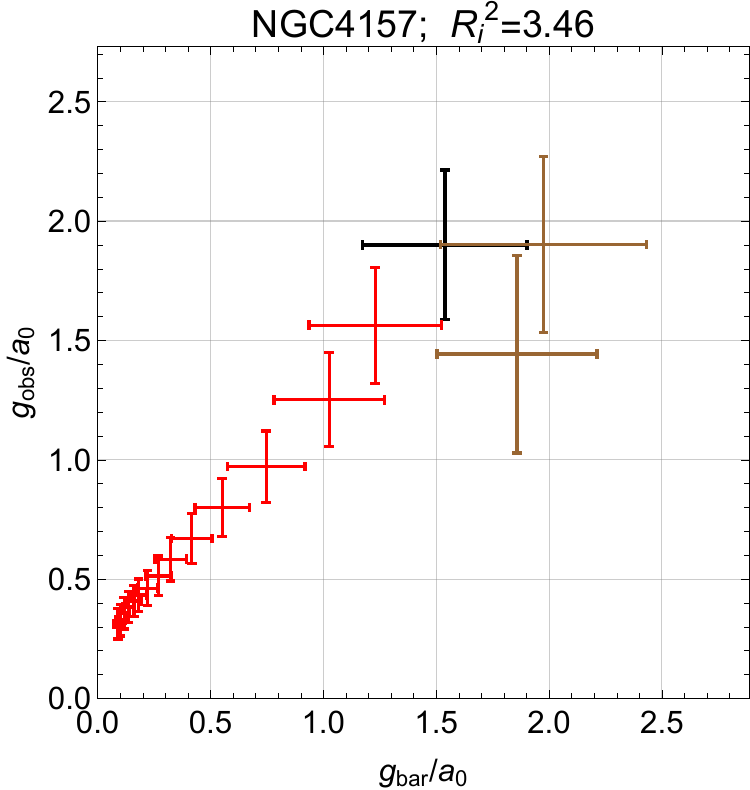}
	\includegraphics[width=0.2\textwidth]{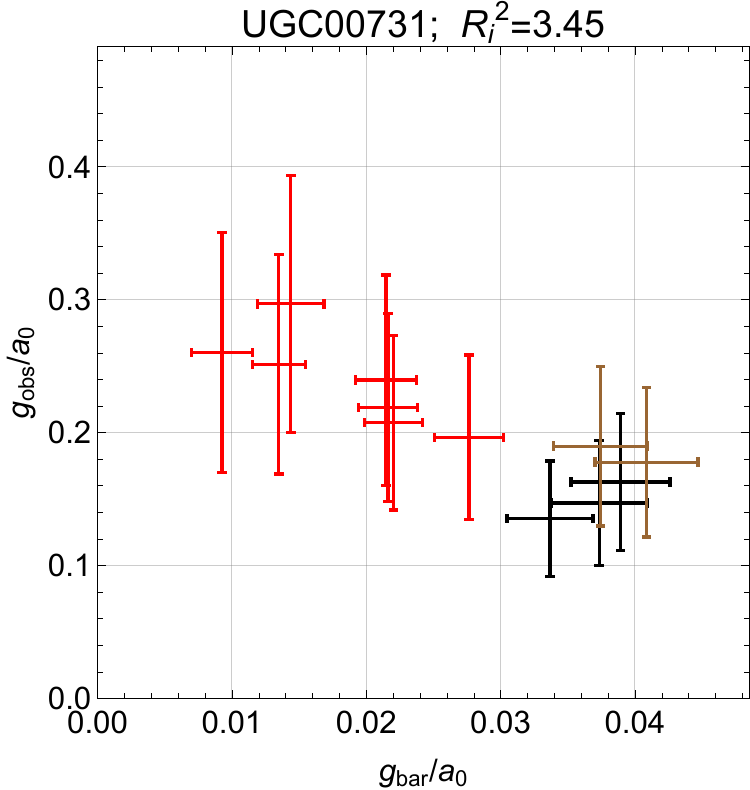}
	\includegraphics[width=0.2\textwidth]{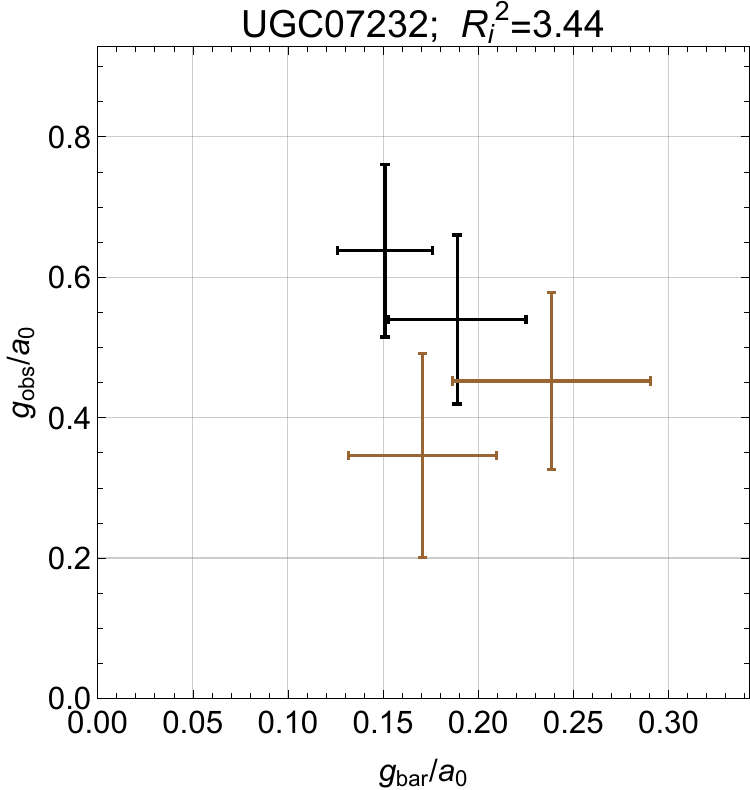}
	\includegraphics[width=0.2\textwidth]{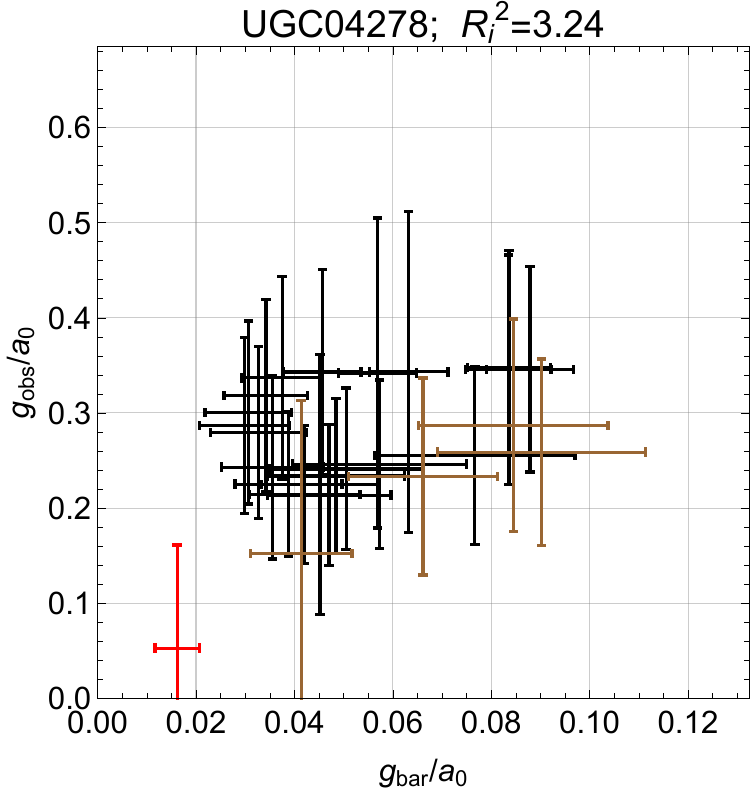}
	\includegraphics[width=0.2\textwidth]{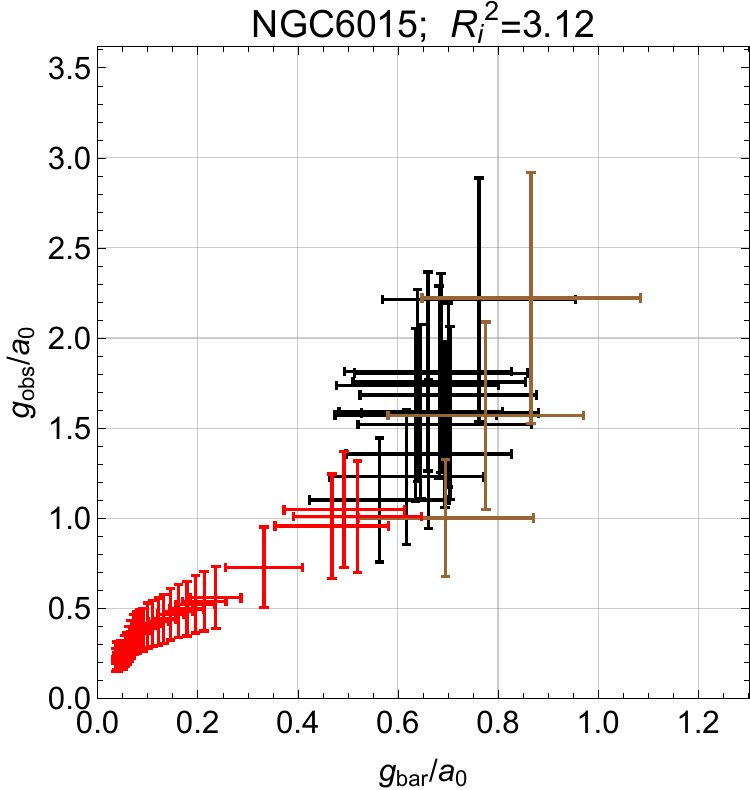}
	\includegraphics[width=0.2\textwidth]{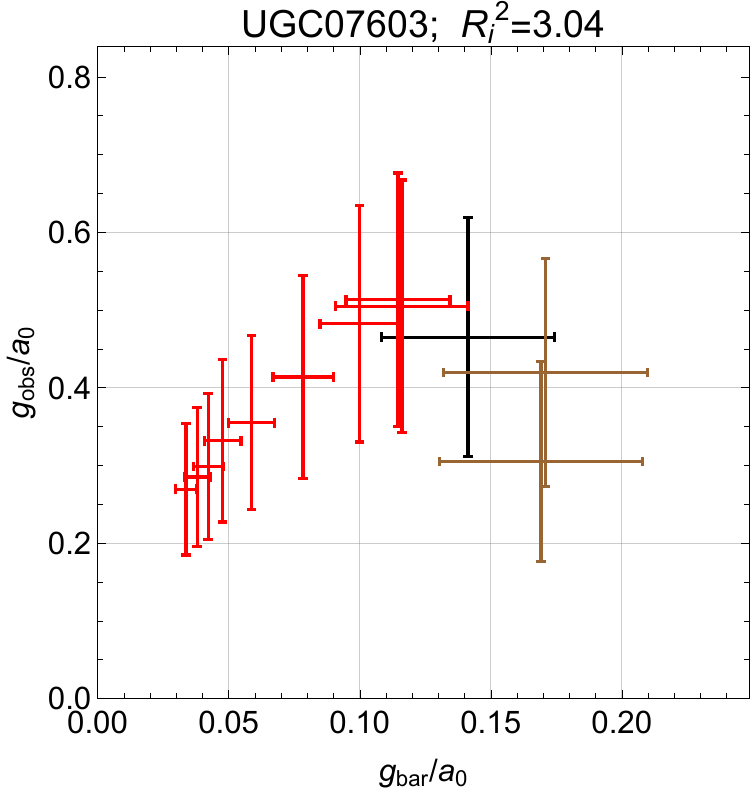}
	\includegraphics[width=0.2\textwidth]{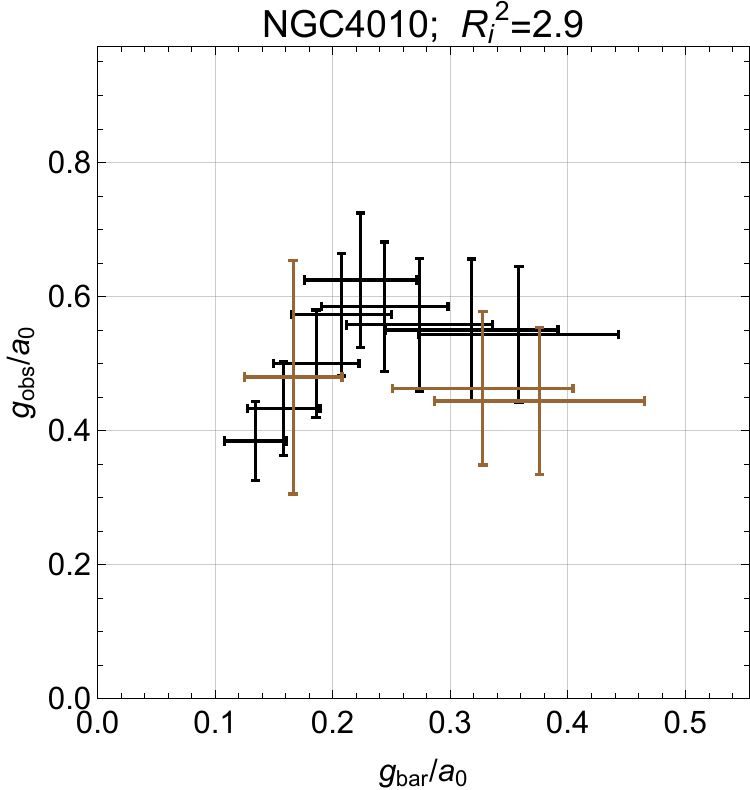}
	\includegraphics[width=0.2\textwidth]{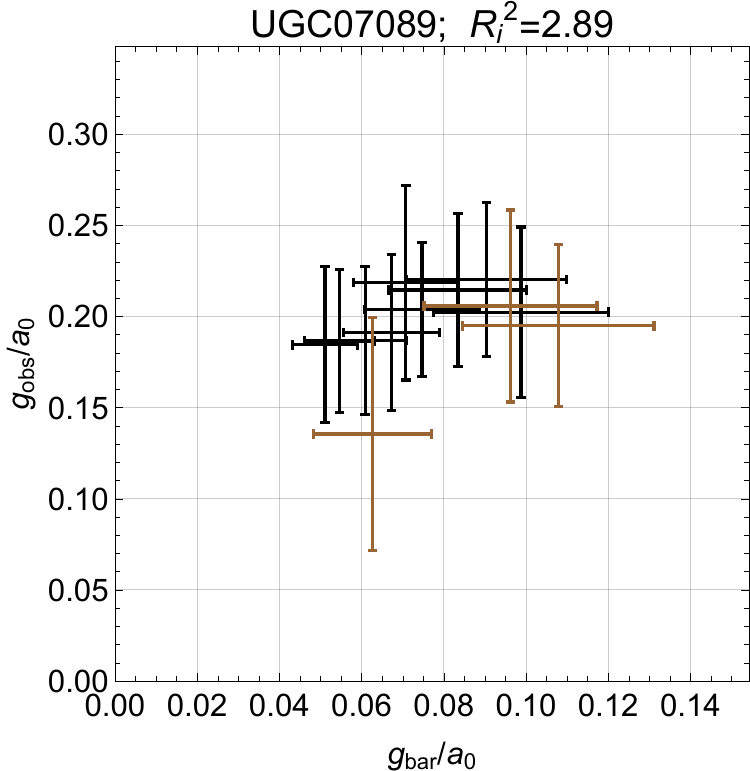}
	\includegraphics[width=0.2\textwidth]{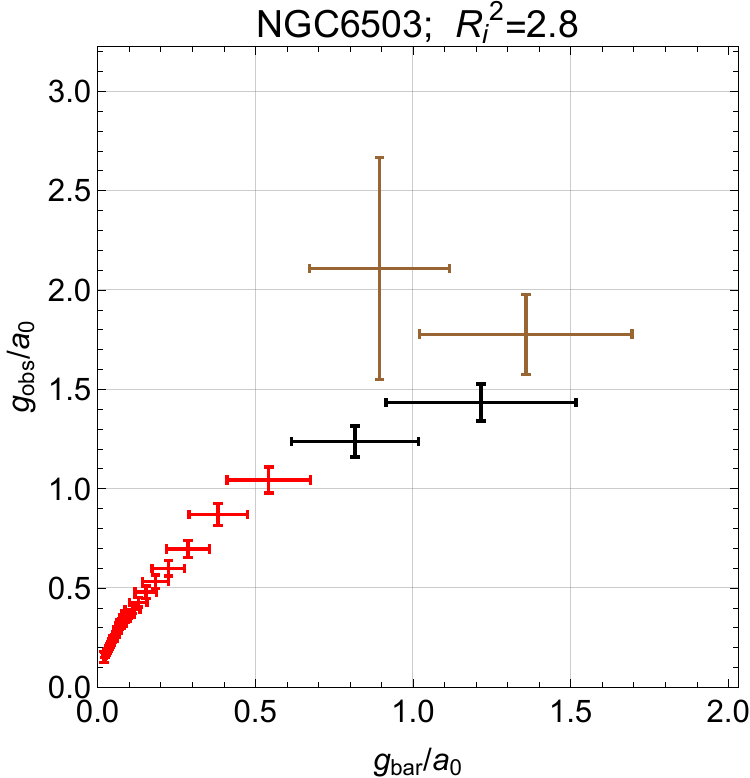}
	\includegraphics[width=0.2\textwidth]{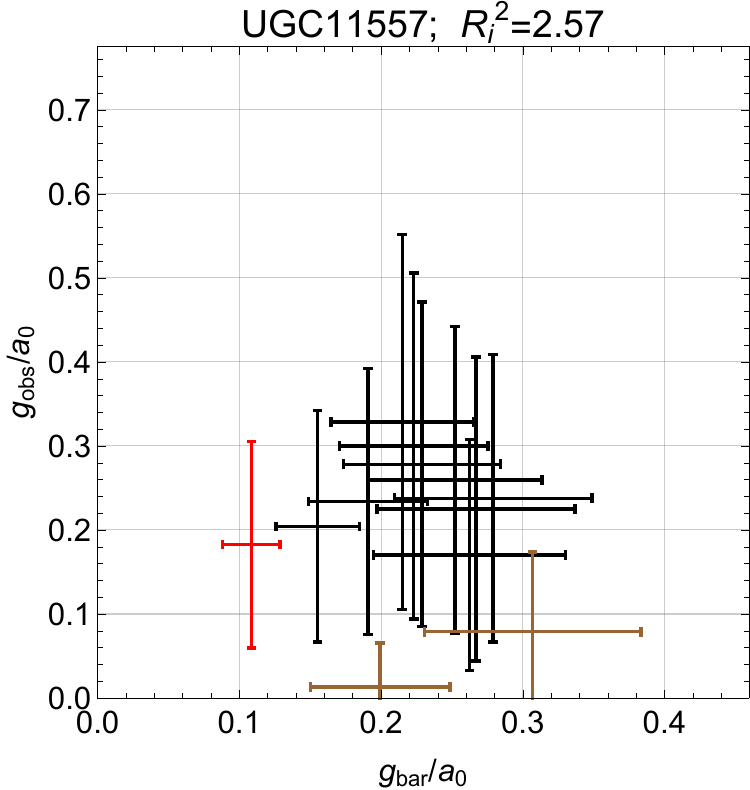}
	\includegraphics[width=0.2\textwidth]{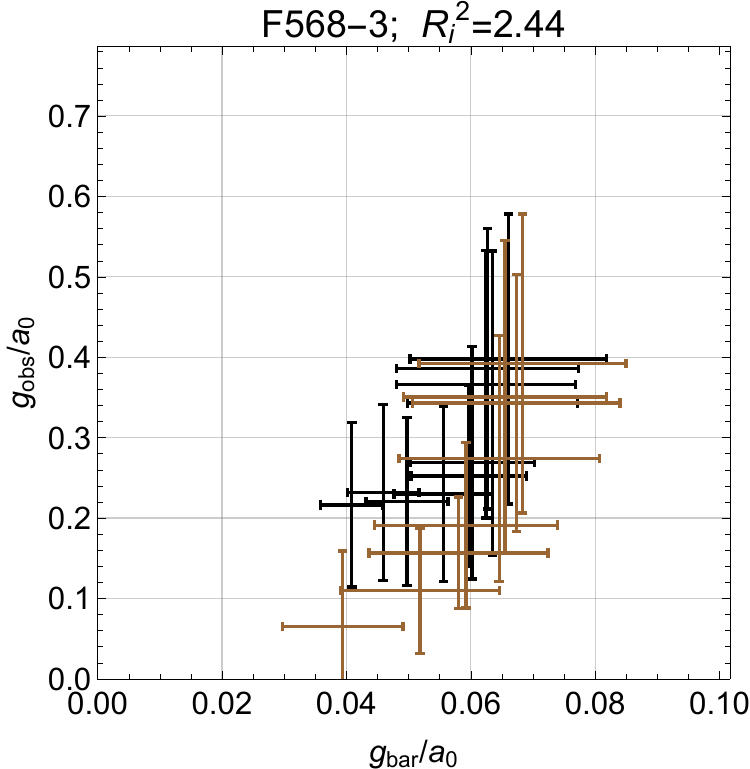}
	\includegraphics[width=0.2\textwidth]{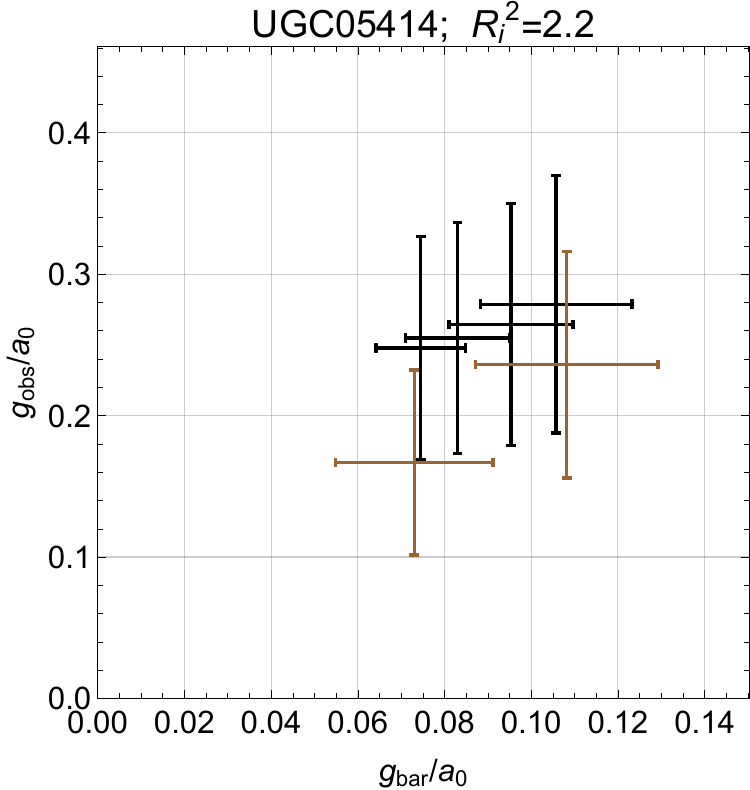}
	\includegraphics[width=0.2\textwidth]{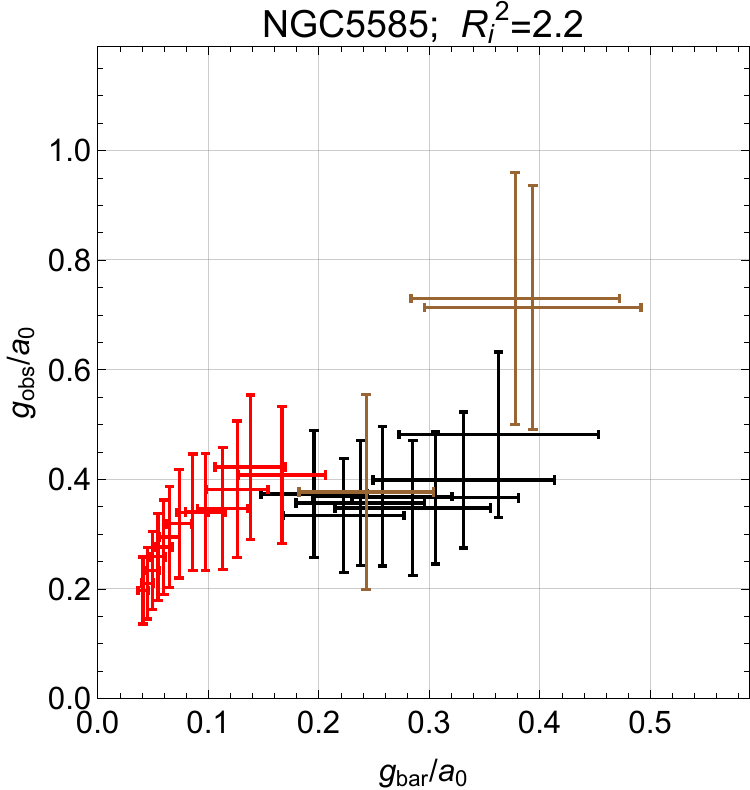}
	\includegraphics[width=0.2\textwidth]{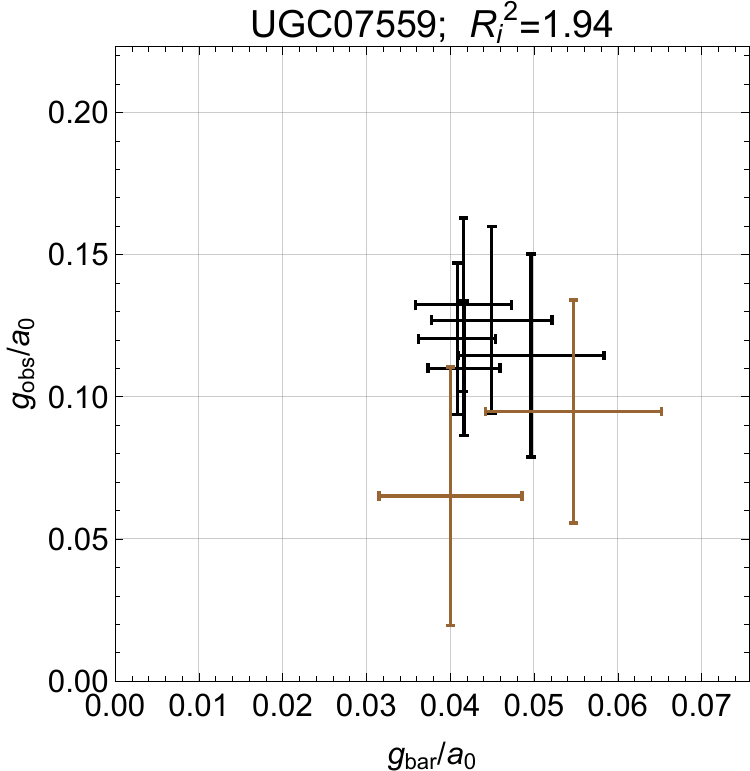}
	\includegraphics[width=0.2\textwidth]{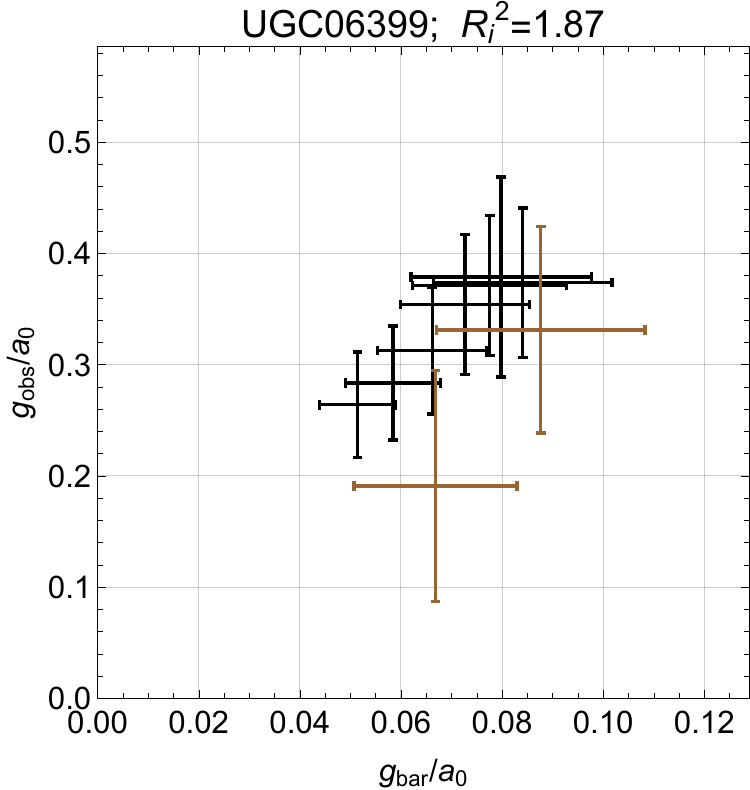}
	\includegraphics[width=0.2\textwidth]{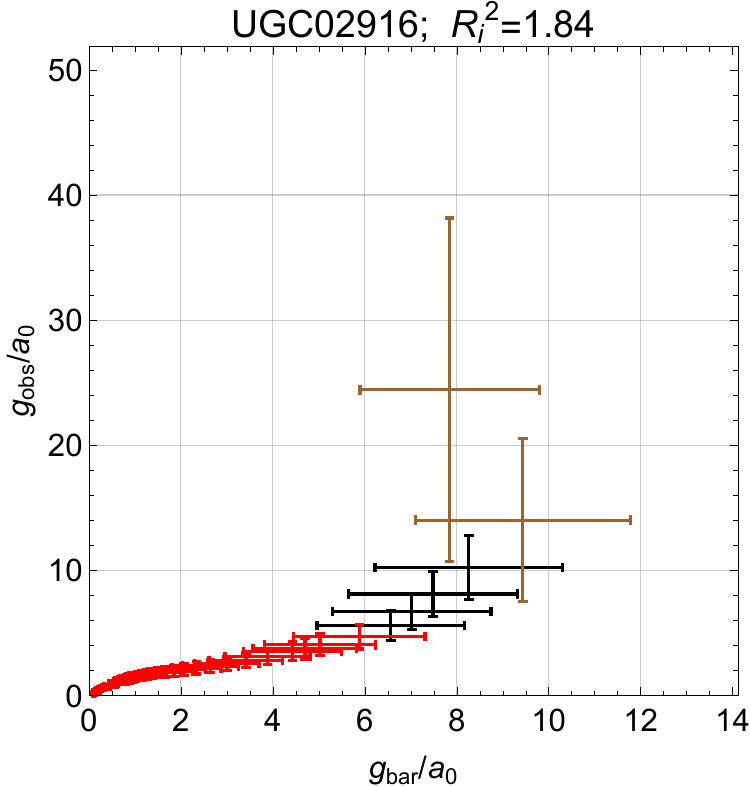}
	\caption{$g2$-space plots corresponding to those of figure \ref{fig:p2}.}
	\label{fig:p4}
\end{figure*}

\begin{figure*}
	\centering
	\includegraphics[width=0.2\textwidth]{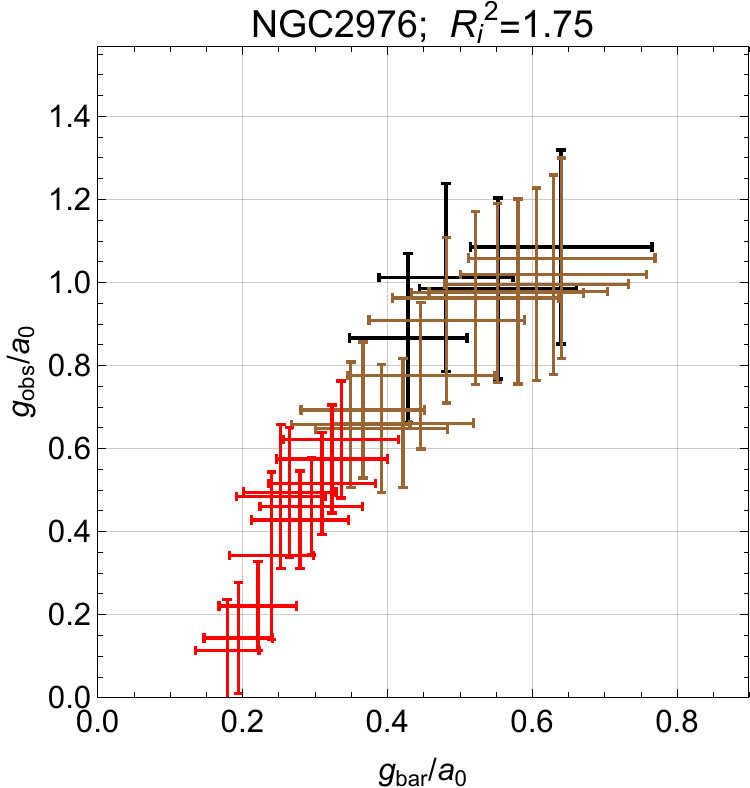}
	\includegraphics[width=0.2\textwidth]{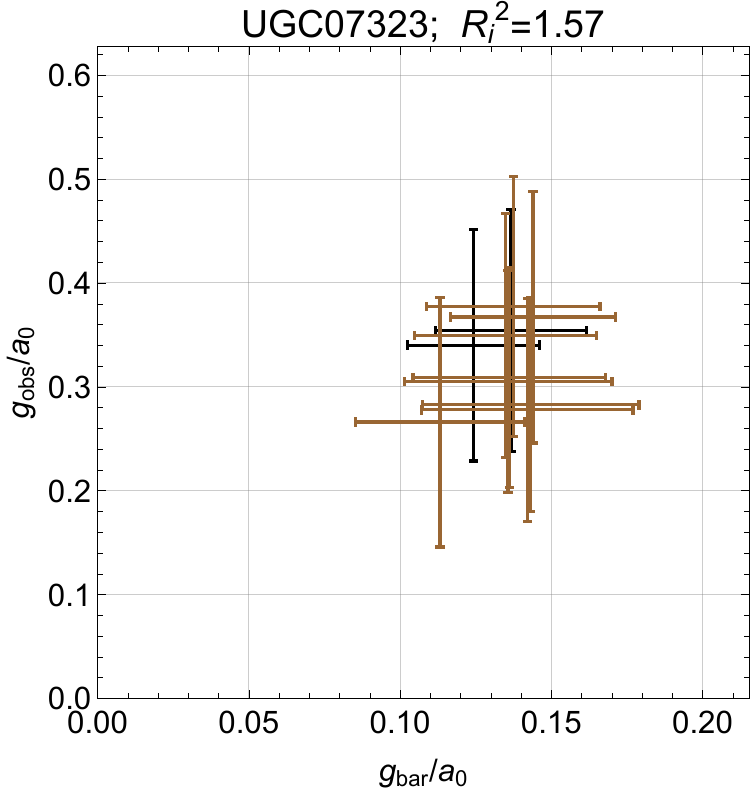}
	\includegraphics[width=0.2\textwidth]{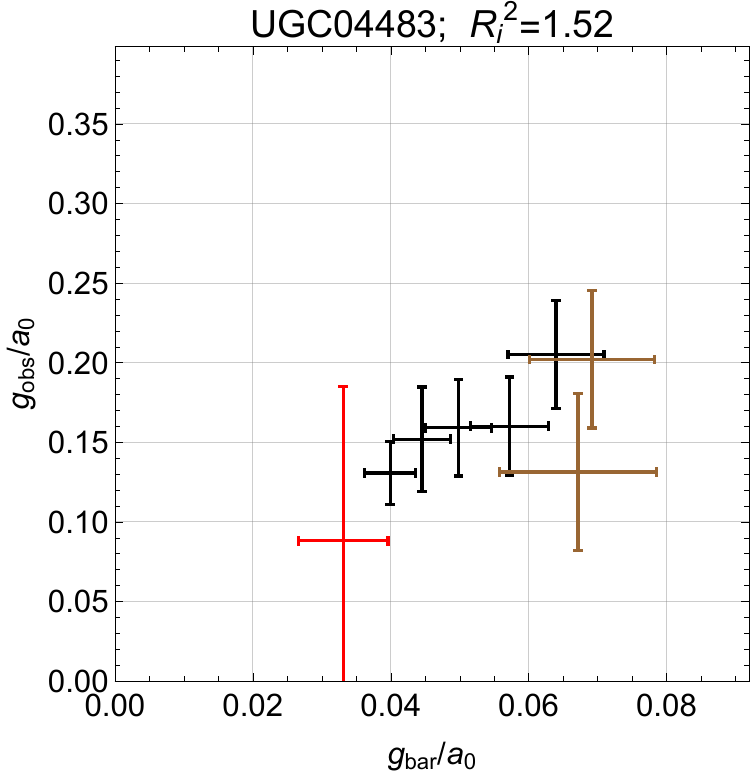}
	\includegraphics[width=0.2\textwidth]{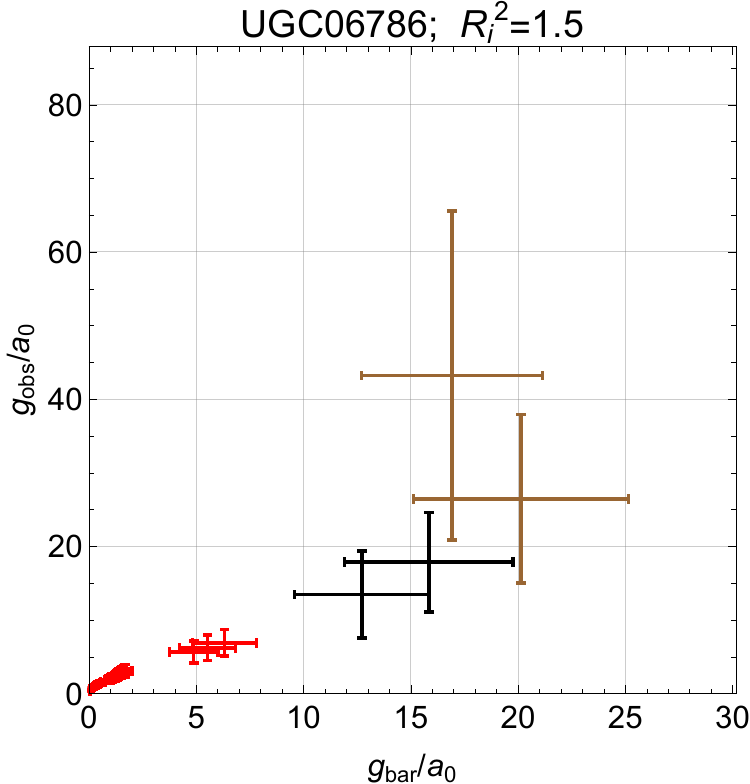}
	\includegraphics[width=0.2\textwidth]{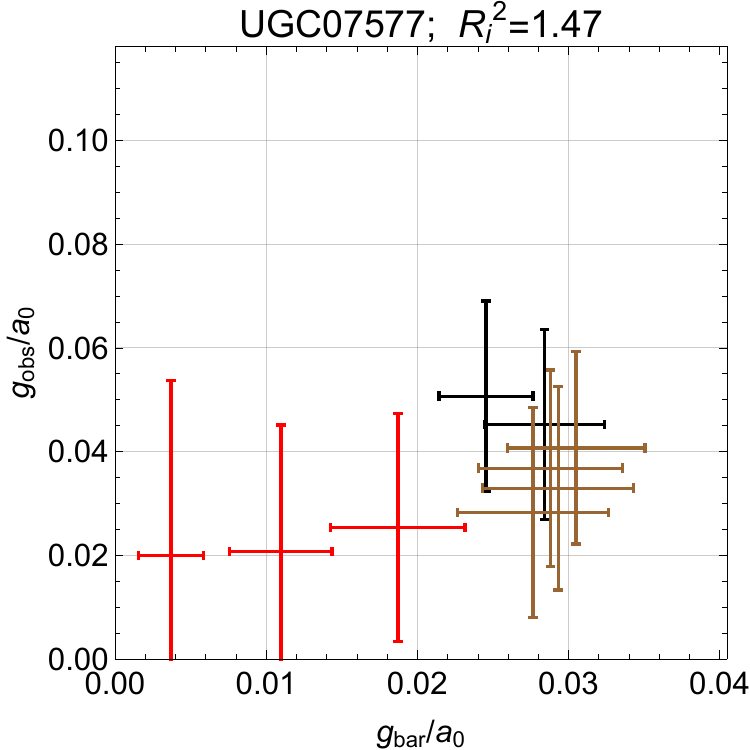}
	\includegraphics[width=0.2\textwidth]{p54}
	\includegraphics[width=0.2\textwidth]{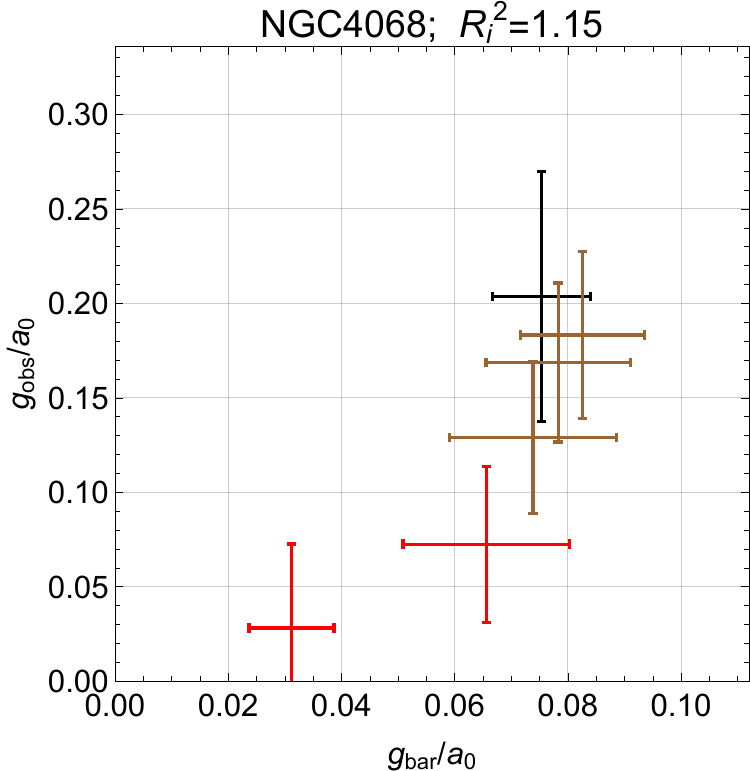}
	\includegraphics[width=0.2\textwidth]{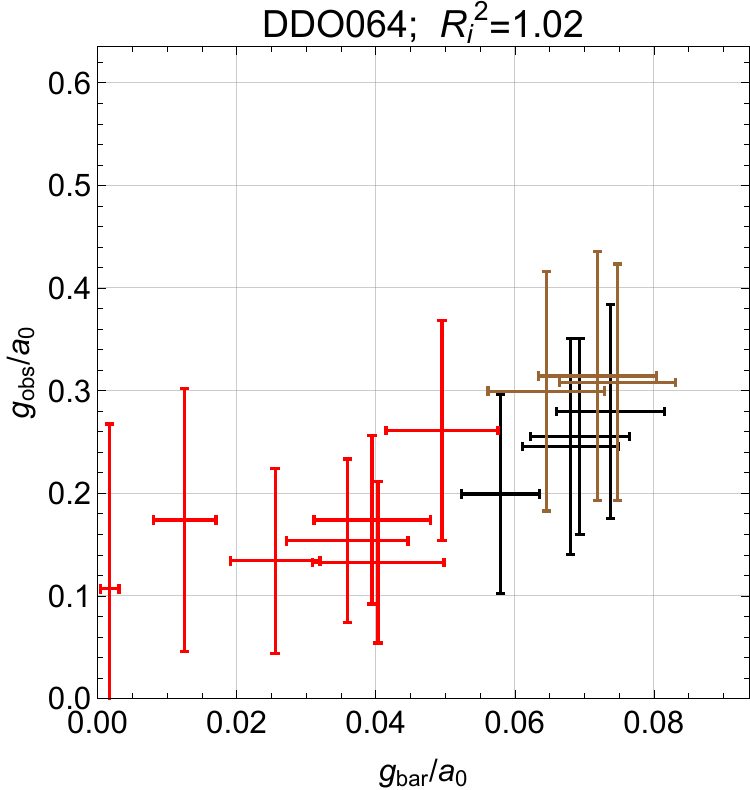}
	\includegraphics[width=0.2\textwidth]{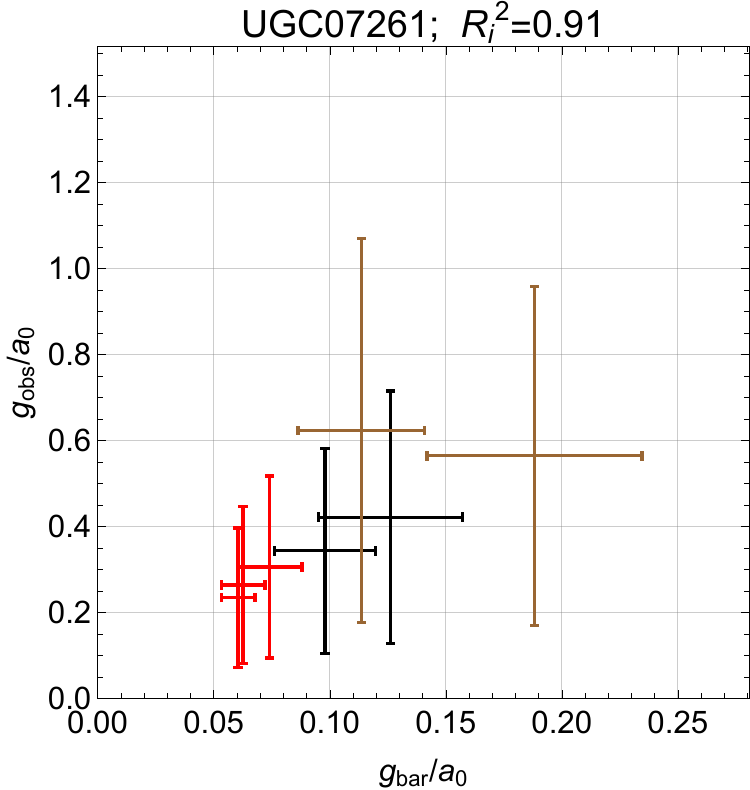}
	\includegraphics[width=0.2\textwidth]{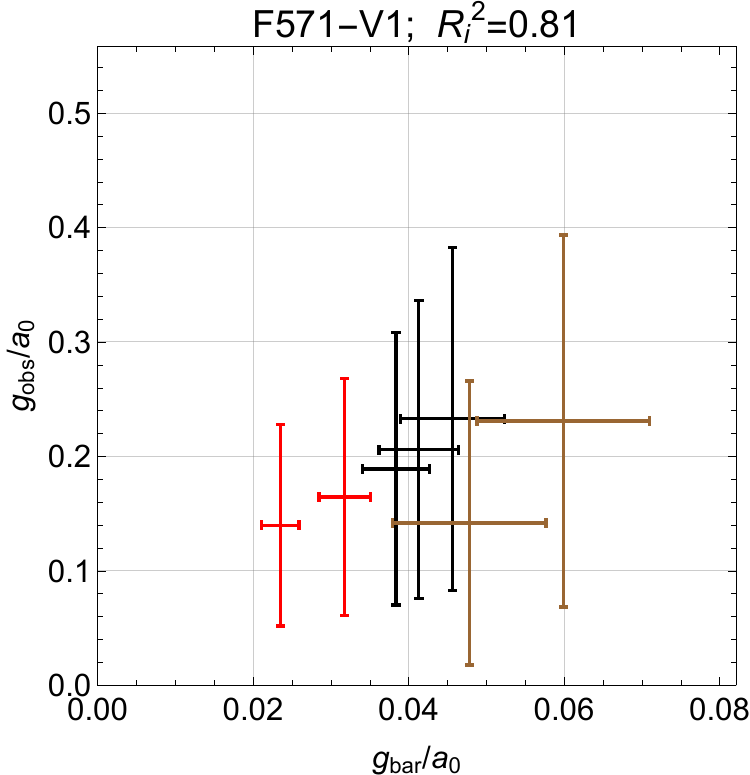}
	\includegraphics[width=0.2\textwidth]{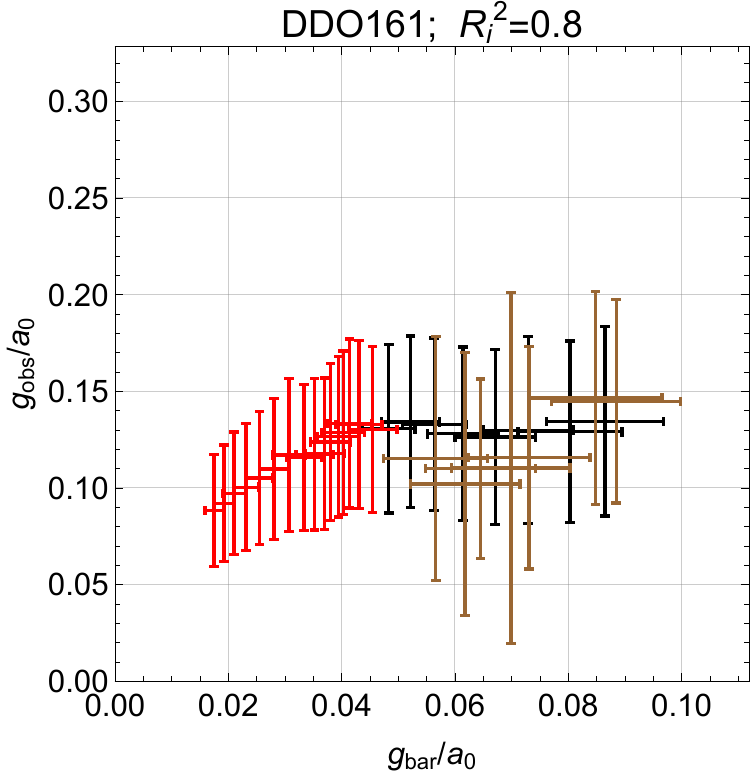}
	\includegraphics[width=0.2\textwidth]{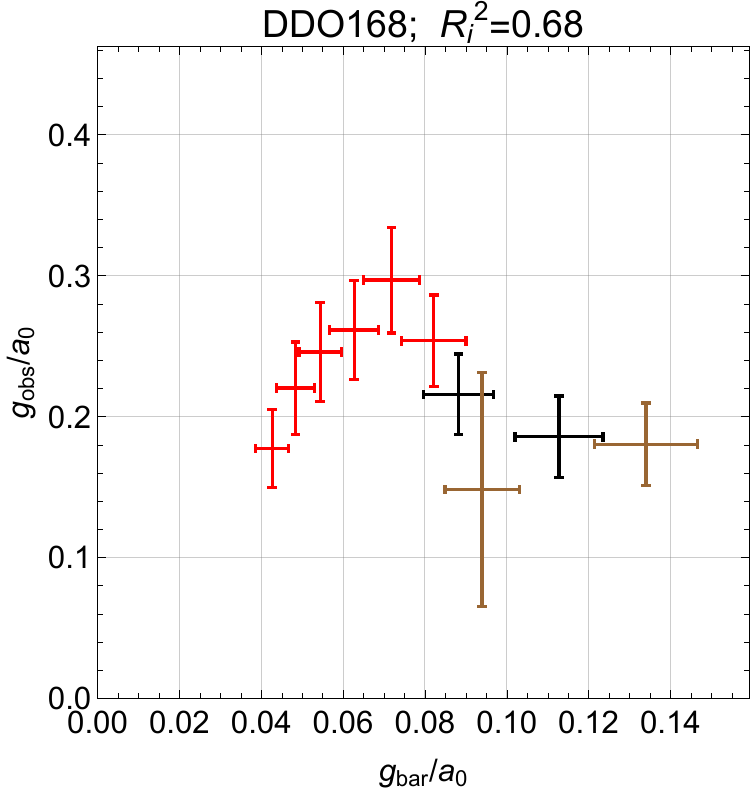}
	\includegraphics[width=0.2\textwidth]{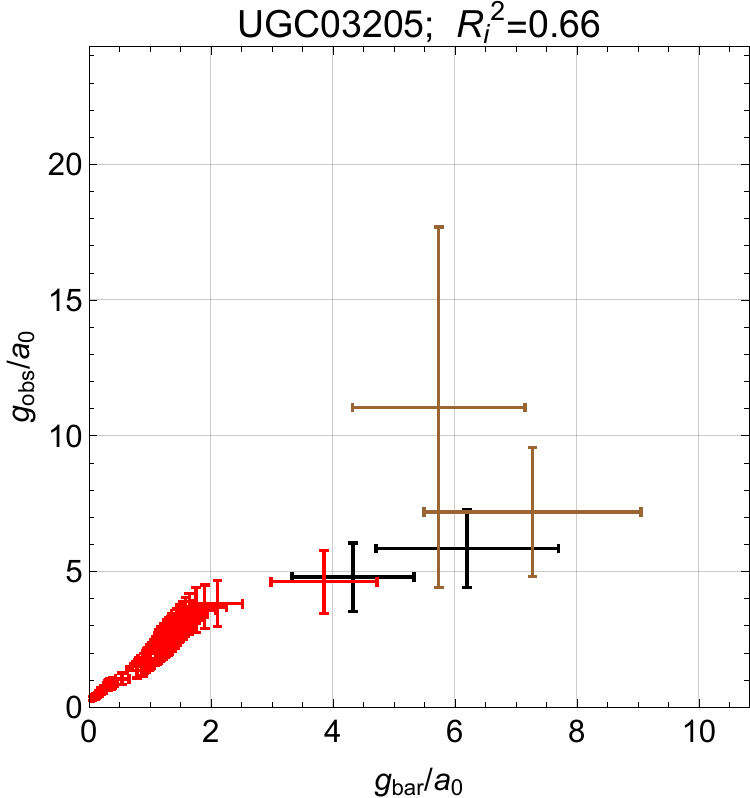}
	\includegraphics[width=0.2\textwidth]{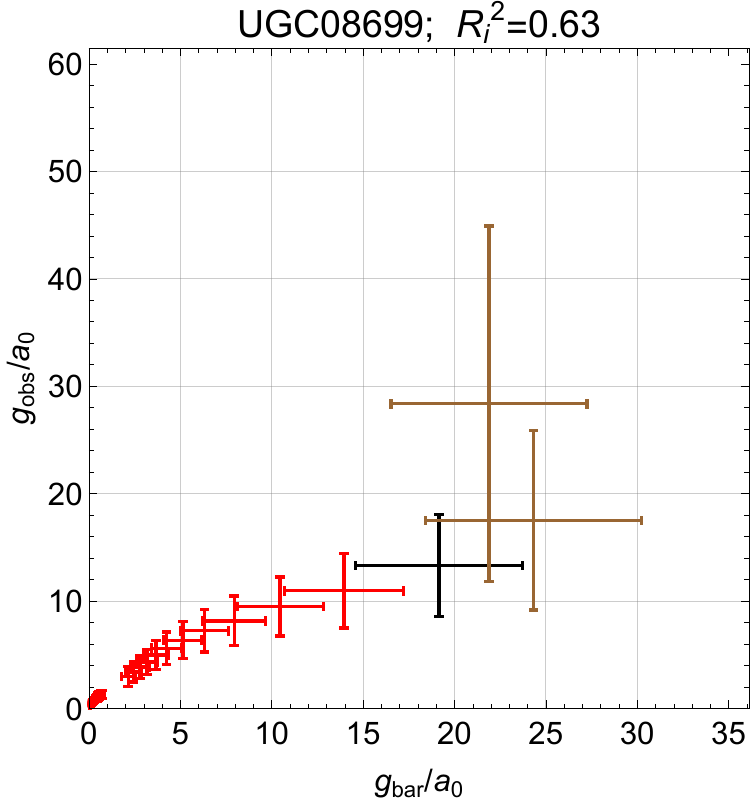}
	\includegraphics[width=0.2\textwidth]{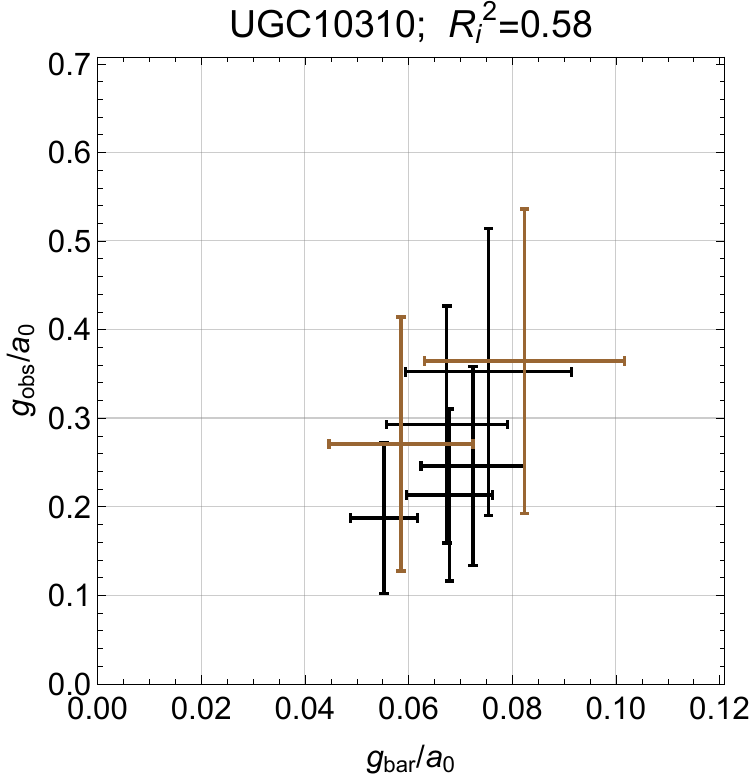}
	\includegraphics[width=0.2\textwidth]{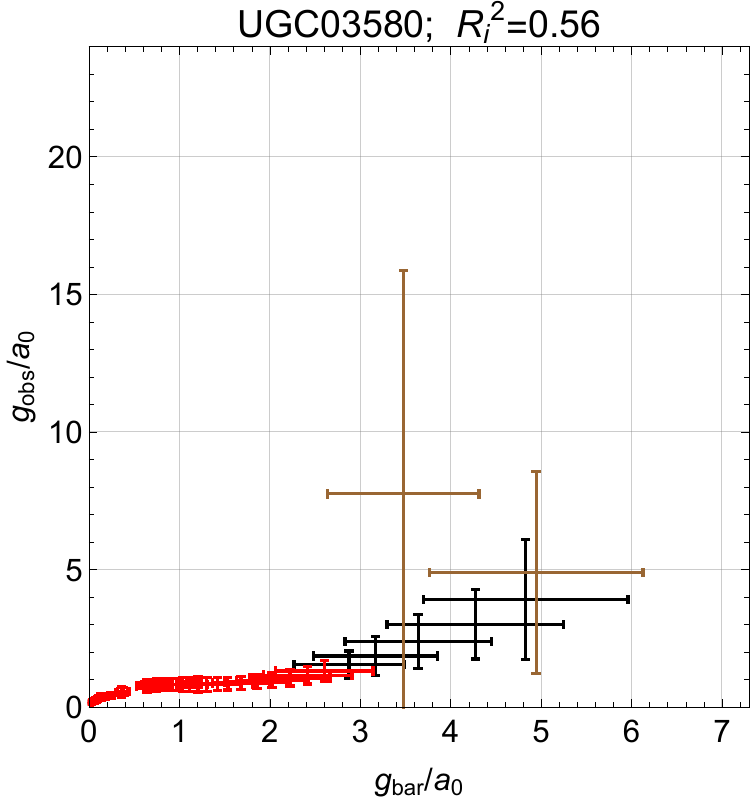}
	\includegraphics[width=0.2\textwidth]{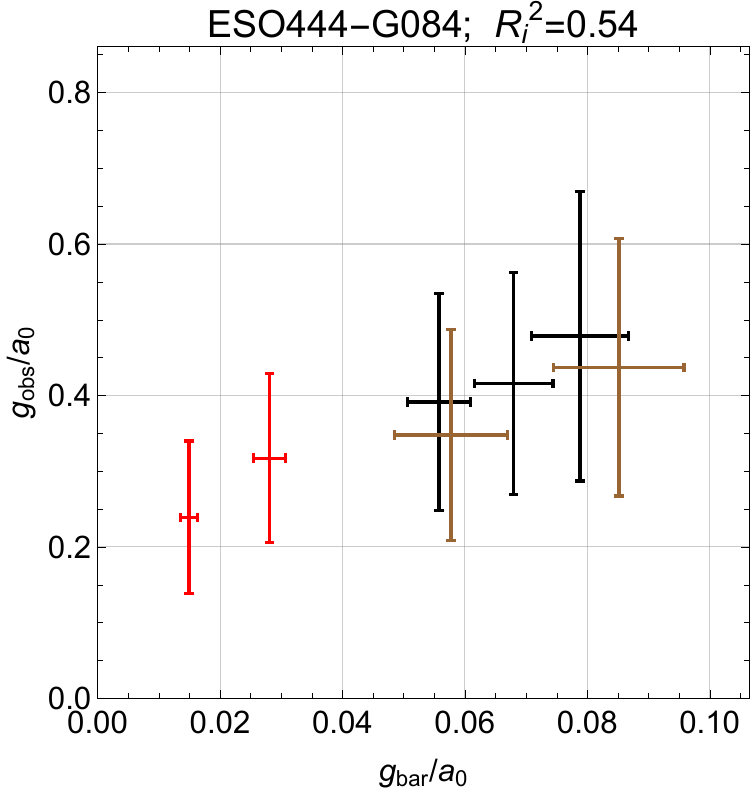}
	\includegraphics[width=0.2\textwidth]{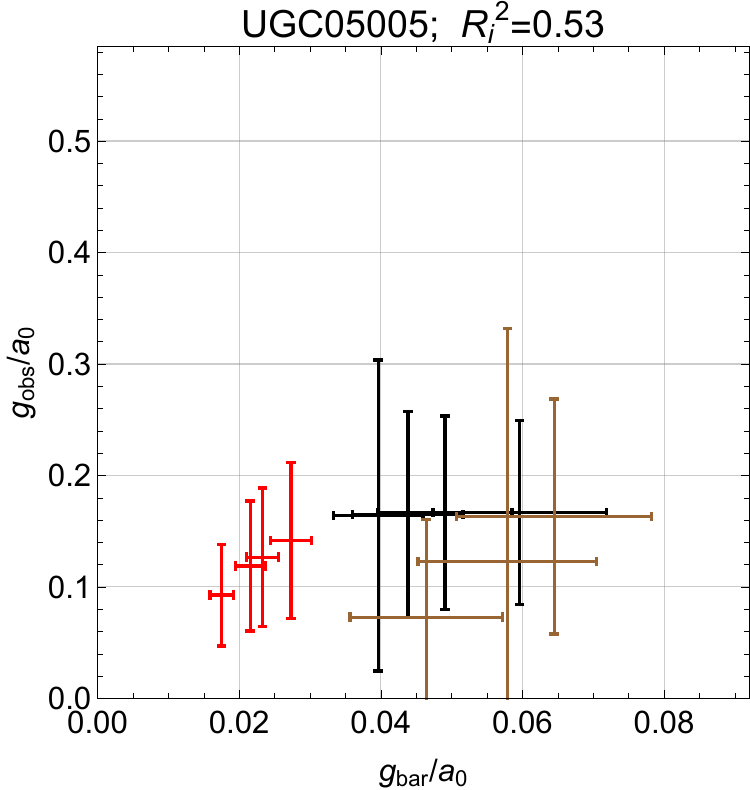}
	\includegraphics[width=0.2\textwidth]{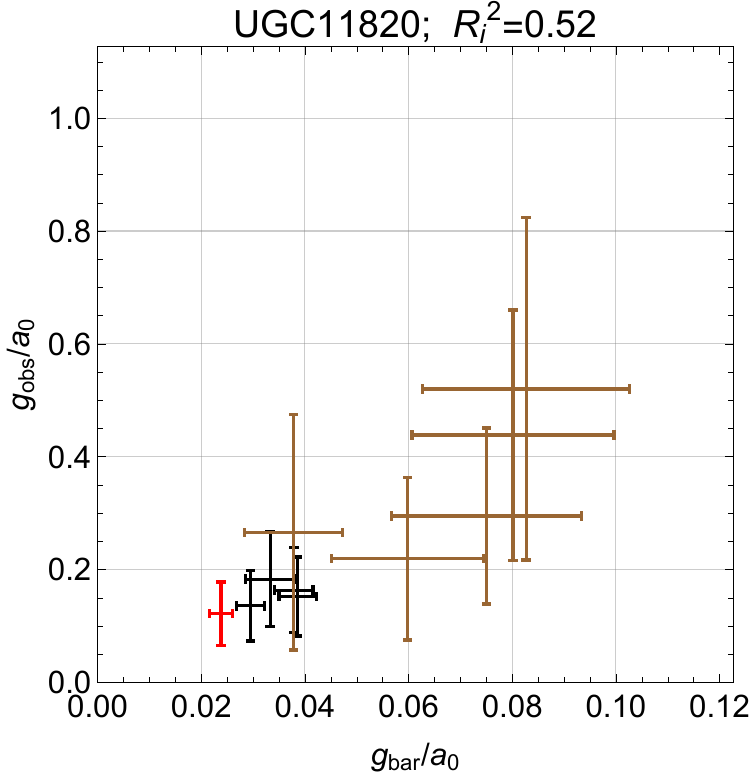}
	\includegraphics[width=0.2\textwidth]{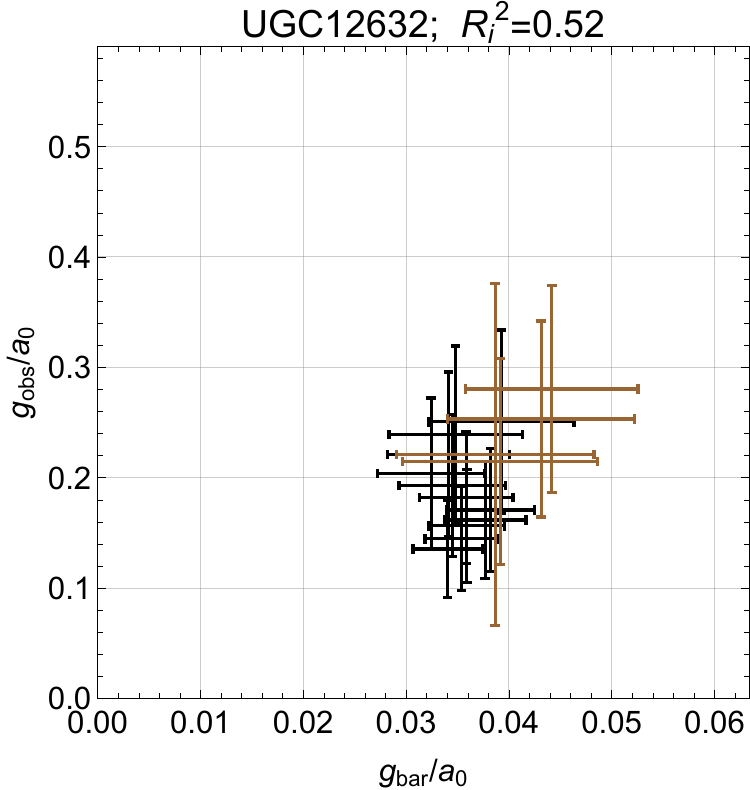}
	\includegraphics[width=0.2\textwidth]{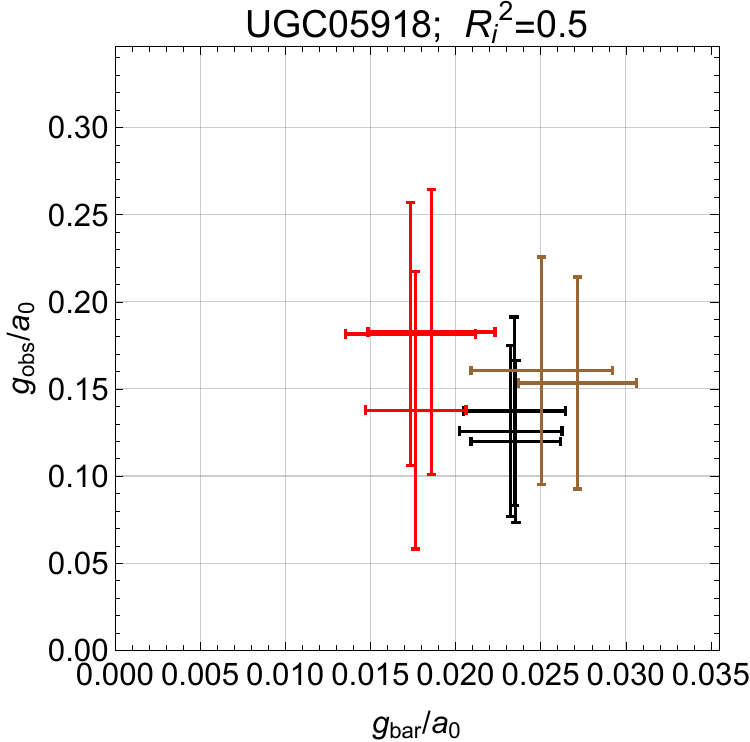}
	\includegraphics[width=0.2\textwidth]{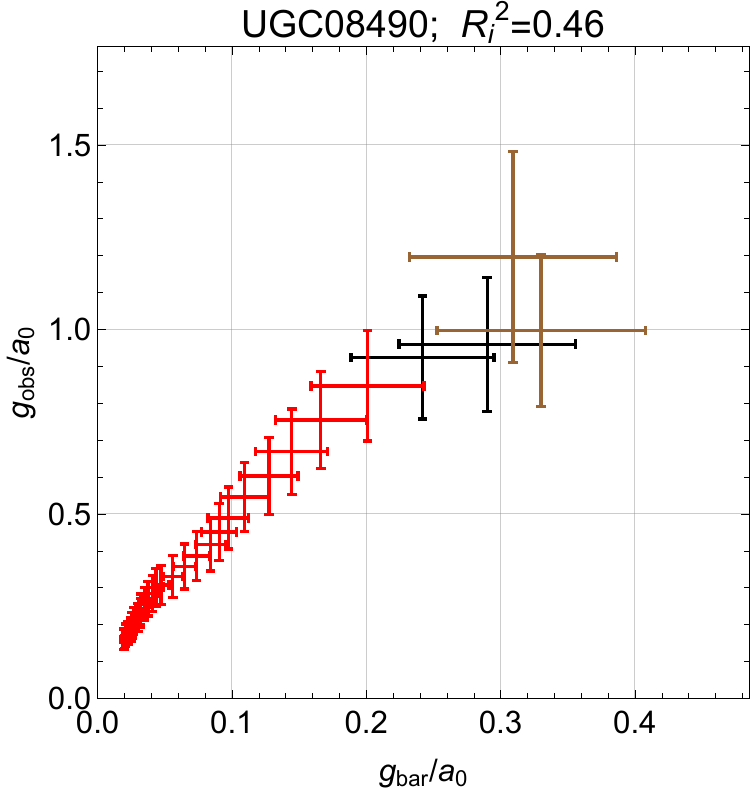}
	\includegraphics[width=0.2\textwidth]{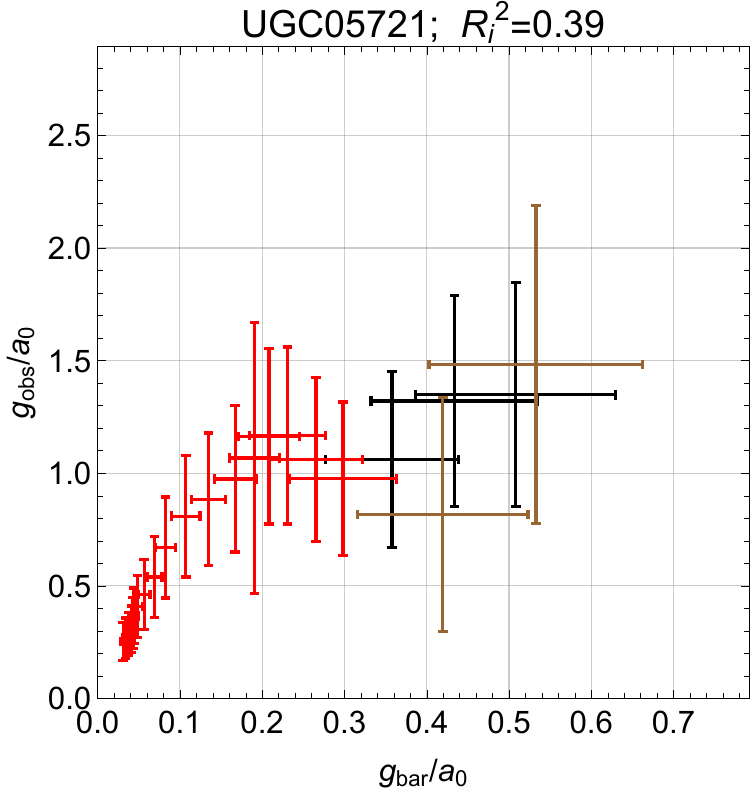}
	\includegraphics[width=0.2\textwidth]{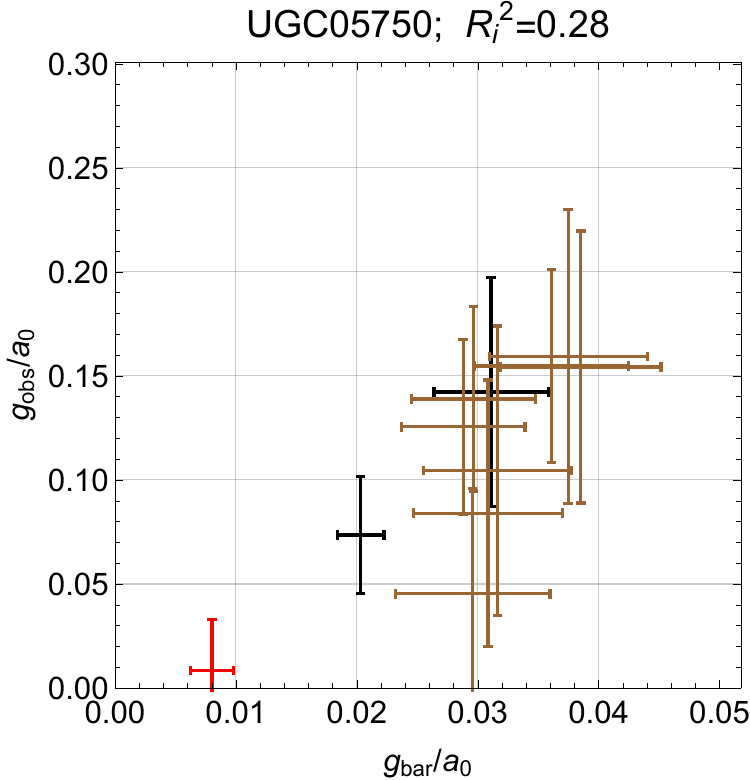}
	\caption{$g2$-space plots corresponding to those of figure \ref{fig:p2}.}
	\label{fig:p5}
\end{figure*}

\begin{figure*}
	\centering
	\includegraphics[width=0.2\textwidth]{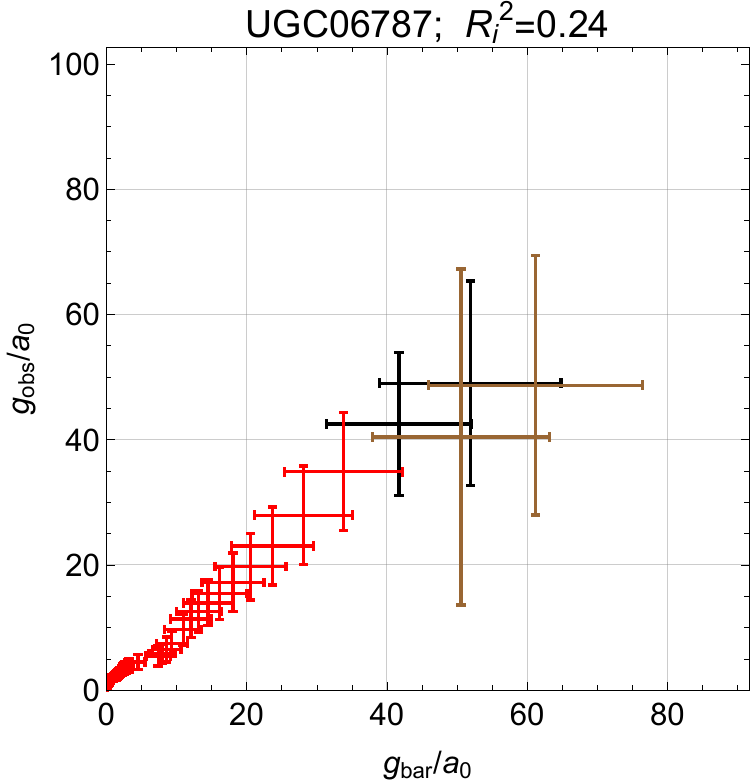}
	\includegraphics[width=0.2\textwidth]{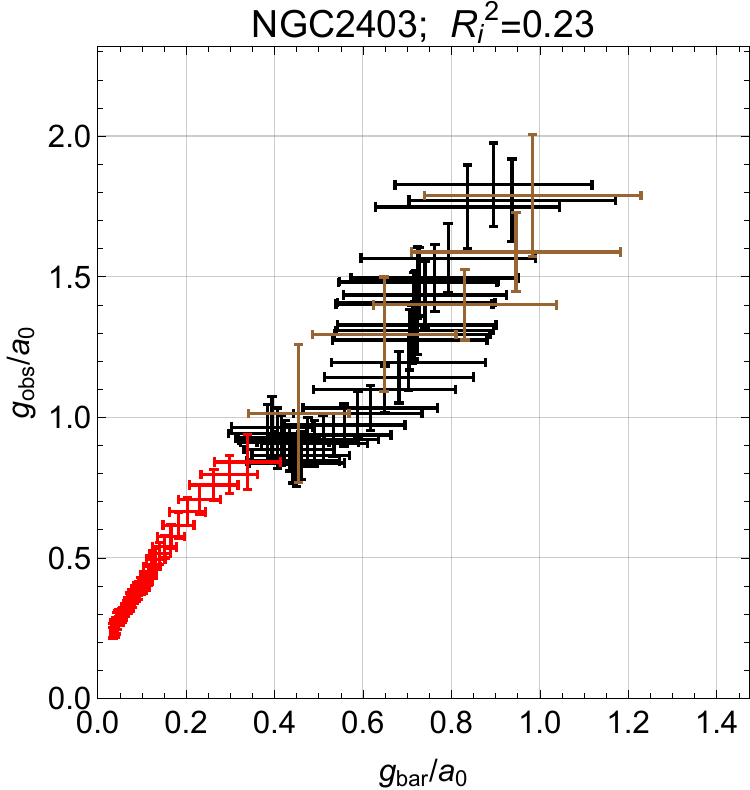}
	\includegraphics[width=0.2\textwidth]{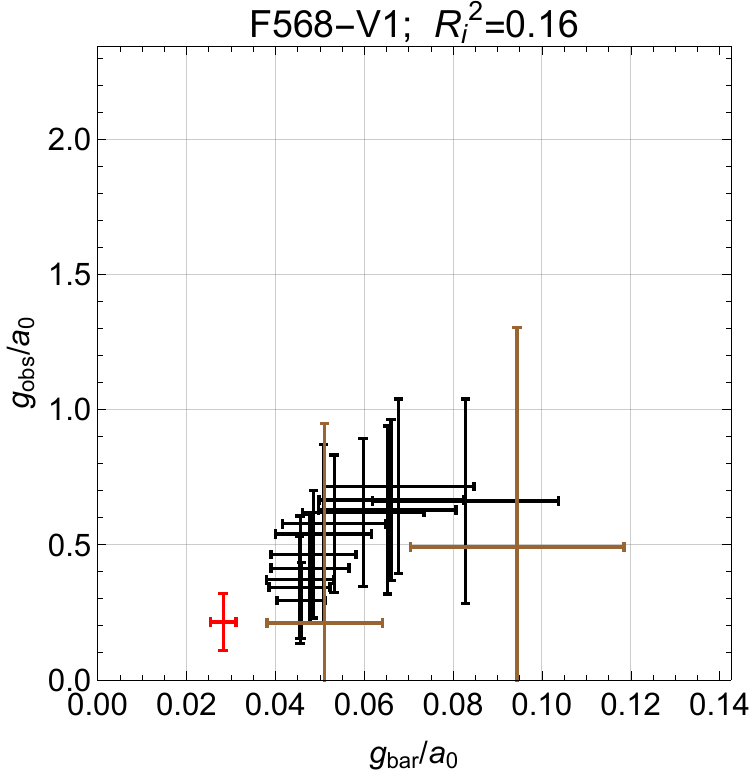}
	\includegraphics[width=0.2\textwidth]{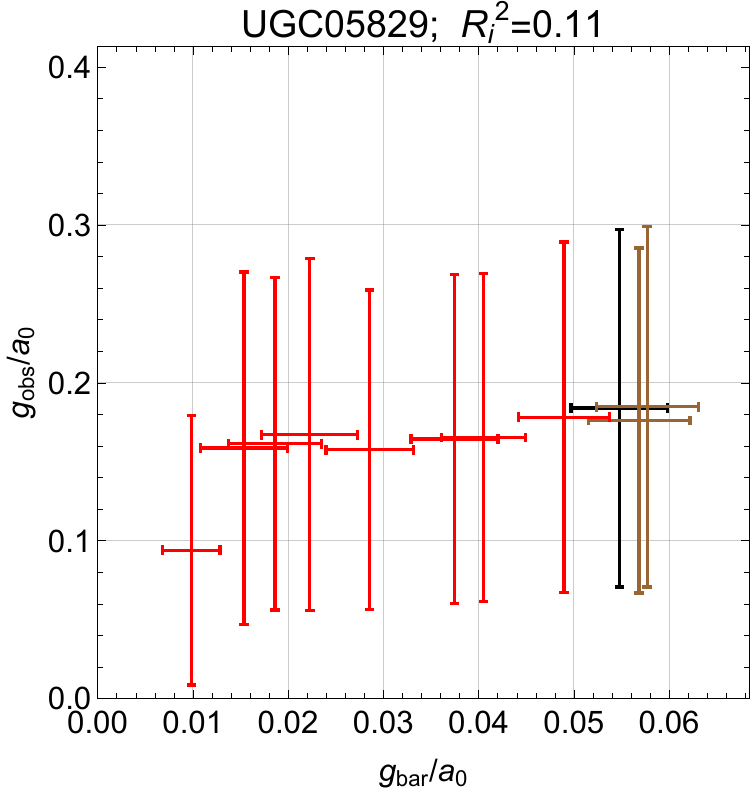}
	\includegraphics[width=0.2\textwidth]{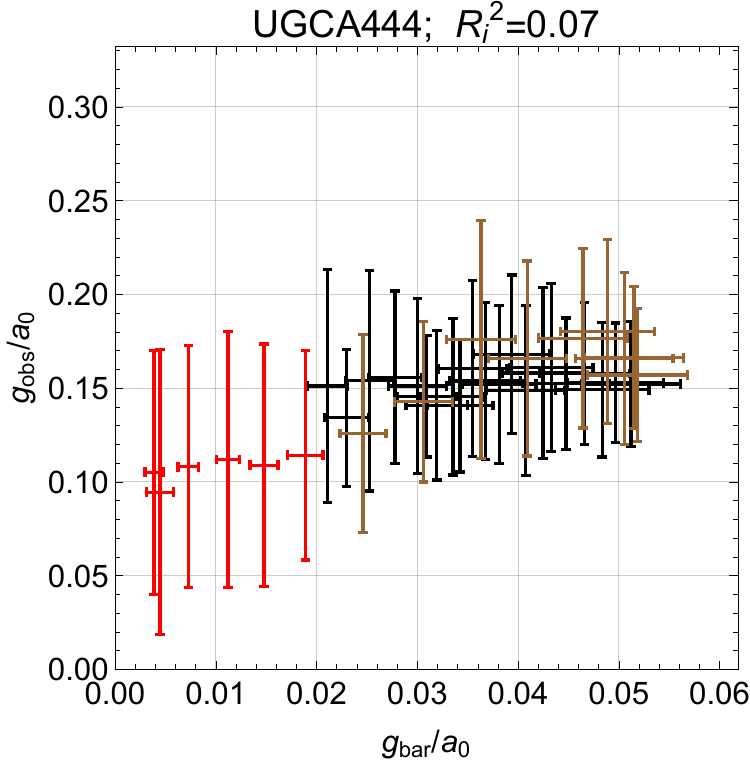}
	\includegraphics[width=0.2\textwidth]{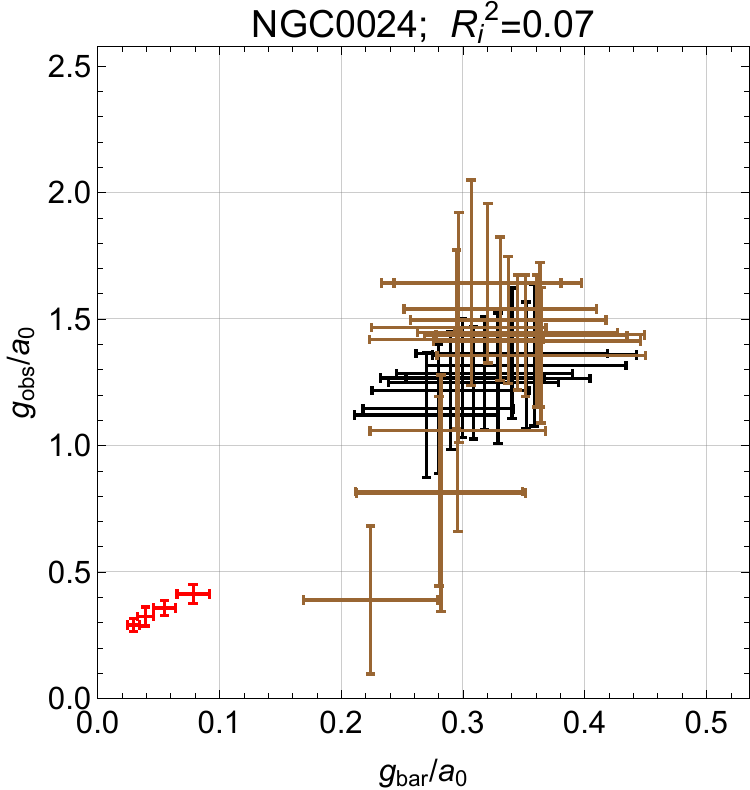}
	\includegraphics[width=0.2\textwidth]{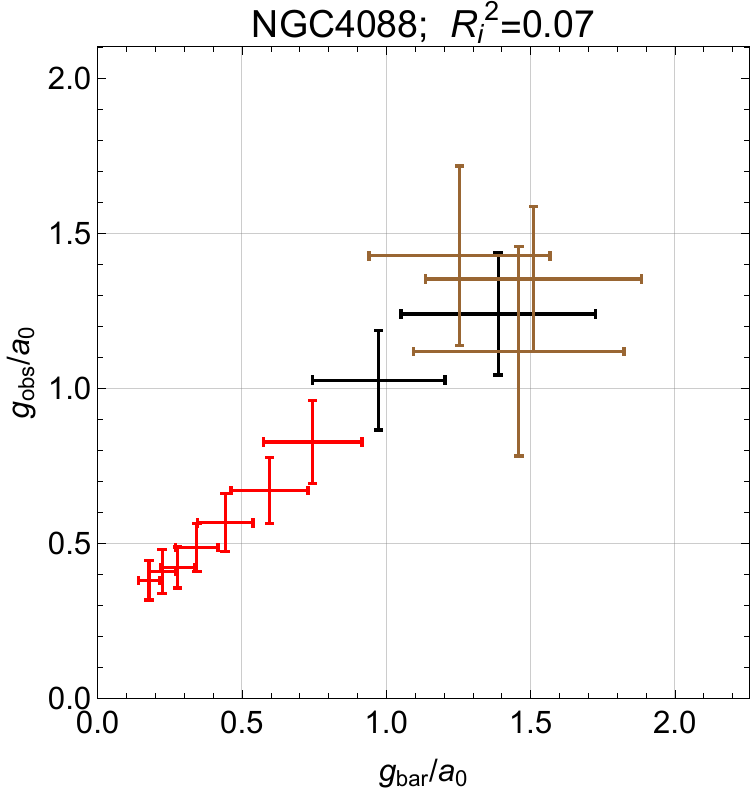}
	\includegraphics[width=0.2\textwidth]{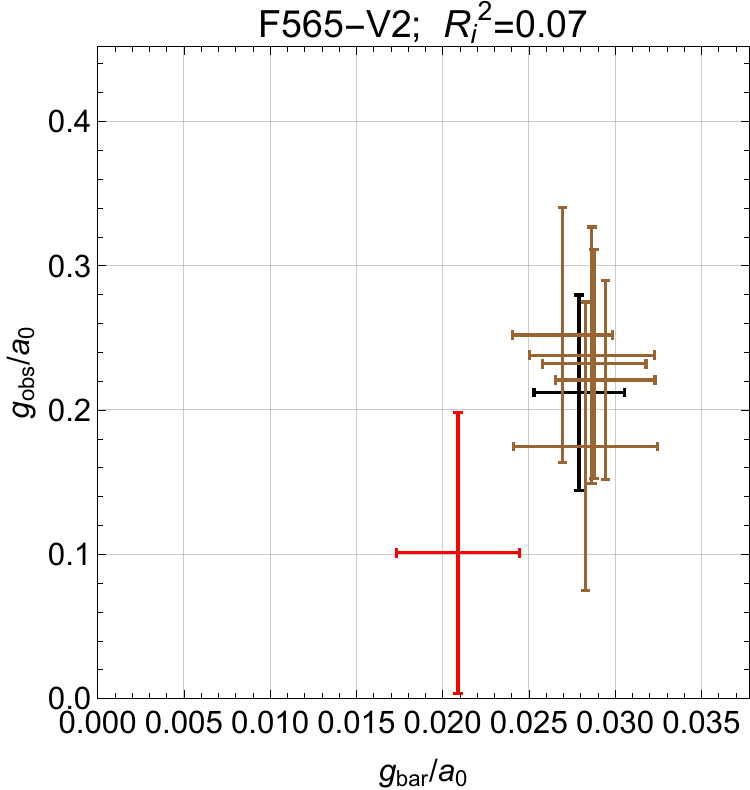}
	\includegraphics[width=0.2\textwidth]{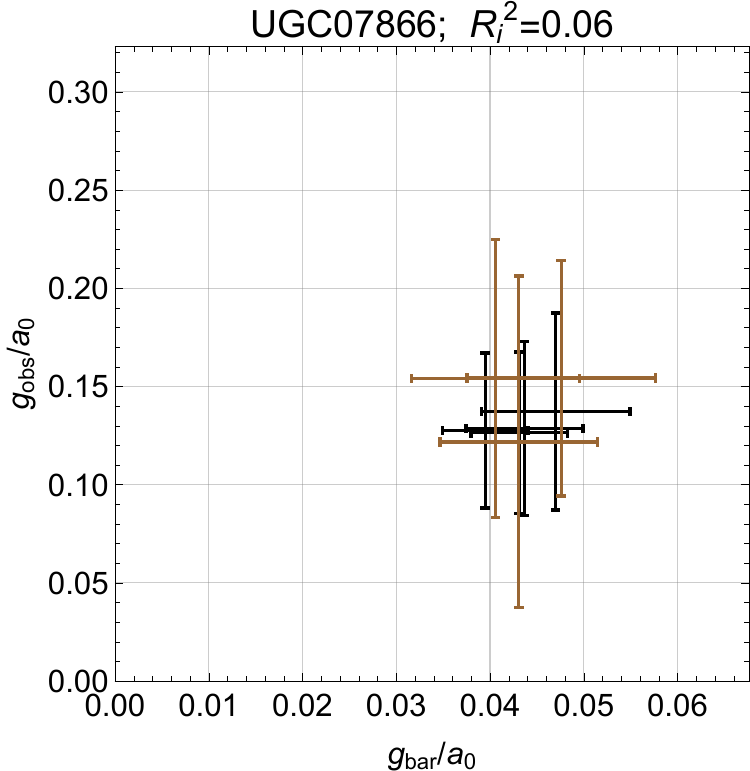}
	\includegraphics[width=0.2\textwidth]{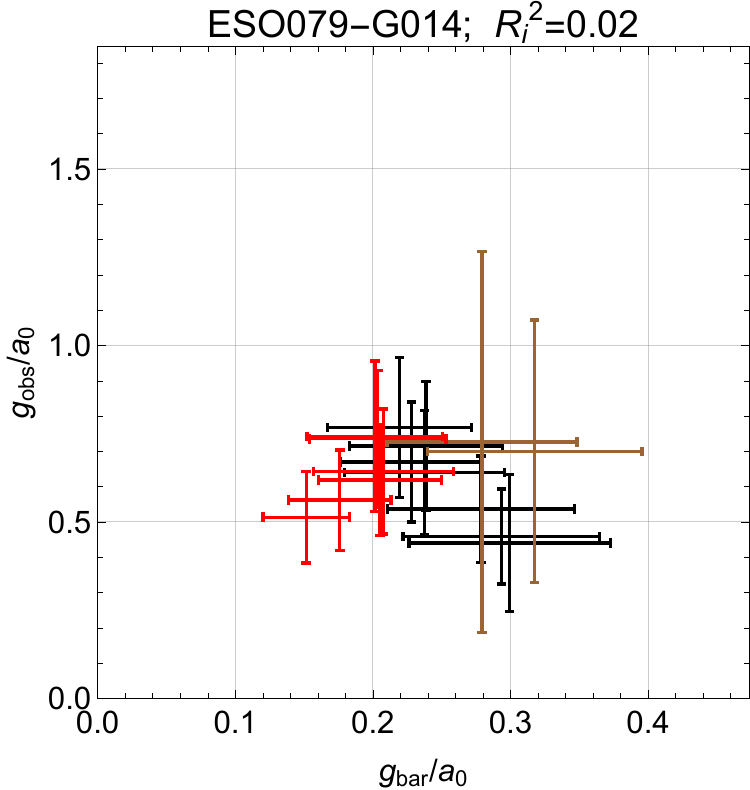}
	\includegraphics[width=0.2\textwidth]{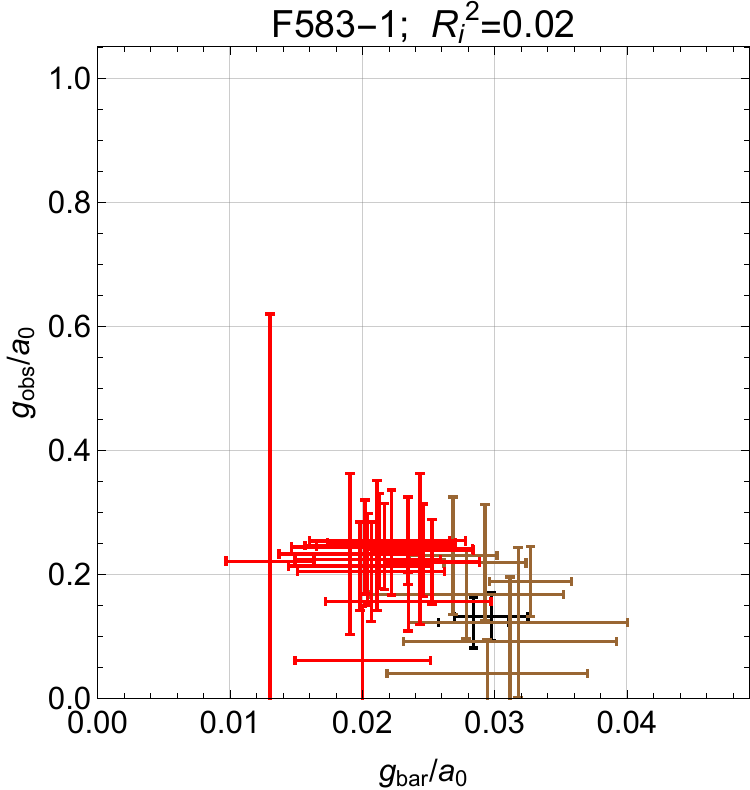}
	\includegraphics[width=0.2\textwidth]{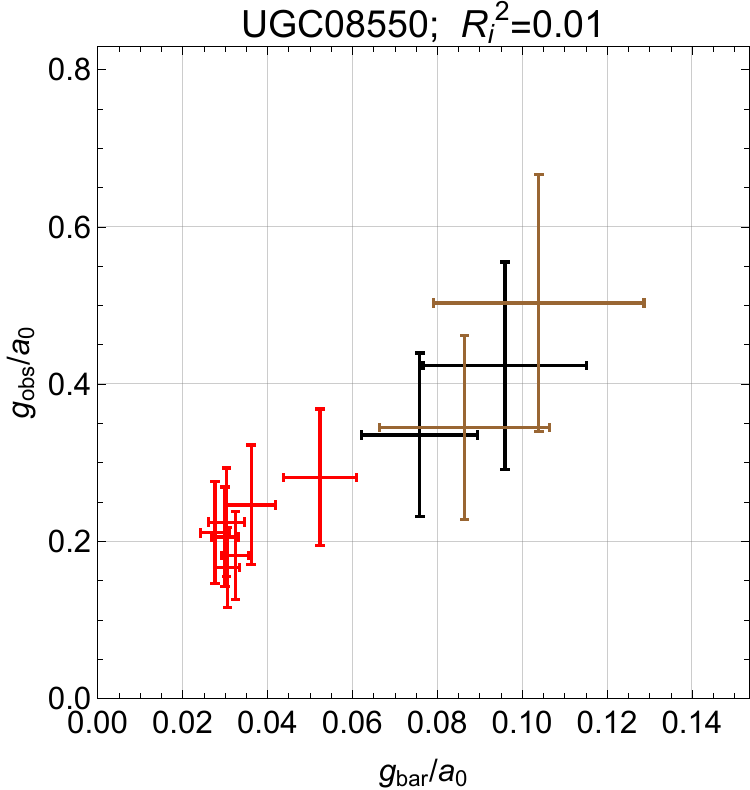}
	\includegraphics[width=0.2\textwidth]{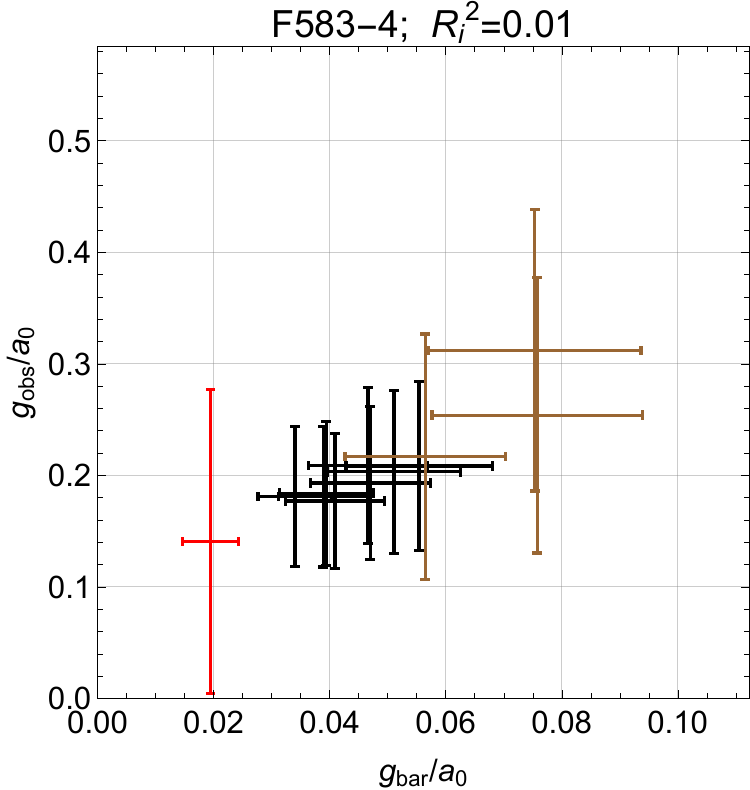}
	\caption{$g2$-space plots corresponding to those of figure \ref{fig:p2}.}
	\label{fig:p6}
\end{figure*}

\begin{figure*}
	\centering
	\includegraphics[width=0.2\textwidth]{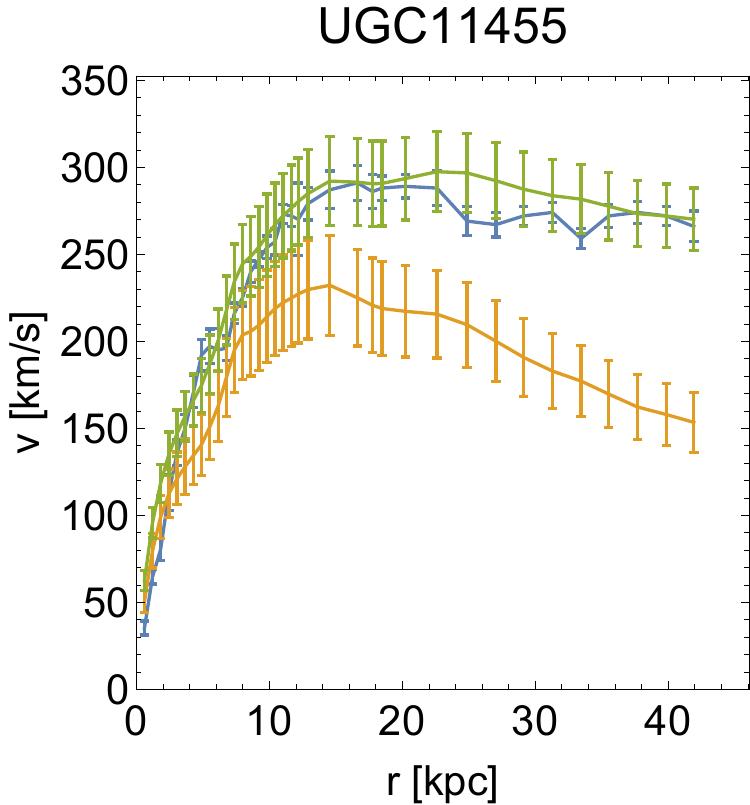}
	\includegraphics[width=0.2\textwidth]{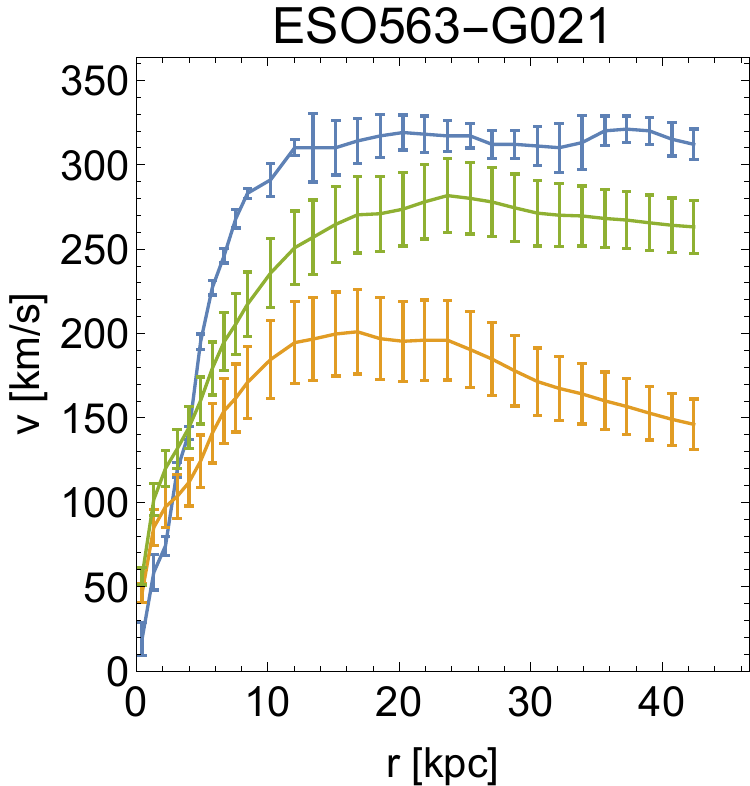}
	\includegraphics[width=0.2\textwidth]{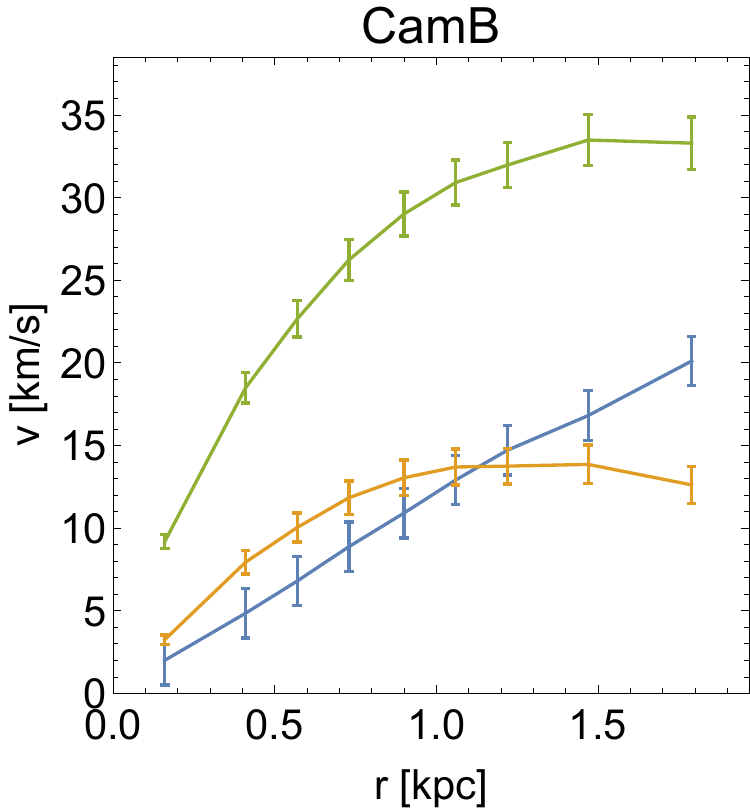}
	\includegraphics[width=0.2\textwidth]{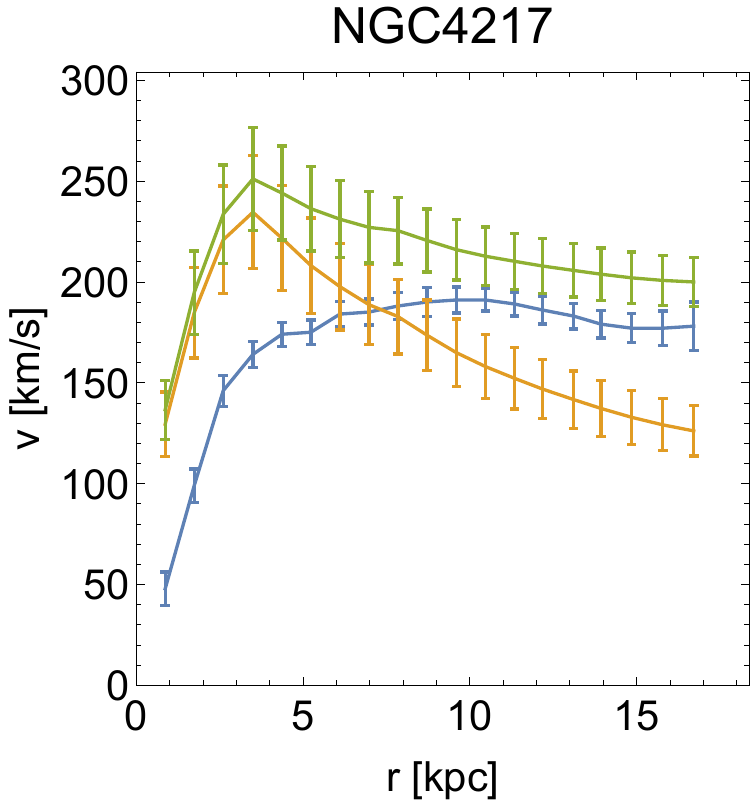}
	\includegraphics[width=0.2\textwidth]{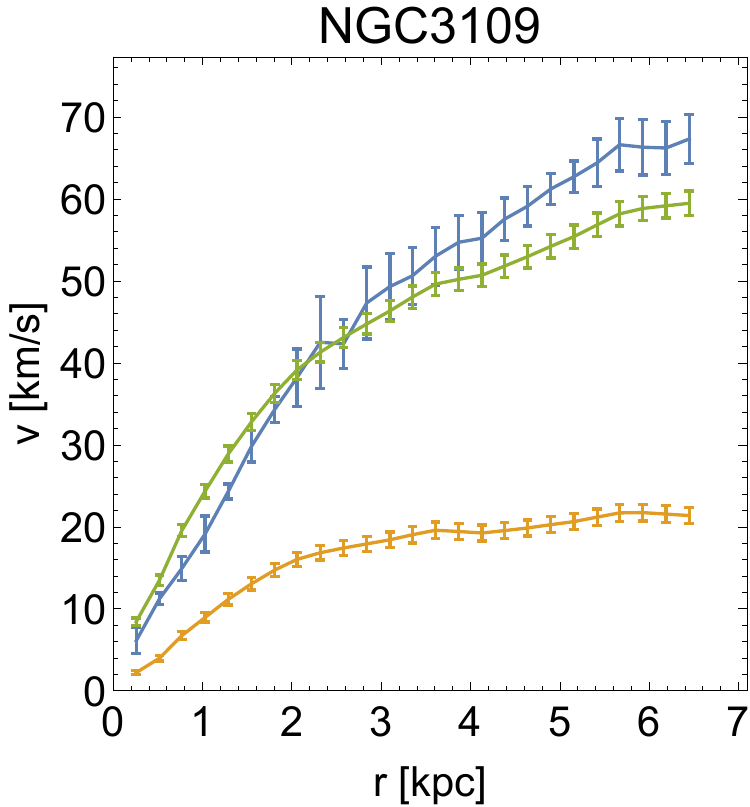}
	\includegraphics[width=0.2\textwidth]{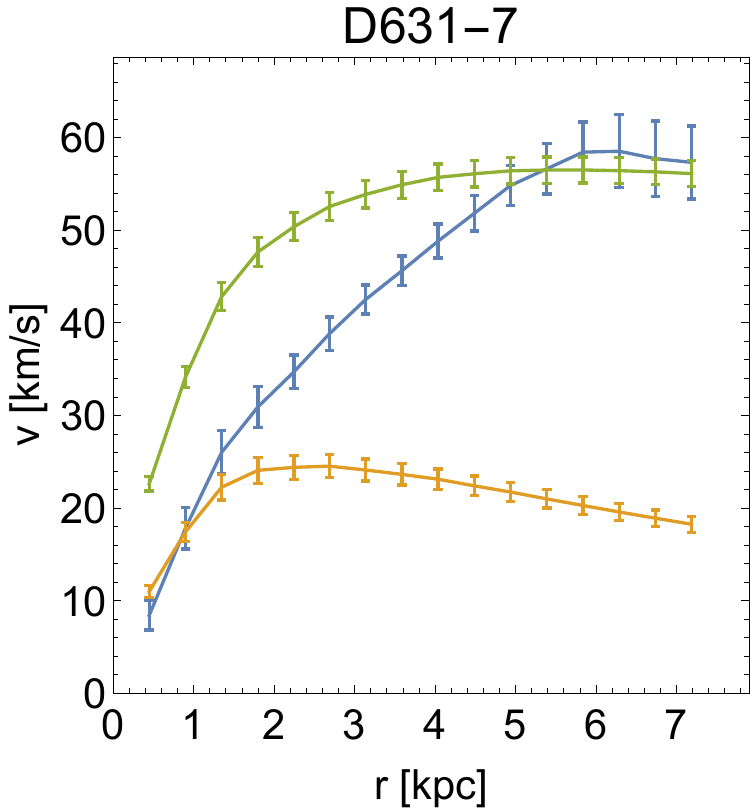}
	\includegraphics[width=0.2\textwidth]{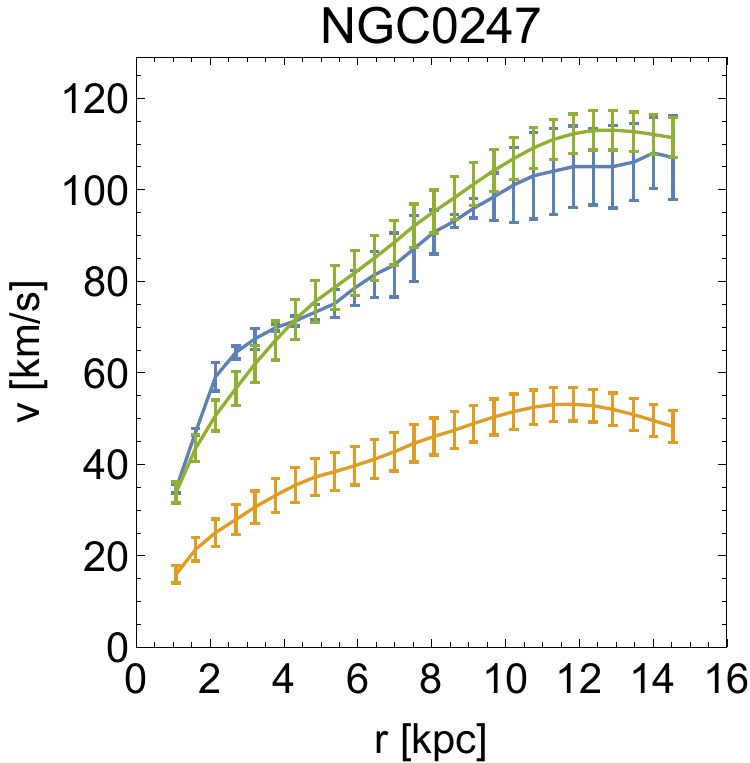}	
	\includegraphics[width=0.2\textwidth]{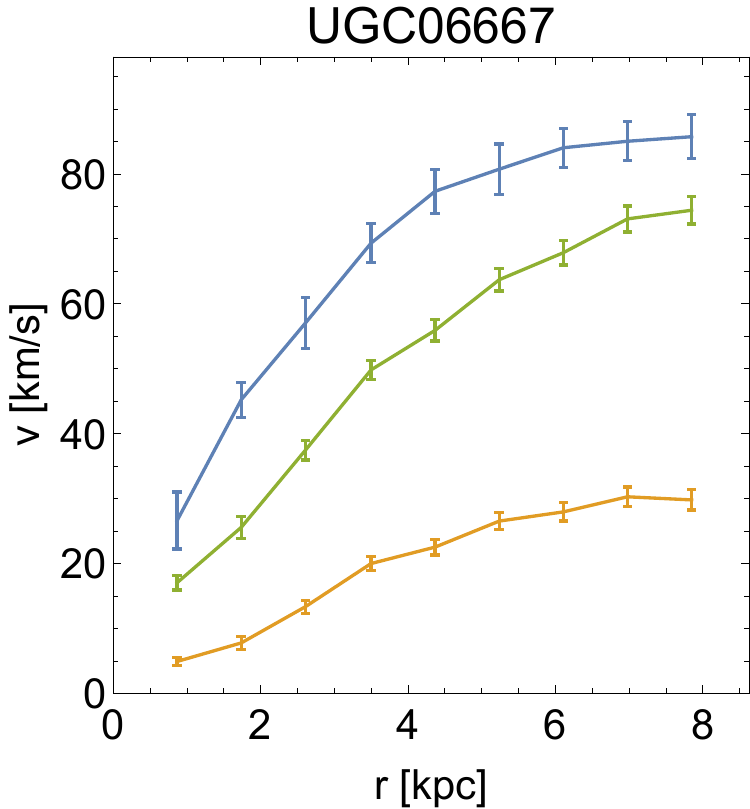}
	\includegraphics[width=0.2\textwidth]{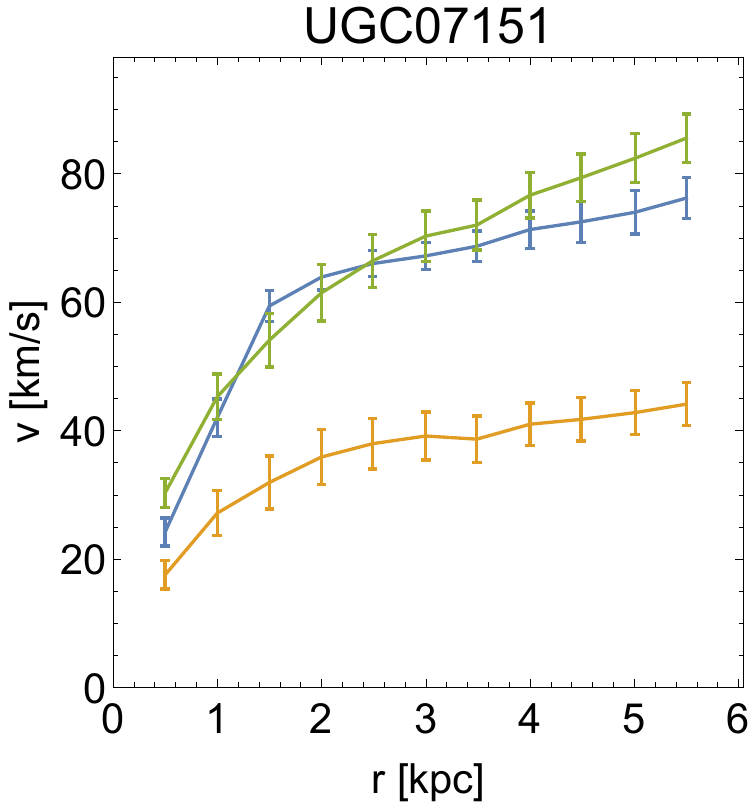}
	\includegraphics[width=0.2\textwidth]{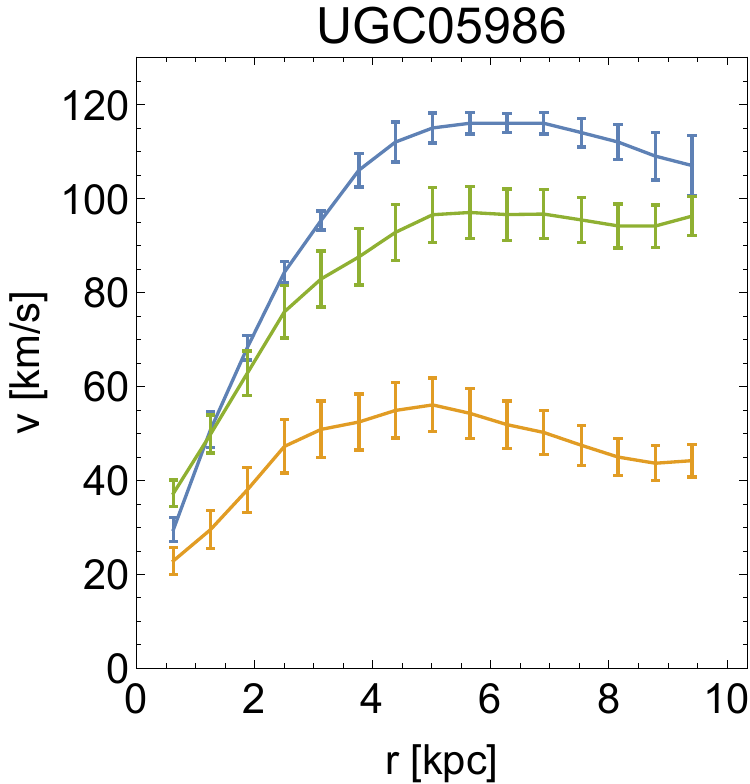}
	\includegraphics[width=0.2\textwidth]{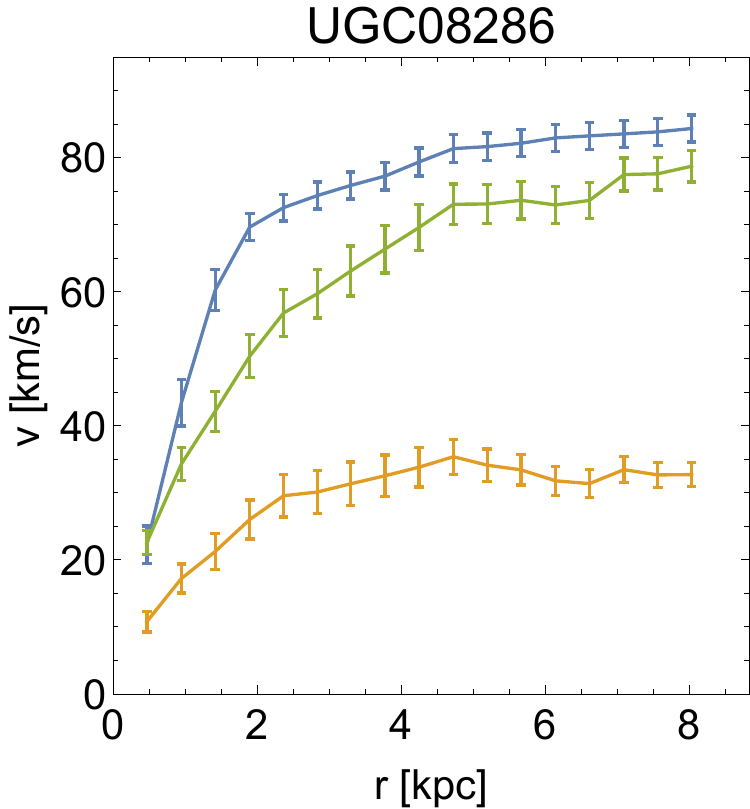}
	\includegraphics[width=0.2\textwidth]{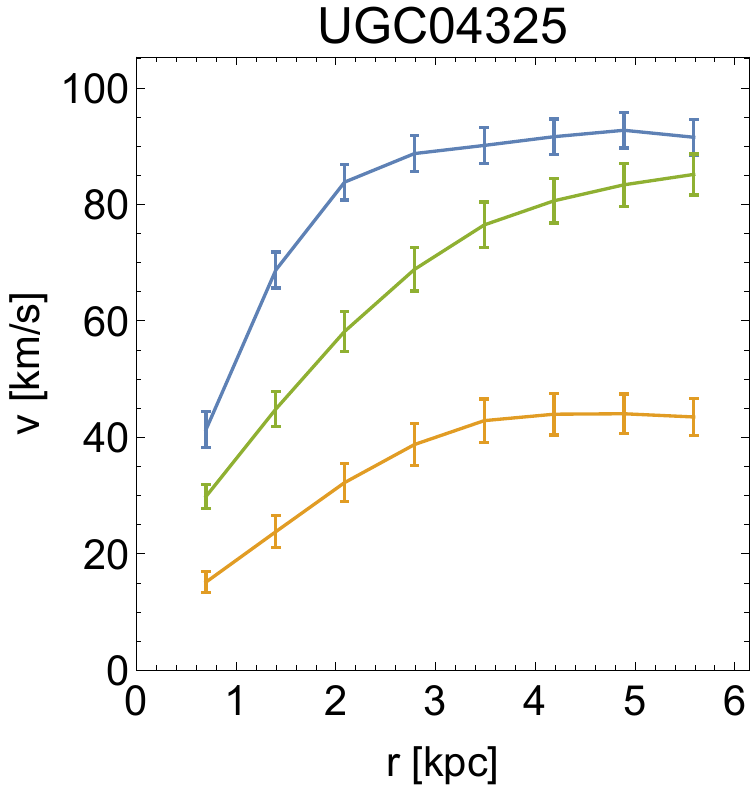}
	\includegraphics[width=0.2\textwidth]{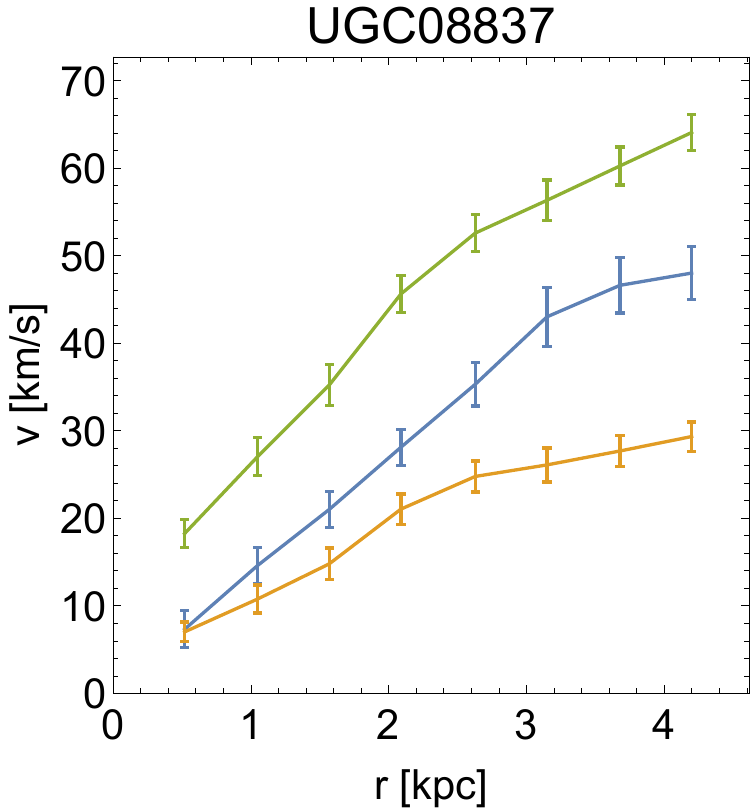}
	\includegraphics[width=0.2\textwidth]{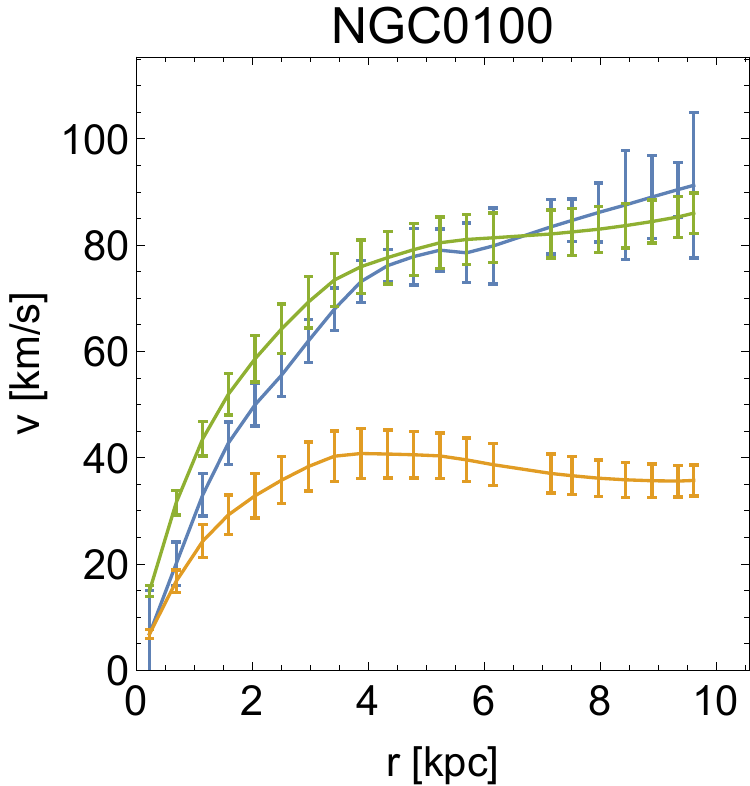}
	\includegraphics[width=0.2\textwidth]{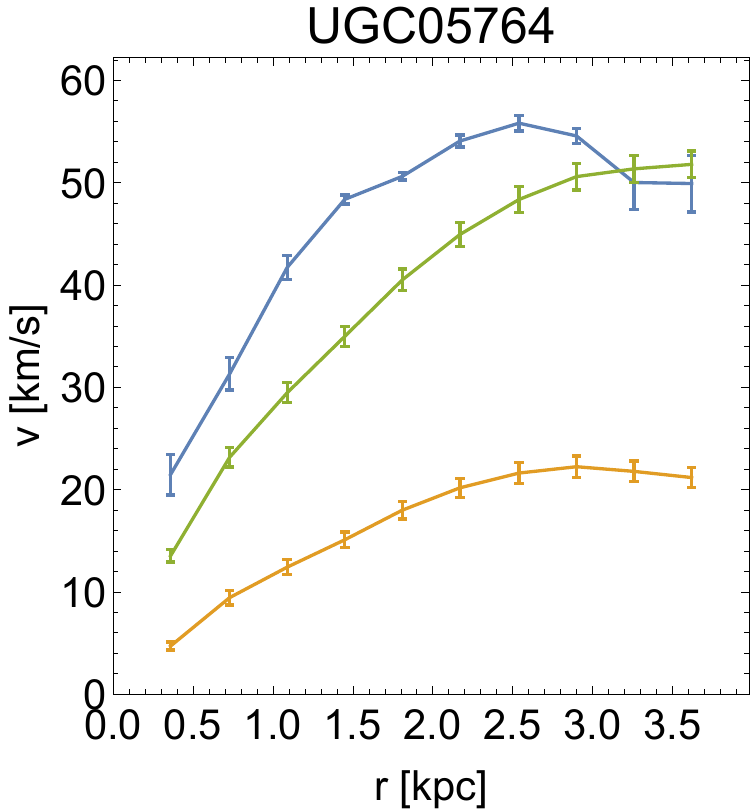}
	\includegraphics[width=0.2\textwidth]{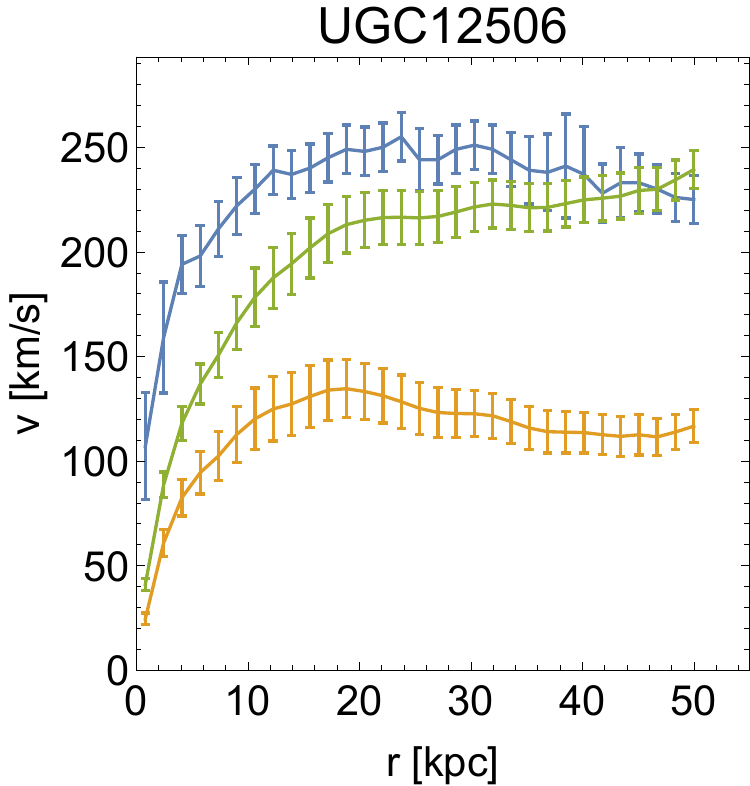}
	\includegraphics[width=0.2\textwidth]{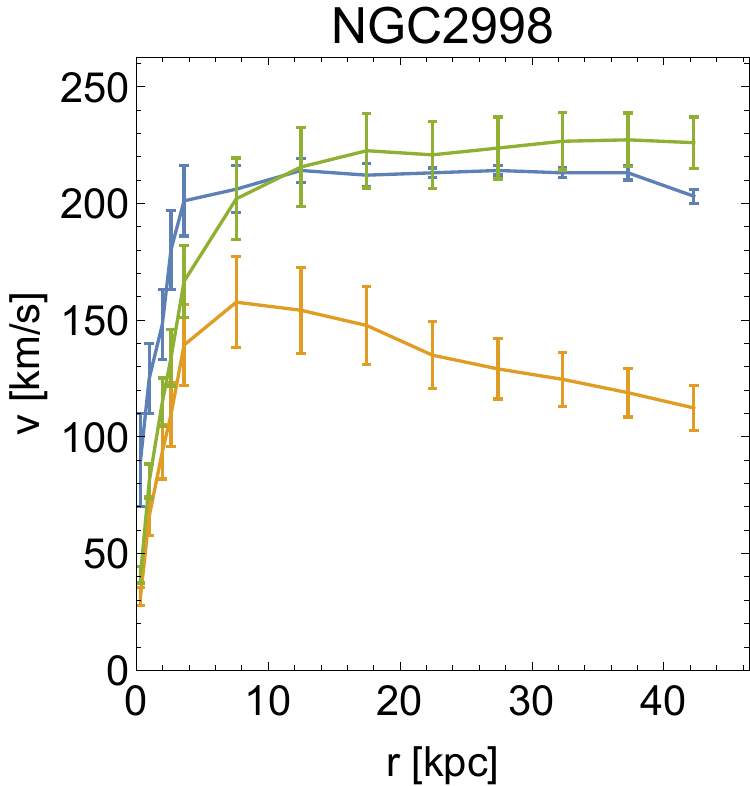}
	\includegraphics[width=0.2\textwidth]{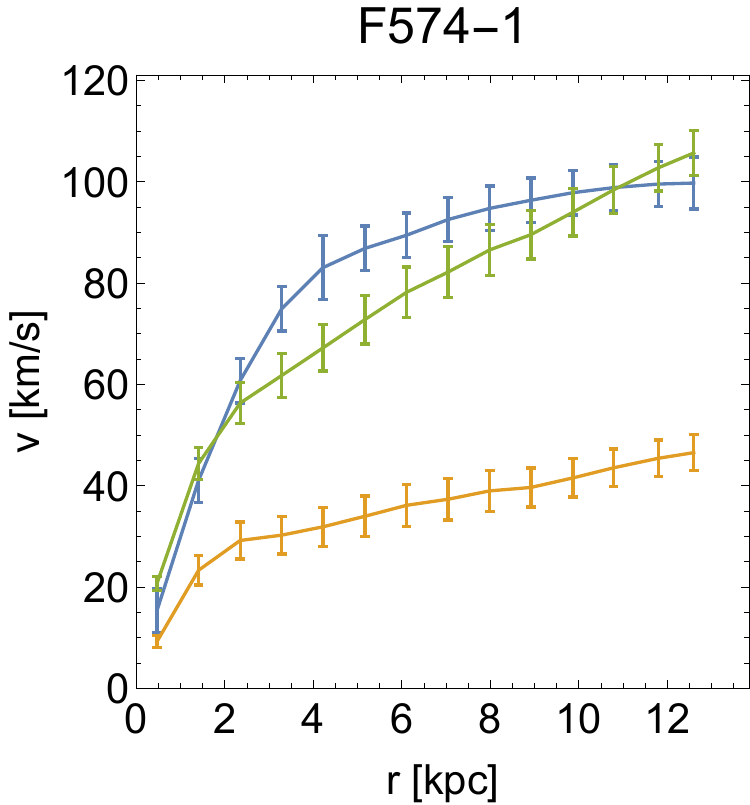}
	\includegraphics[width=0.2\textwidth]{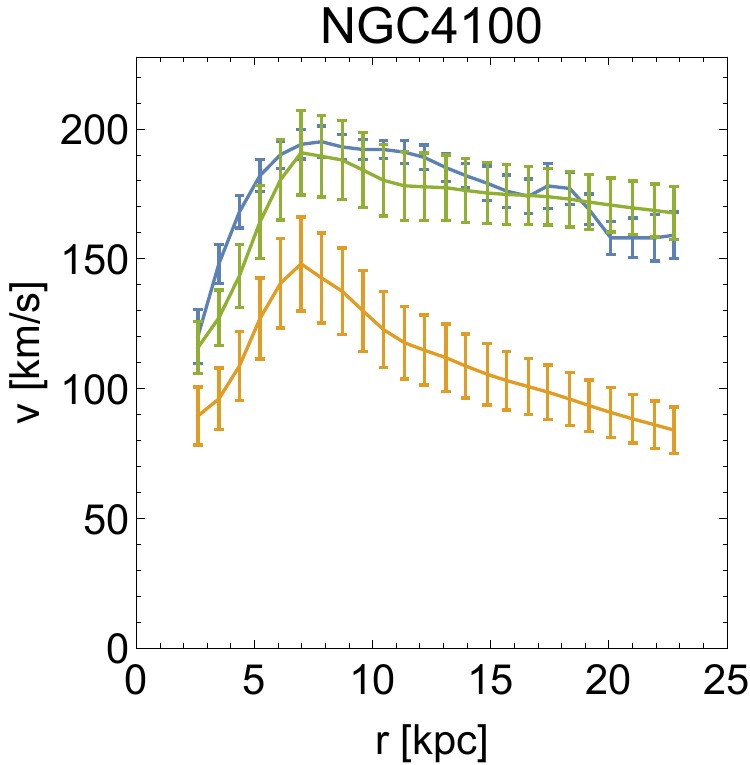}
	\includegraphics[width=0.2\textwidth]{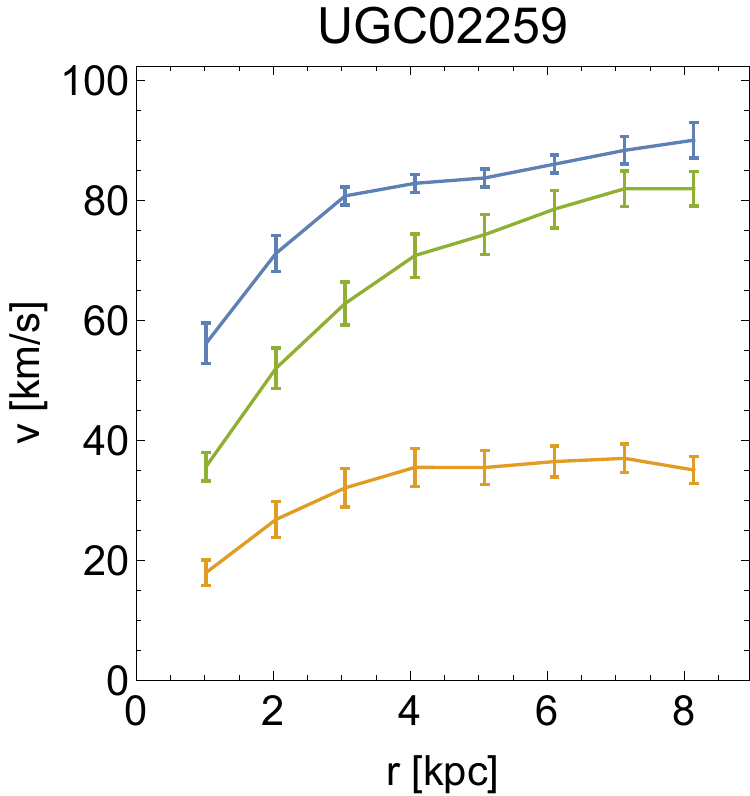}
	\includegraphics[width=0.2\textwidth]{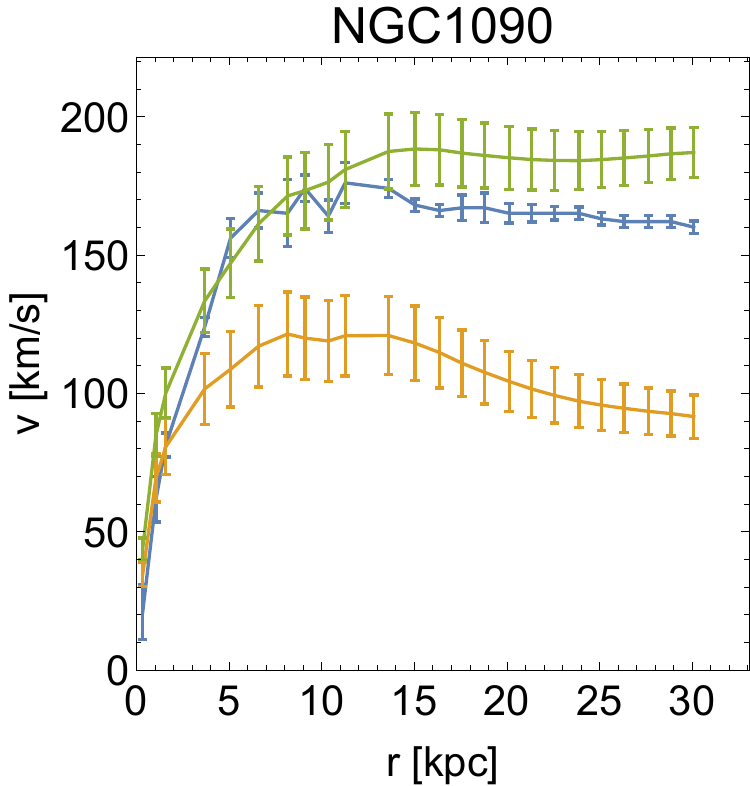}
	\includegraphics[width=0.2\textwidth]{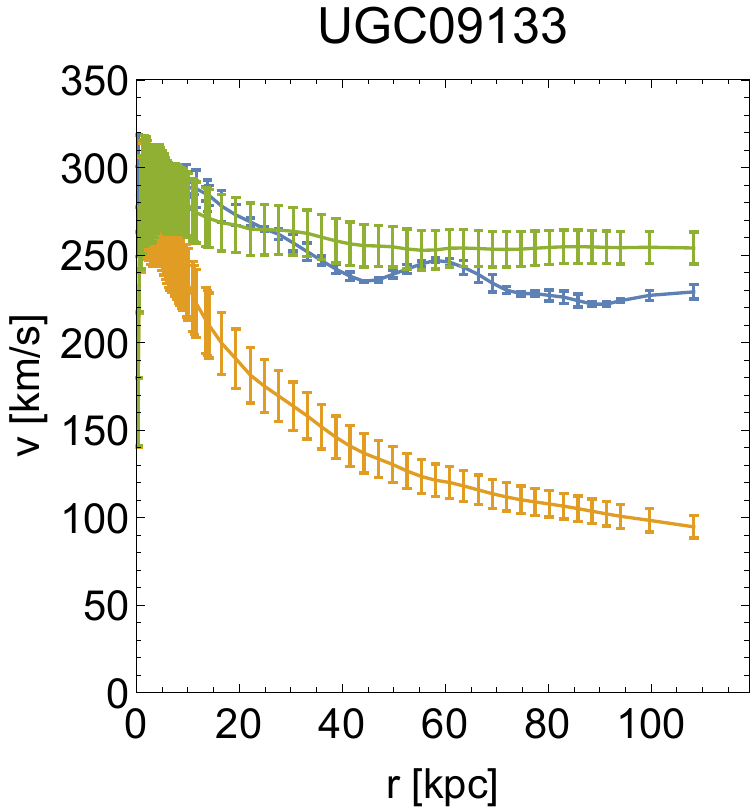}
	\includegraphics[width=0.2\textwidth]{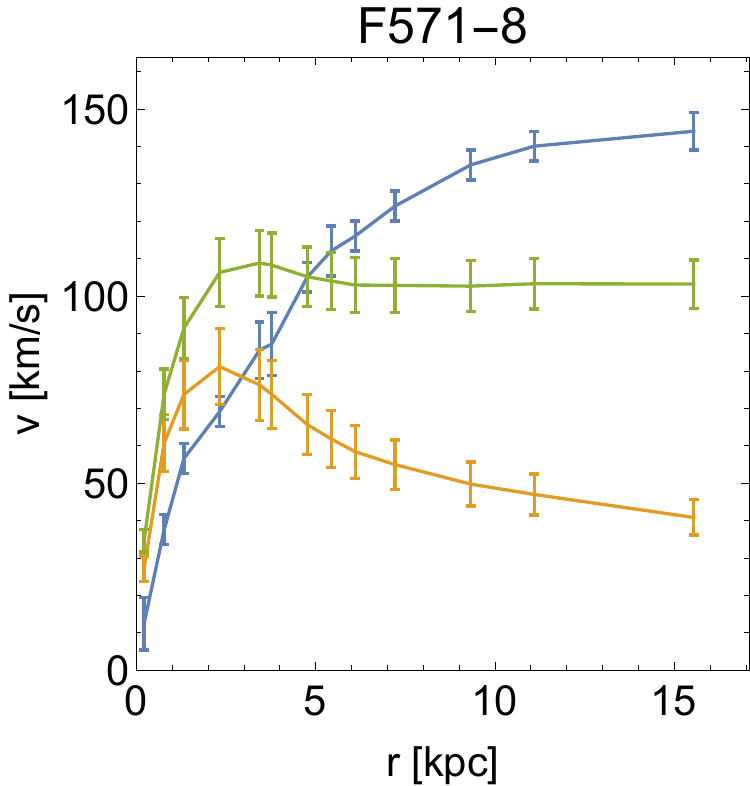}
	\includegraphics[width=0.2\textwidth]{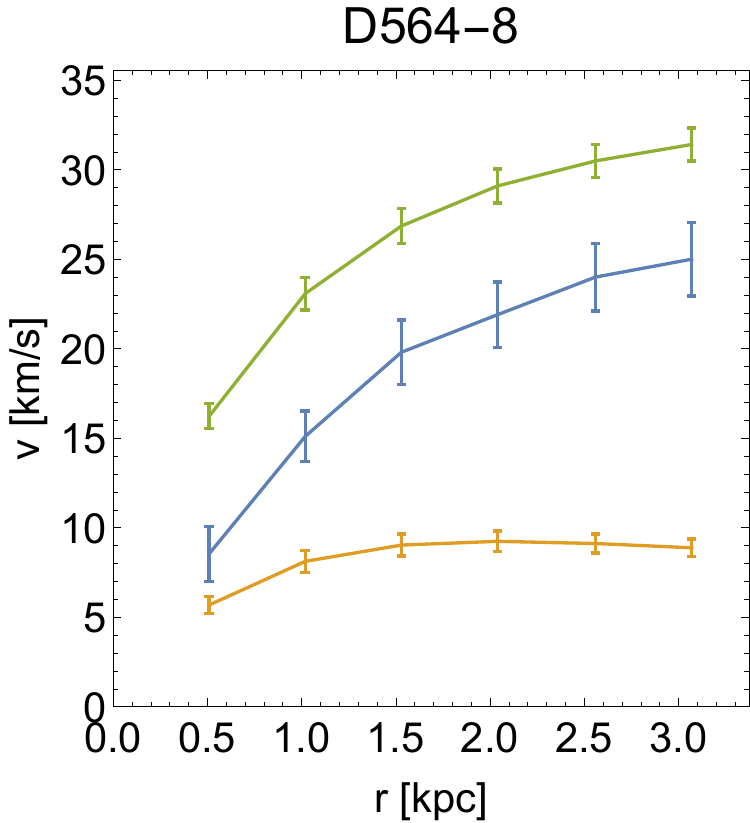}
	\caption{Rotation curve plots of galaxies corresponding to figure \ref{fig:p3} (same ordering as well). Blue denote $v_{obs}$, yellow denote $v_{bar}$ and green denote $v_{tot}^{RAR}$ based on observed $v_{bar}$.}
	\label{fig:p7}
\end{figure*}

\begin{figure*}
	\centering
	\includegraphics[width=0.2\textwidth]{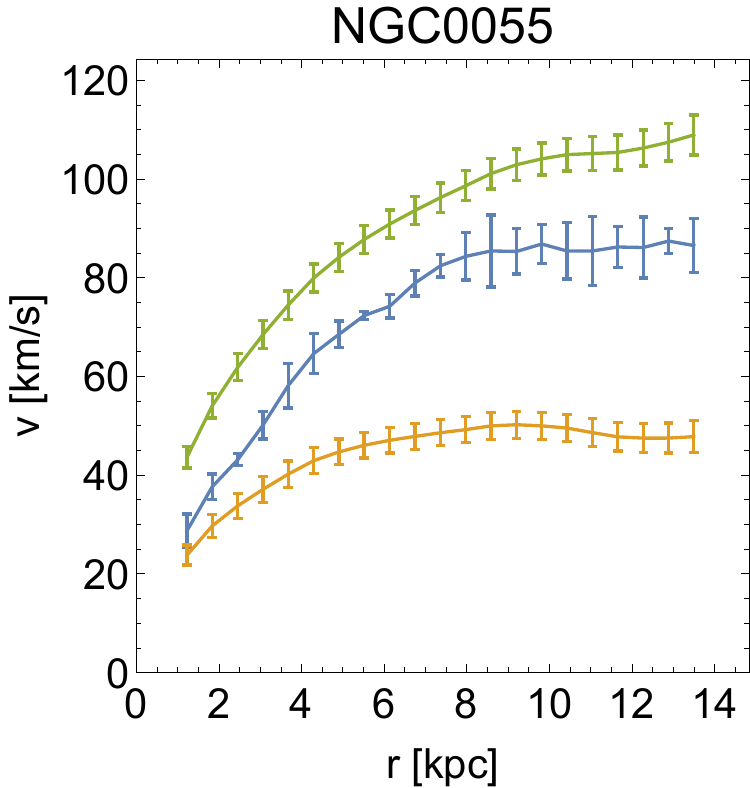}
	\includegraphics[width=0.2\textwidth]{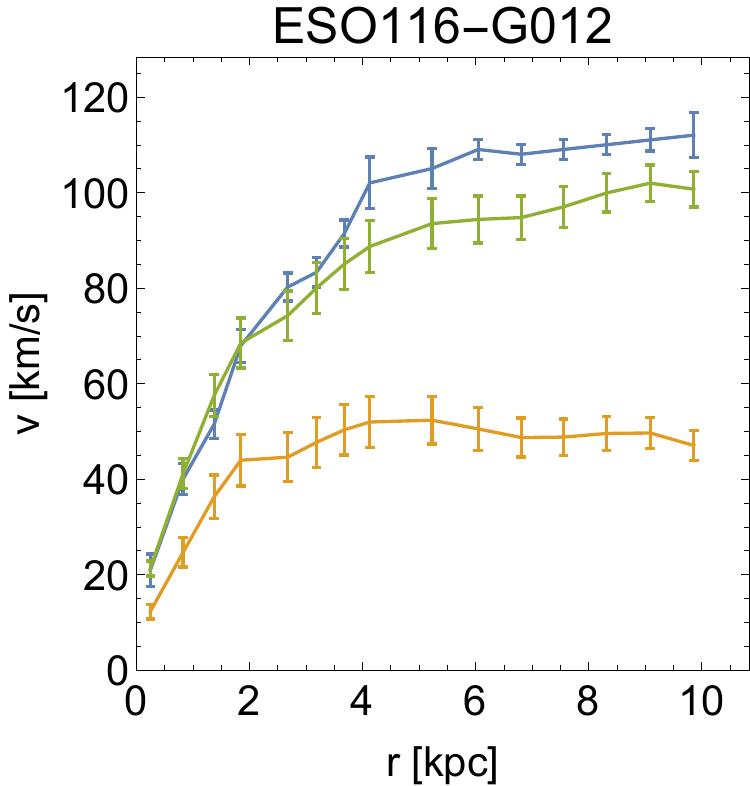}
	\includegraphics[width=0.2\textwidth]{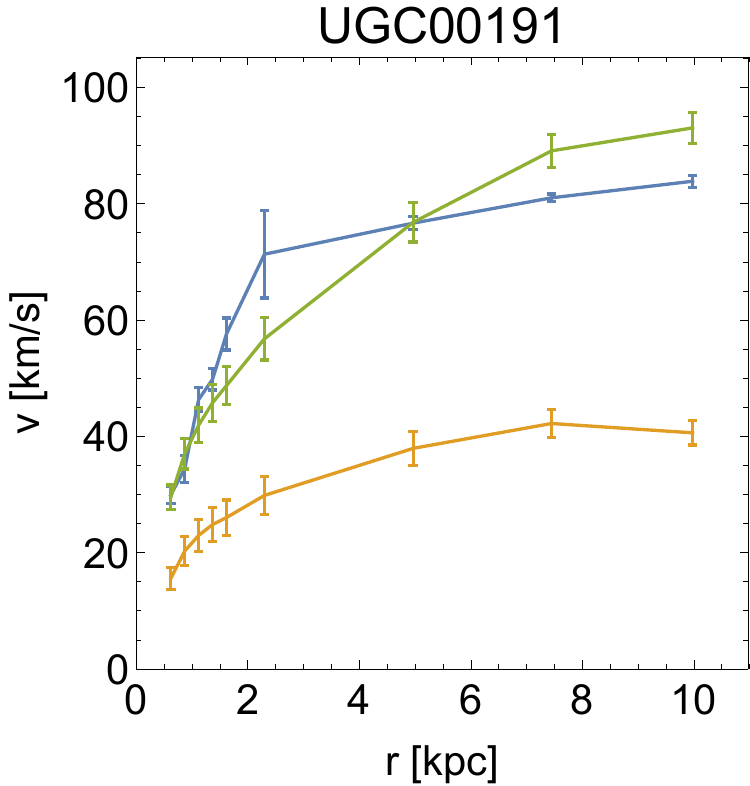}
	\includegraphics[width=0.2\textwidth]{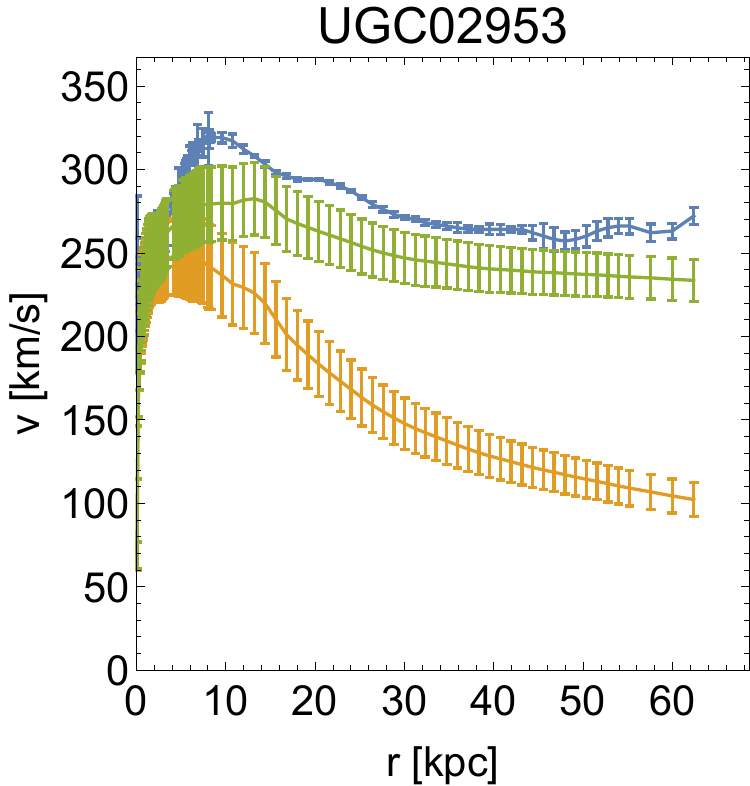}
	\includegraphics[width=0.2\textwidth]{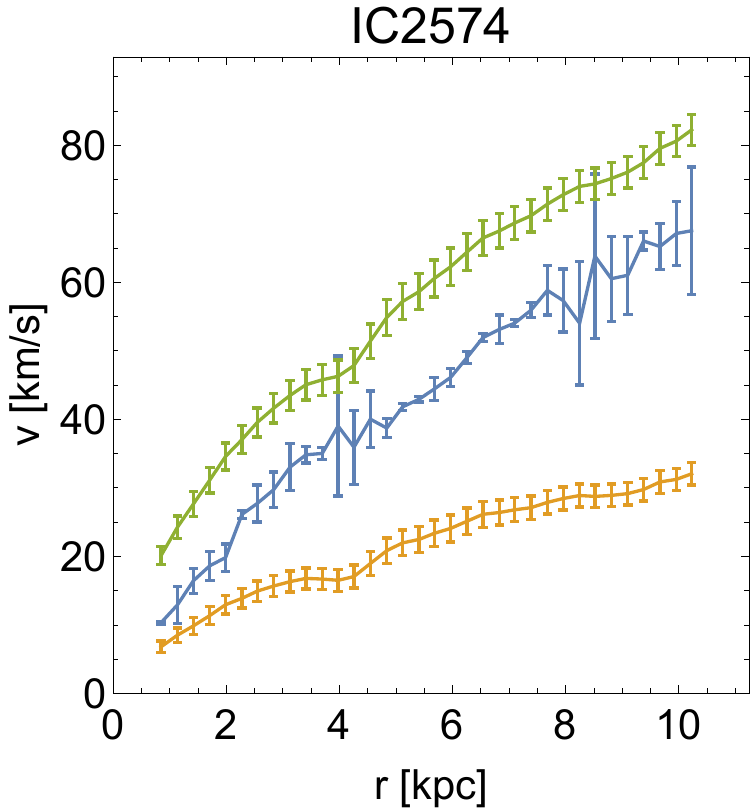}
	\includegraphics[width=0.2\textwidth]{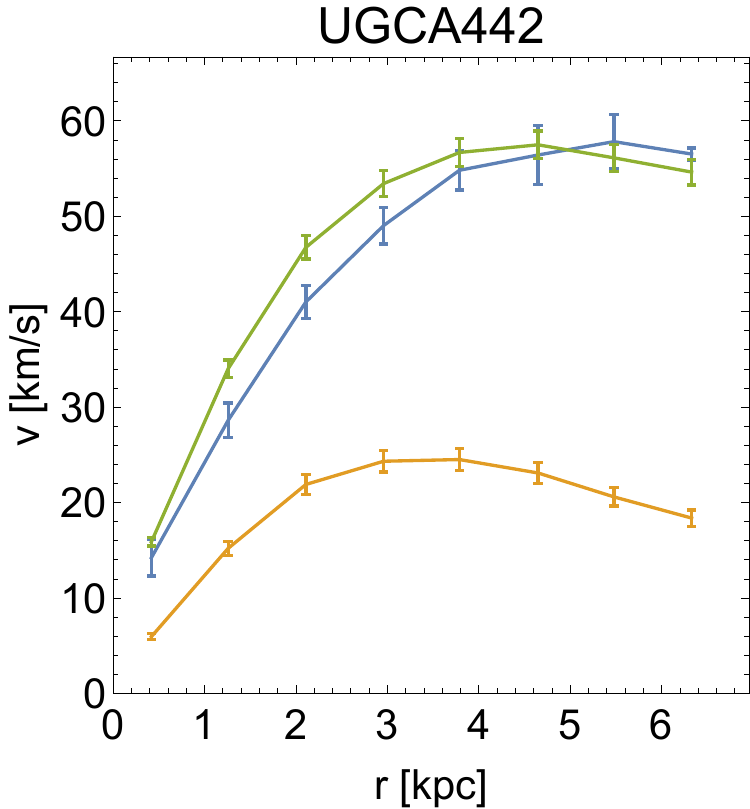}
	\includegraphics[width=0.2\textwidth]{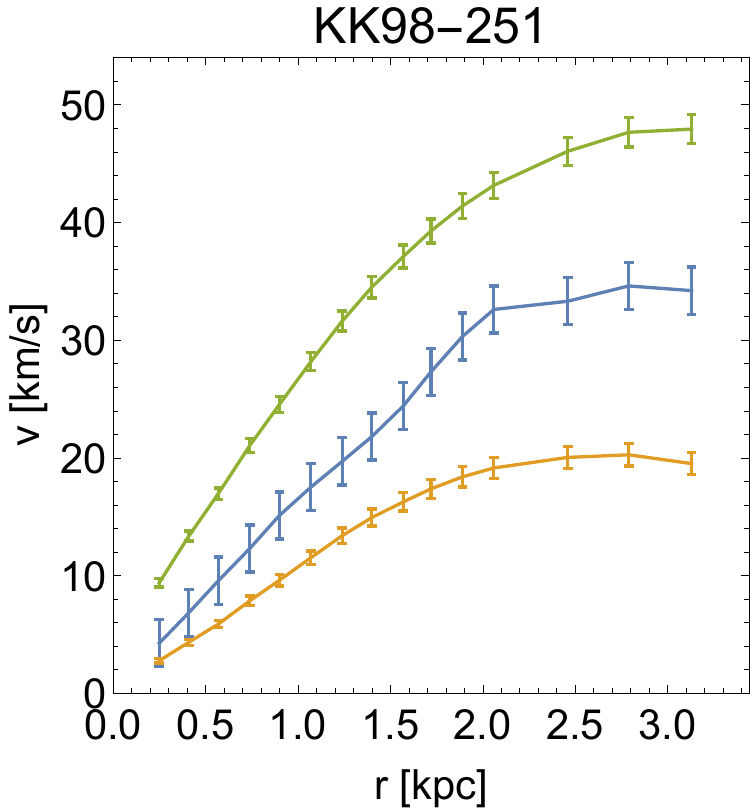}
	\includegraphics[width=0.2\textwidth]{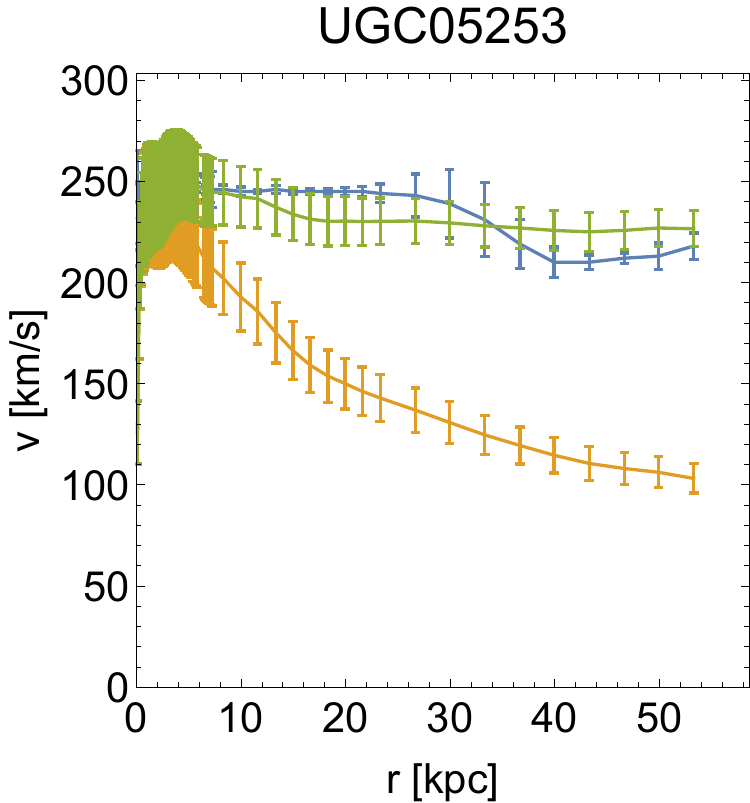}
	\includegraphics[width=0.2\textwidth]{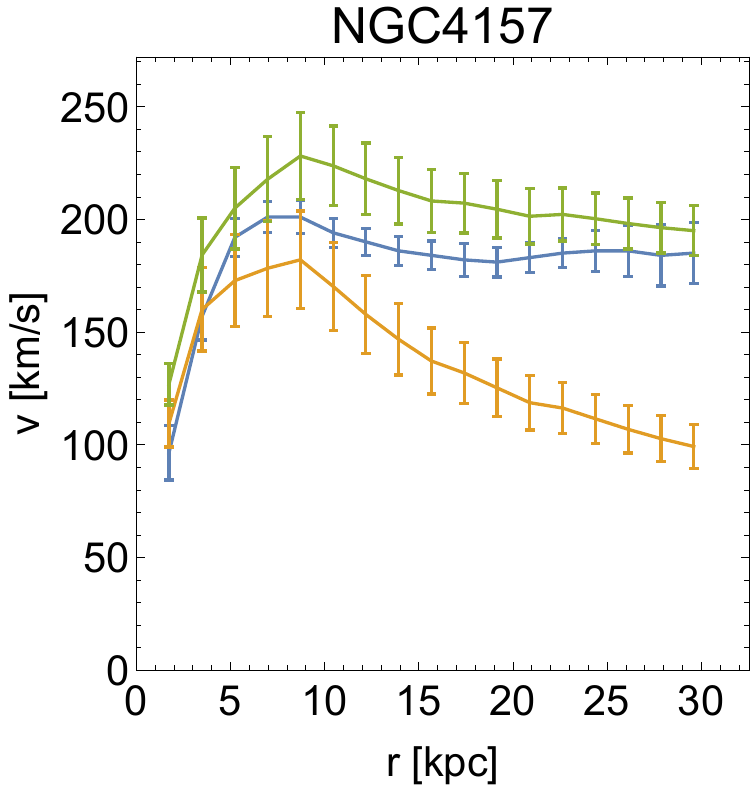}
	\includegraphics[width=0.2\textwidth]{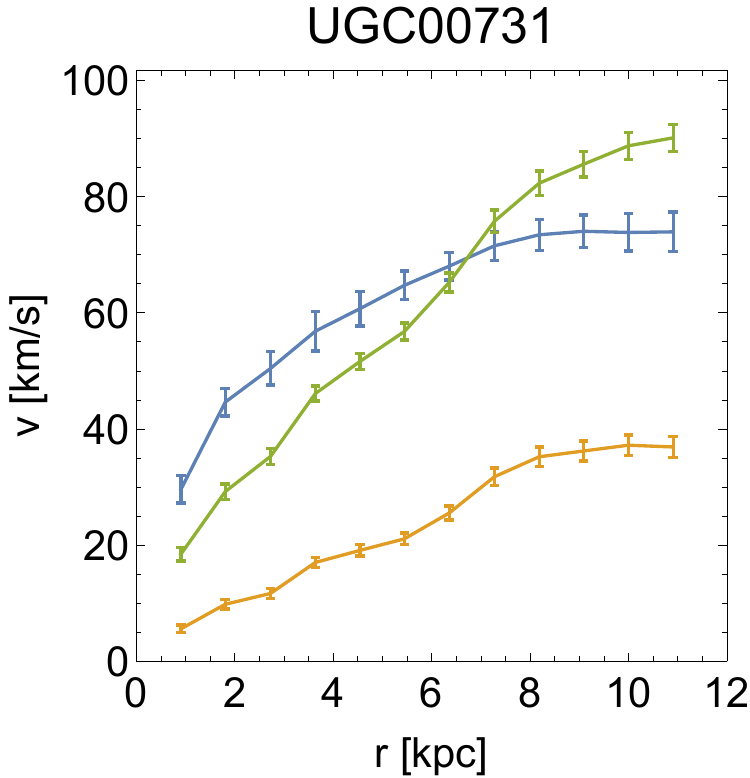}
	\includegraphics[width=0.2\textwidth]{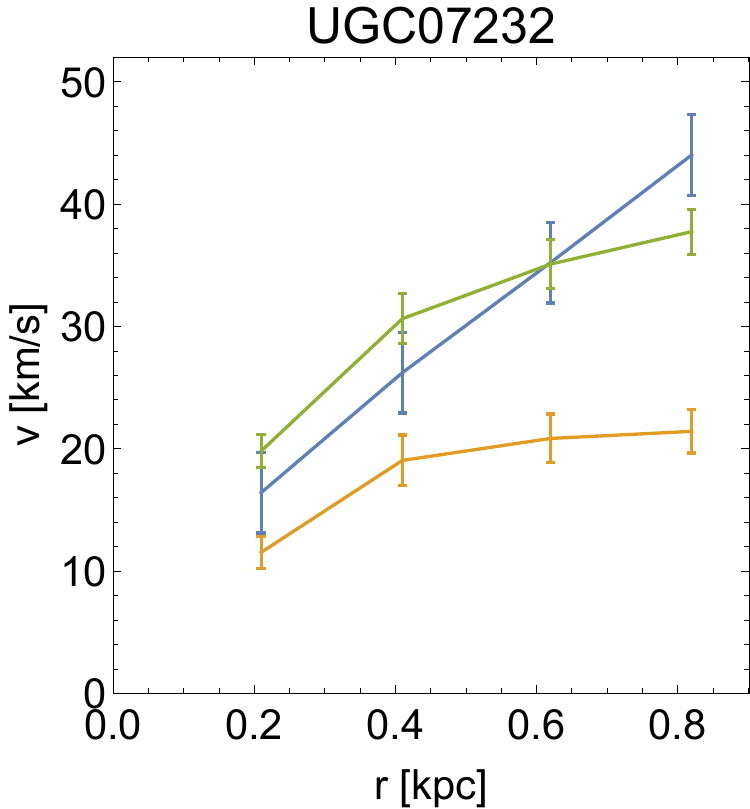}
	\includegraphics[width=0.2\textwidth]{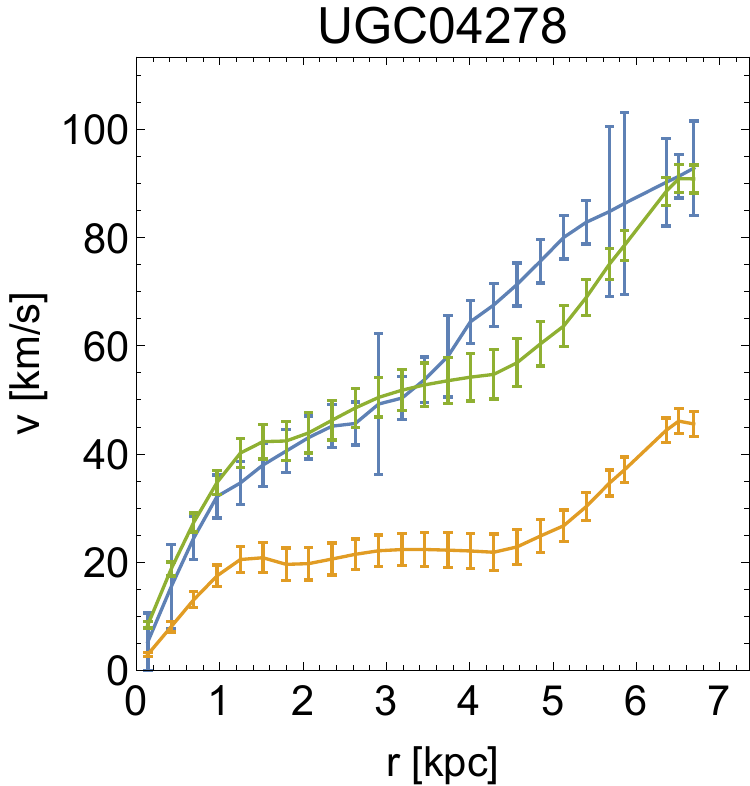}
	\includegraphics[width=0.2\textwidth]{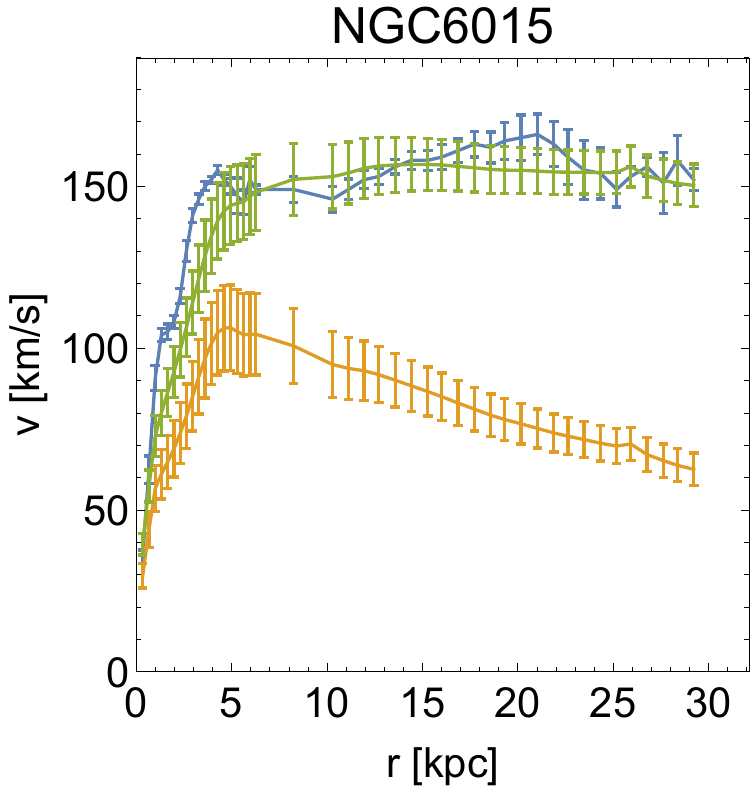}
	\includegraphics[width=0.2\textwidth]{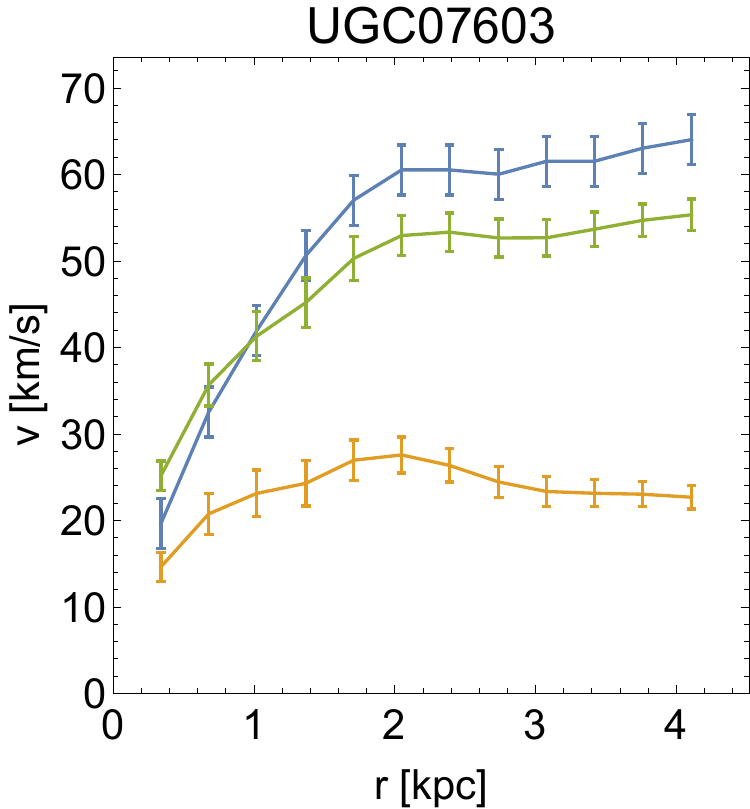}
	\includegraphics[width=0.2\textwidth]{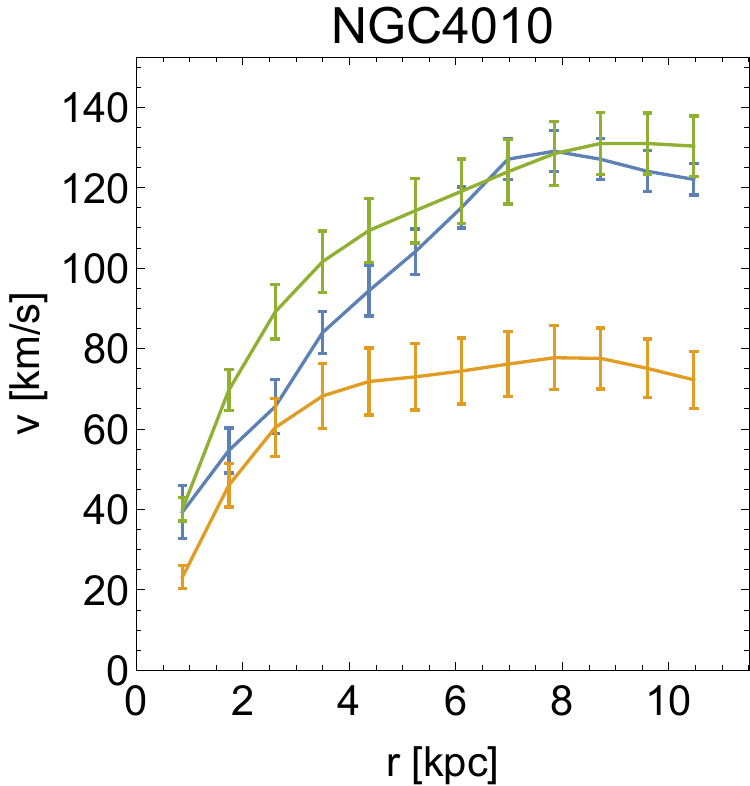}
	\includegraphics[width=0.2\textwidth]{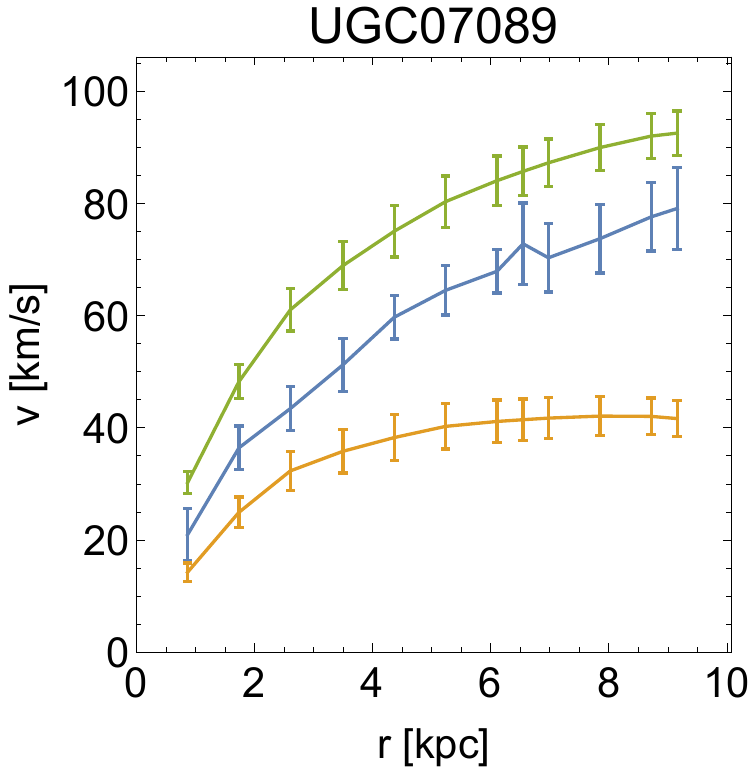}
	\includegraphics[width=0.2\textwidth]{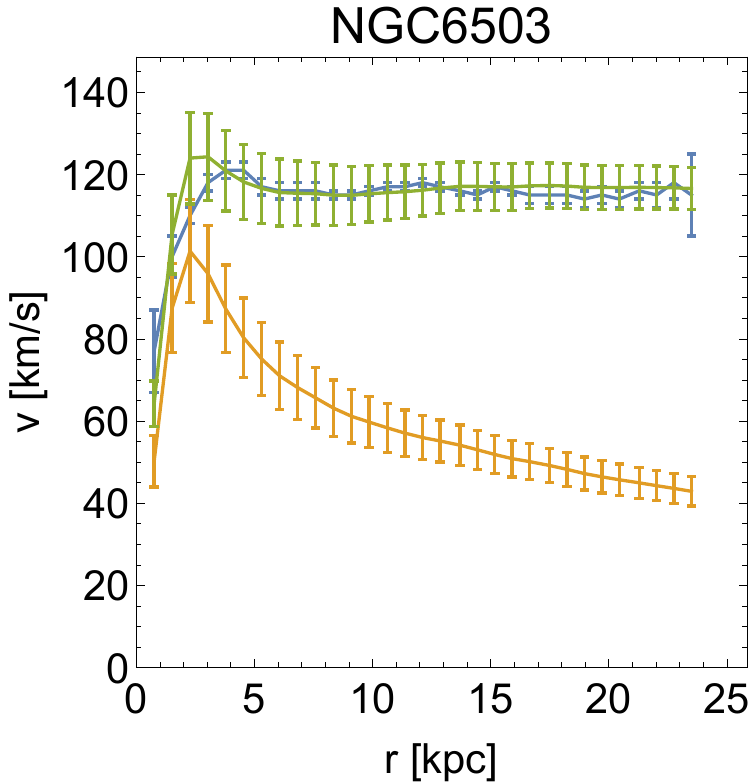}
	\includegraphics[width=0.2\textwidth]{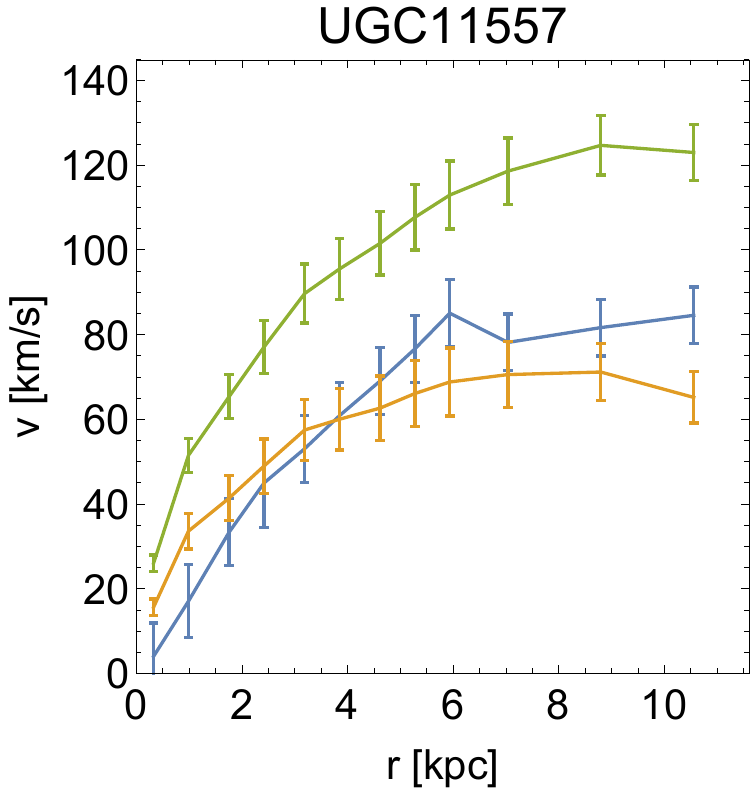}
	\includegraphics[width=0.2\textwidth]{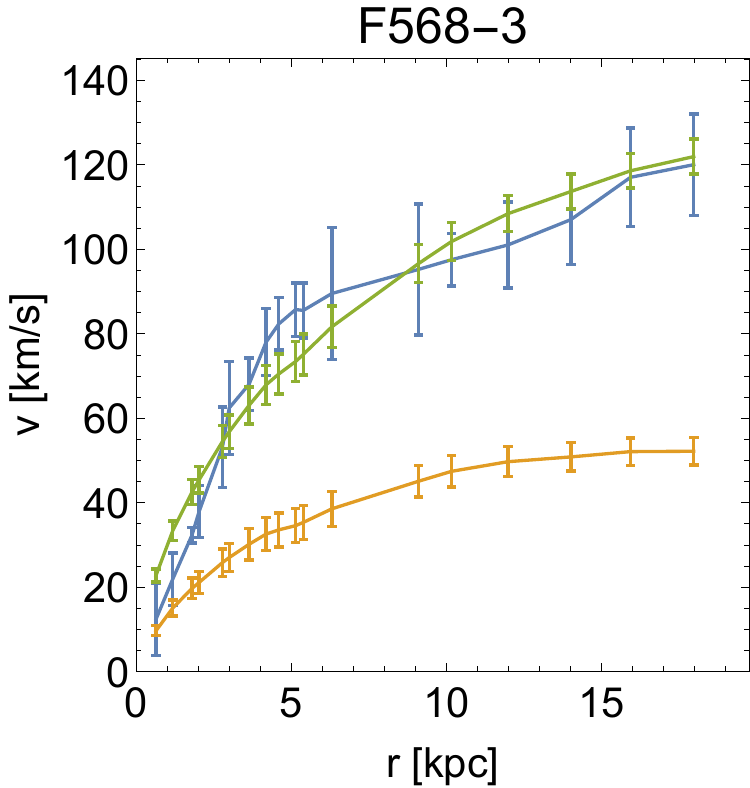}
	\includegraphics[width=0.2\textwidth]{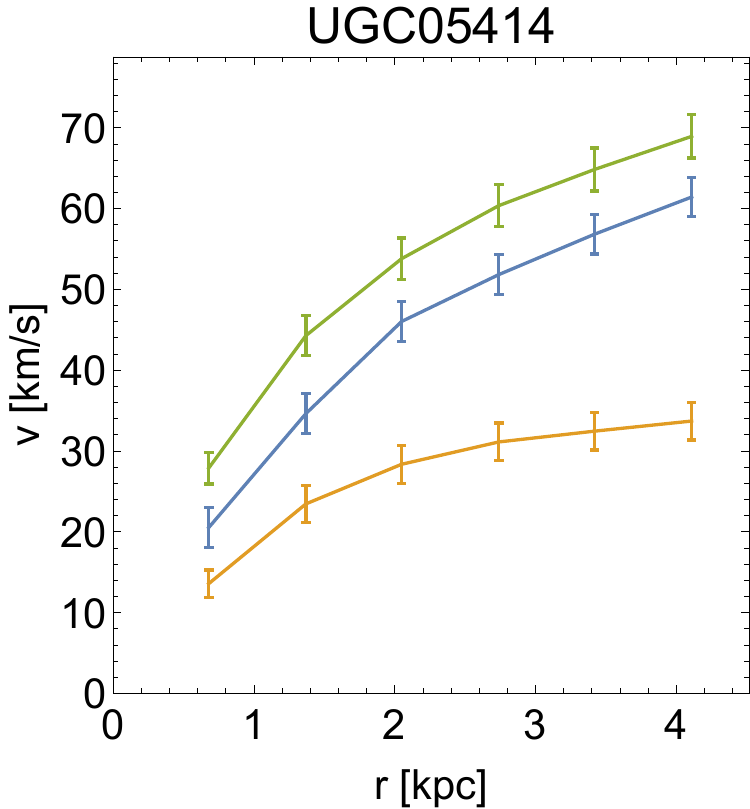}
	\includegraphics[width=0.2\textwidth]{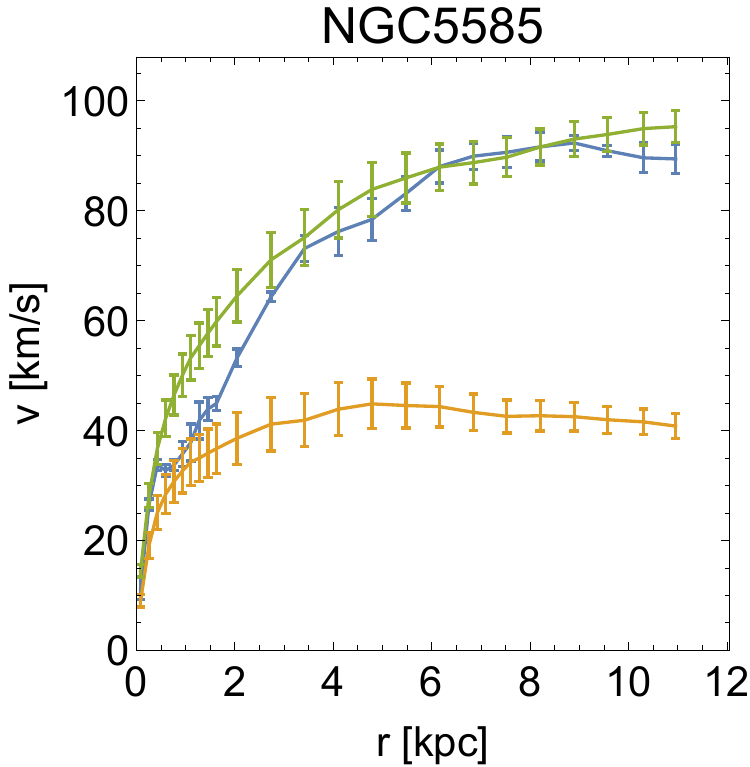}
	\includegraphics[width=0.2\textwidth]{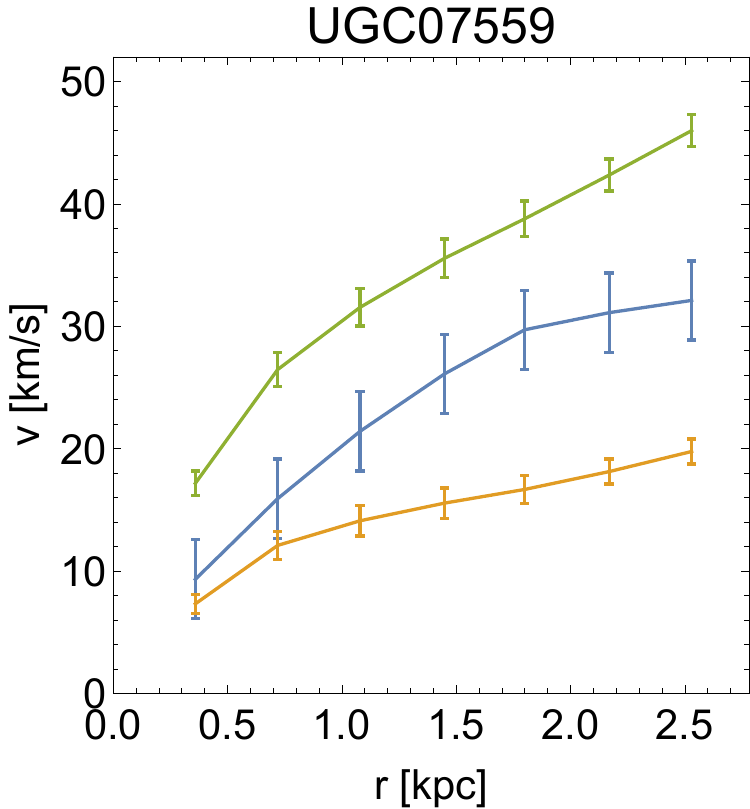}
	\includegraphics[width=0.2\textwidth]{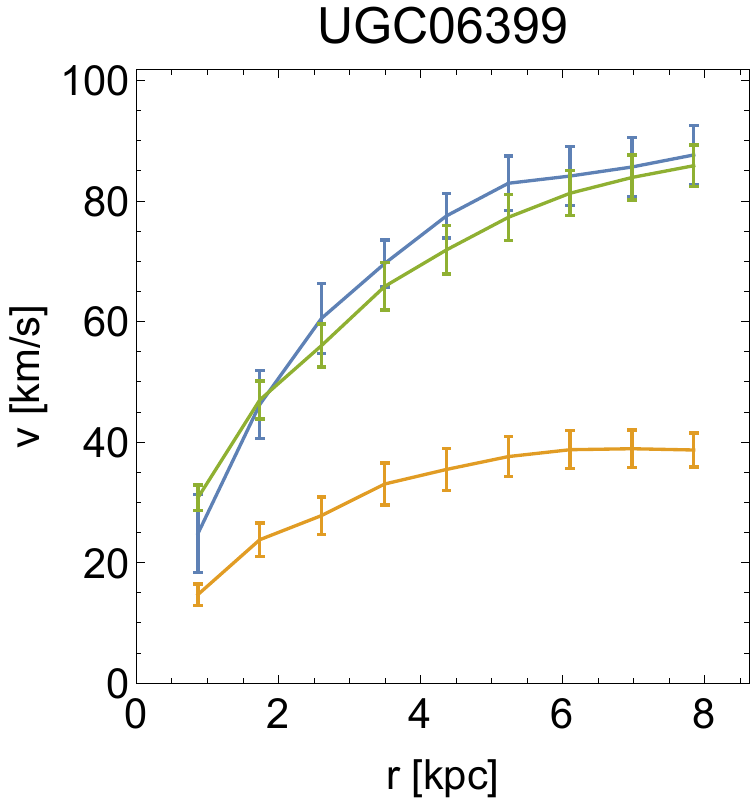}
	\includegraphics[width=0.2\textwidth]{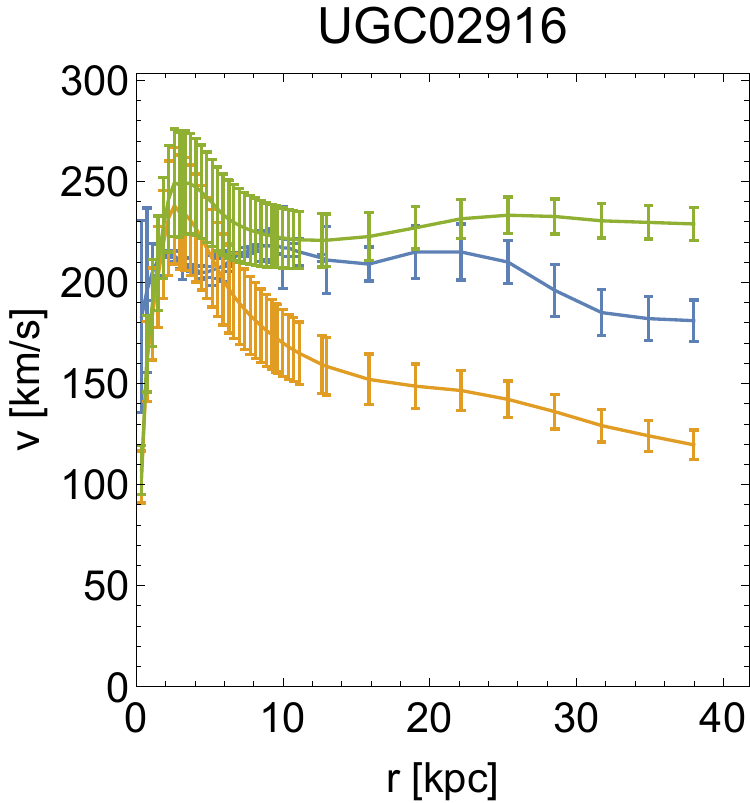}
	\caption{Rotation curve plots of galaxies corresponding to figure \ref{fig:p4} (same ordering as well). Blue denote $v_{obs}$, yellow denote $v_{bar}$ and green denote $v_{tot}^{RAR}$ based on observed $v_{bar}$.}
	\label{fig:p8}
\end{figure*}

\begin{figure*}
	\centering
	\includegraphics[width=0.2\textwidth]{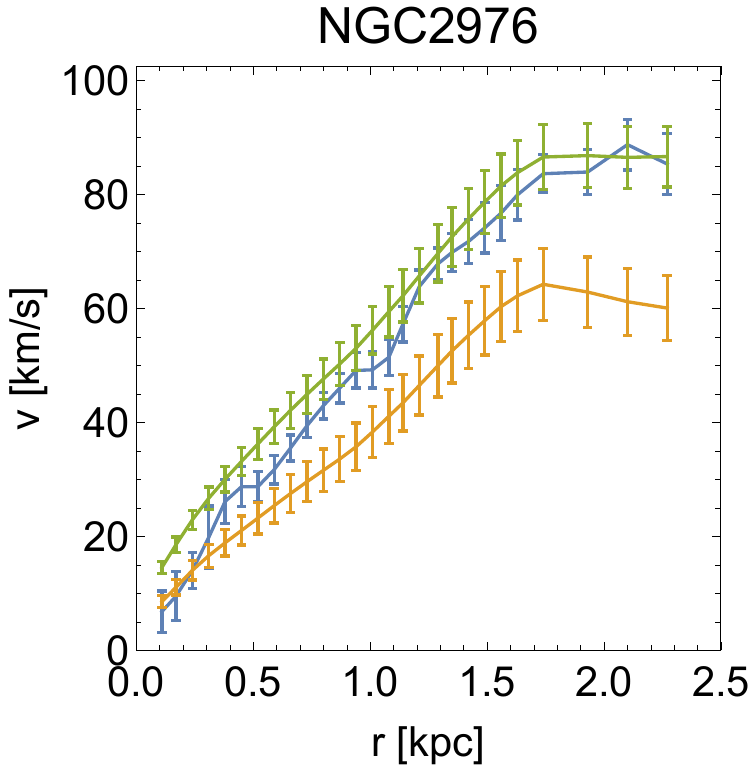}
	\includegraphics[width=0.2\textwidth]{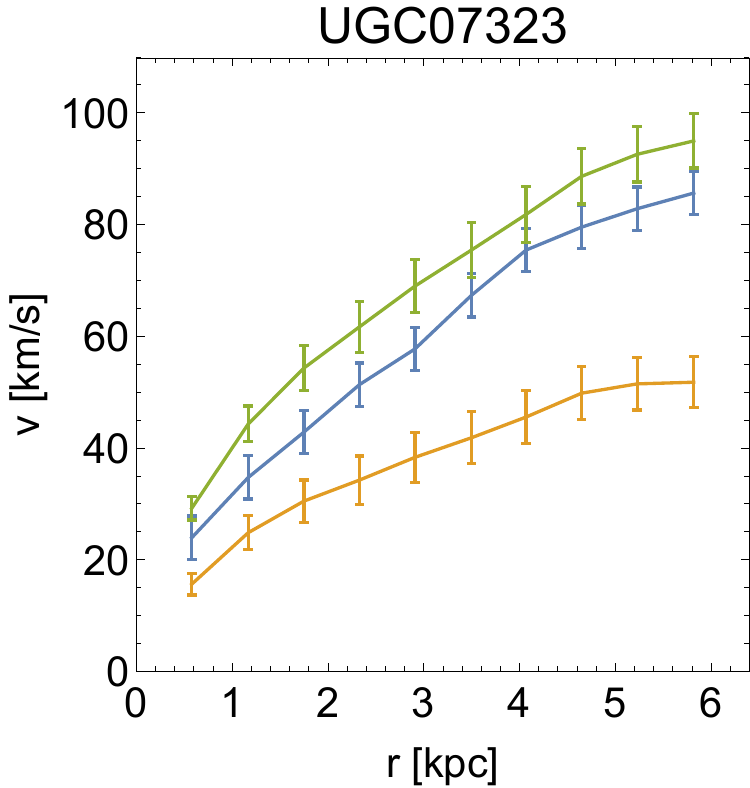}
	\includegraphics[width=0.2\textwidth]{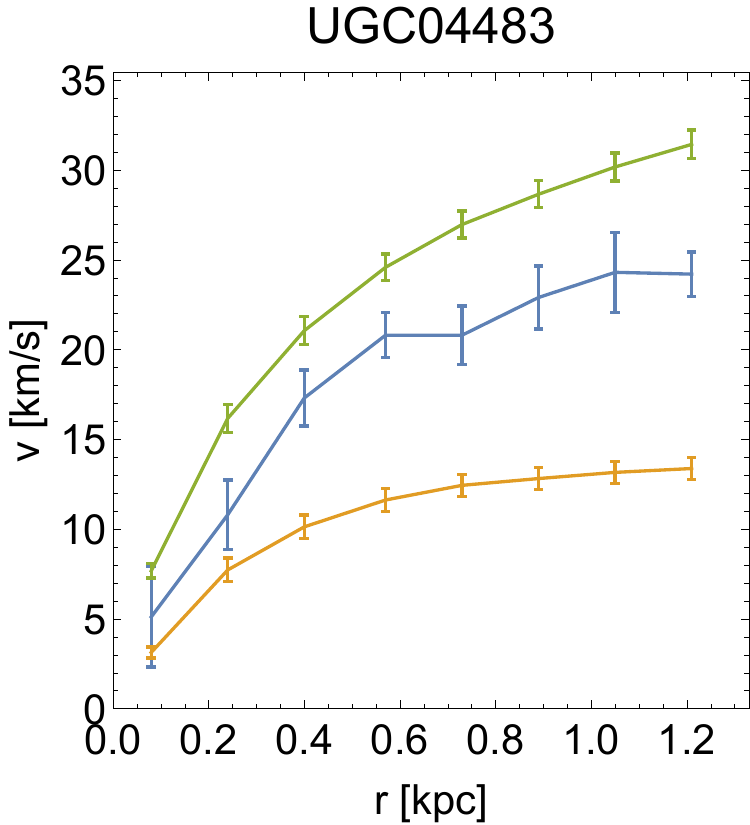}
	\includegraphics[width=0.2\textwidth]{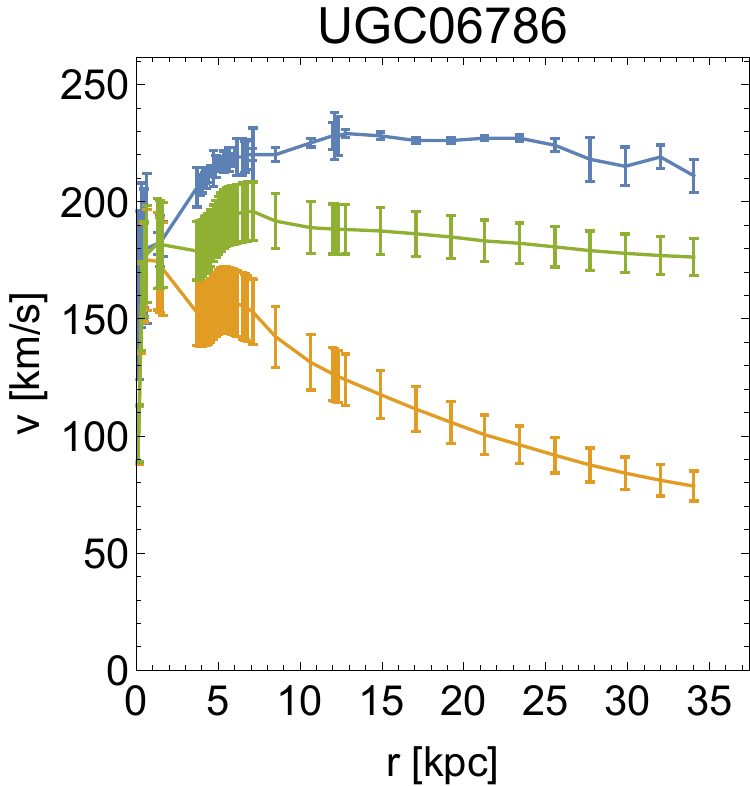}
	\includegraphics[width=0.2\textwidth]{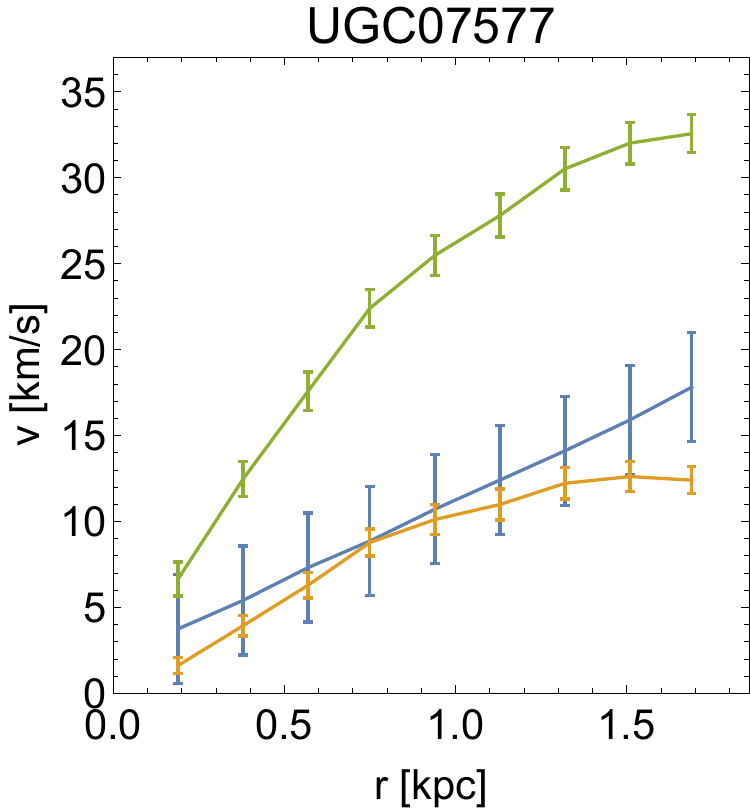}
	\includegraphics[width=0.2\textwidth]{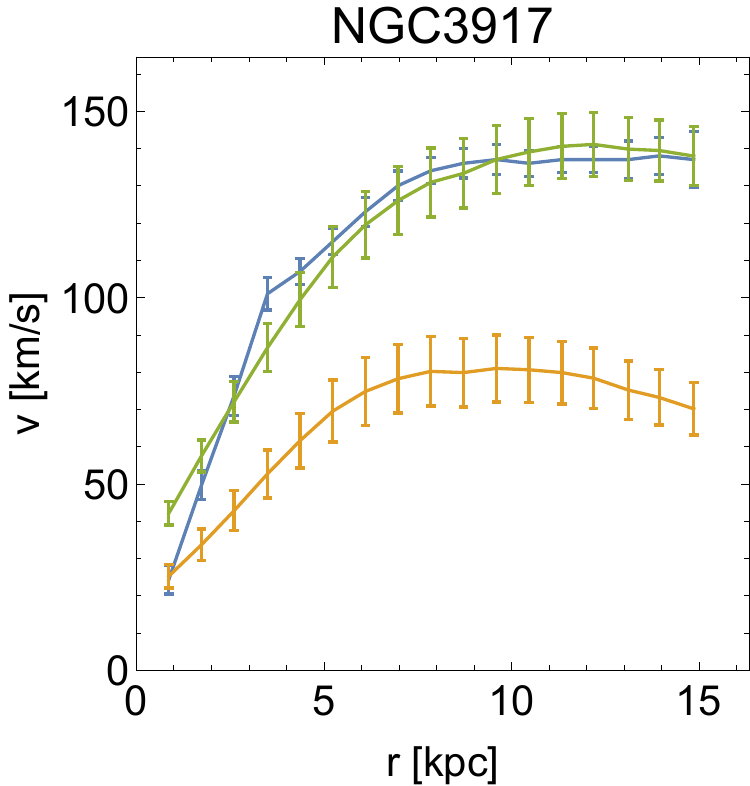}
	\includegraphics[width=0.2\textwidth]{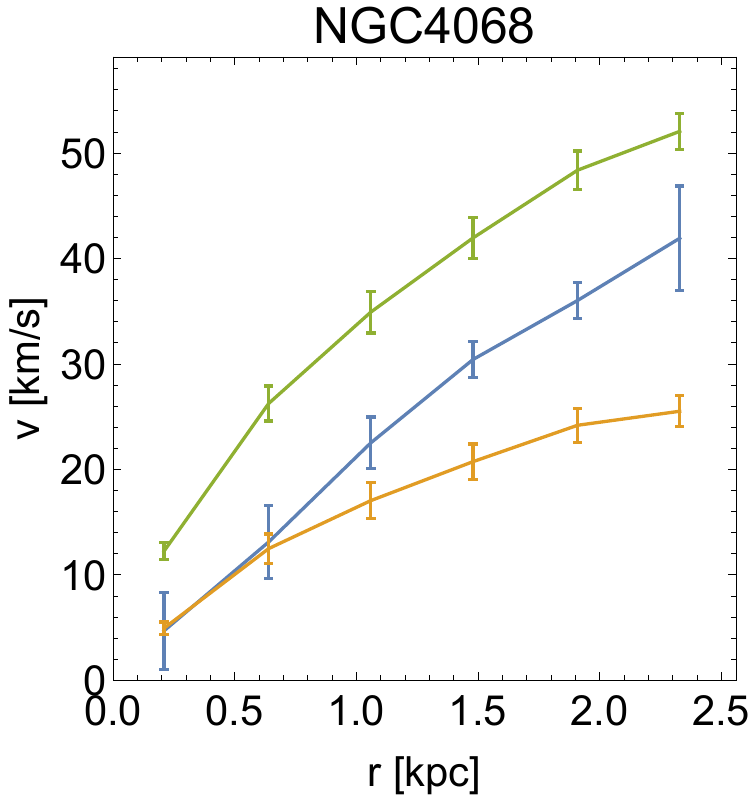}
	\includegraphics[width=0.2\textwidth]{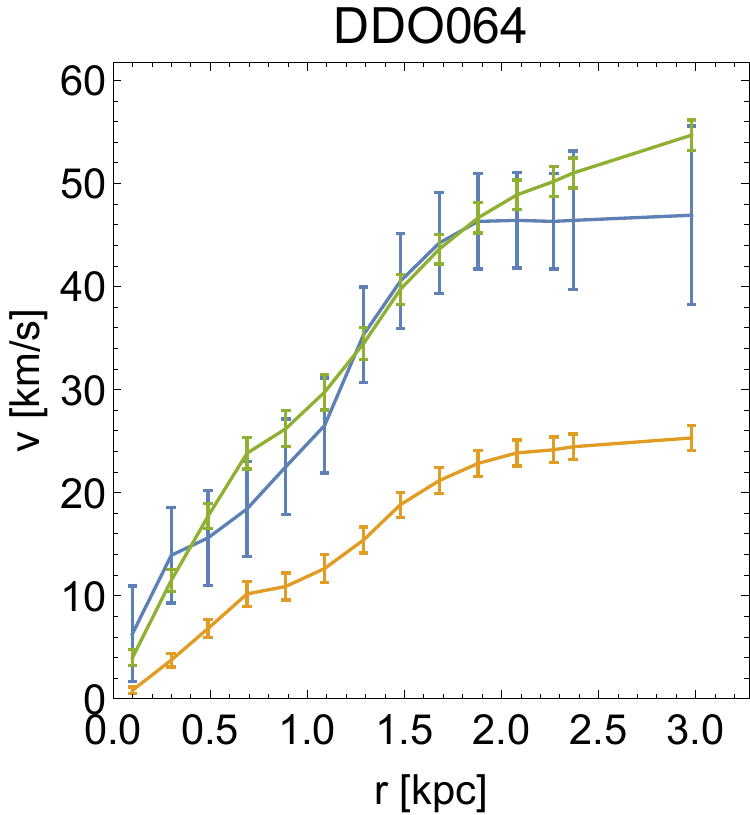}
	\includegraphics[width=0.2\textwidth]{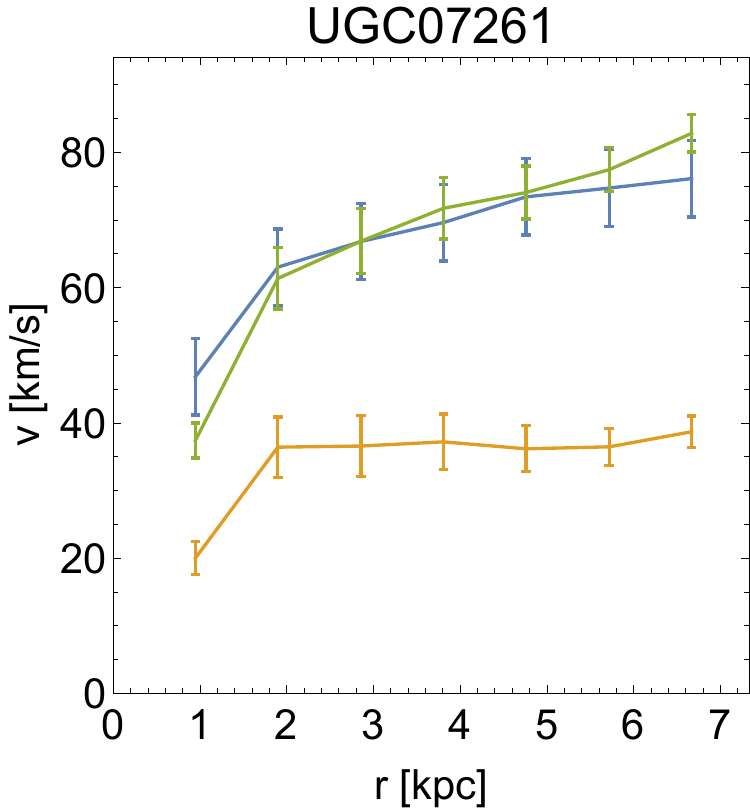}
	\includegraphics[width=0.2\textwidth]{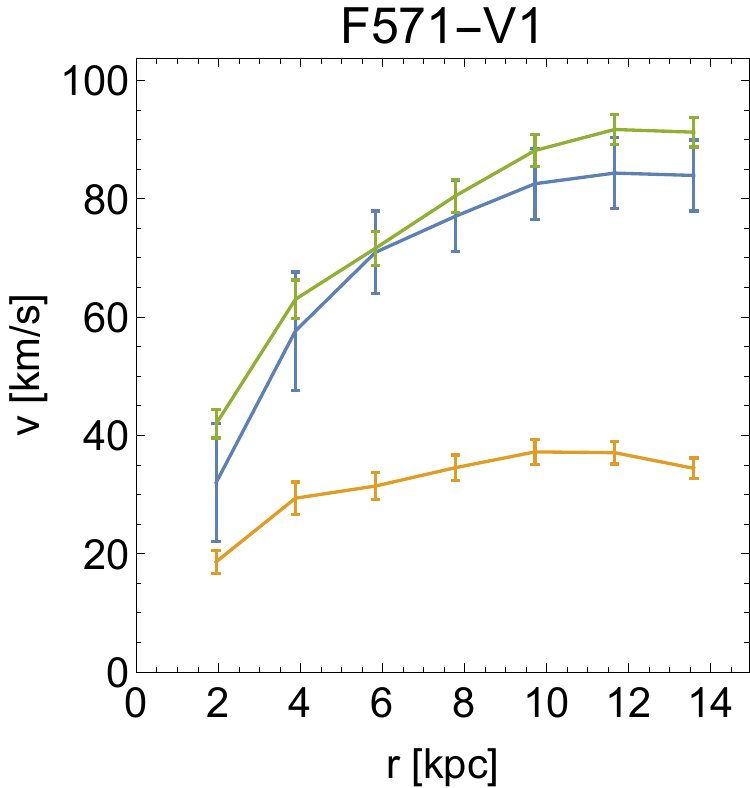}
	\includegraphics[width=0.2\textwidth]{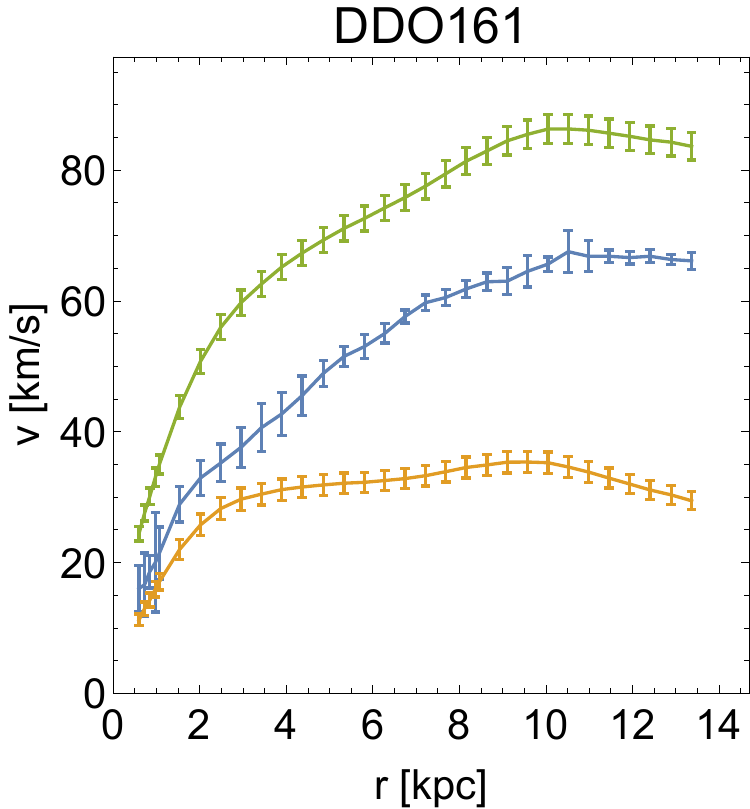}
	\includegraphics[width=0.2\textwidth]{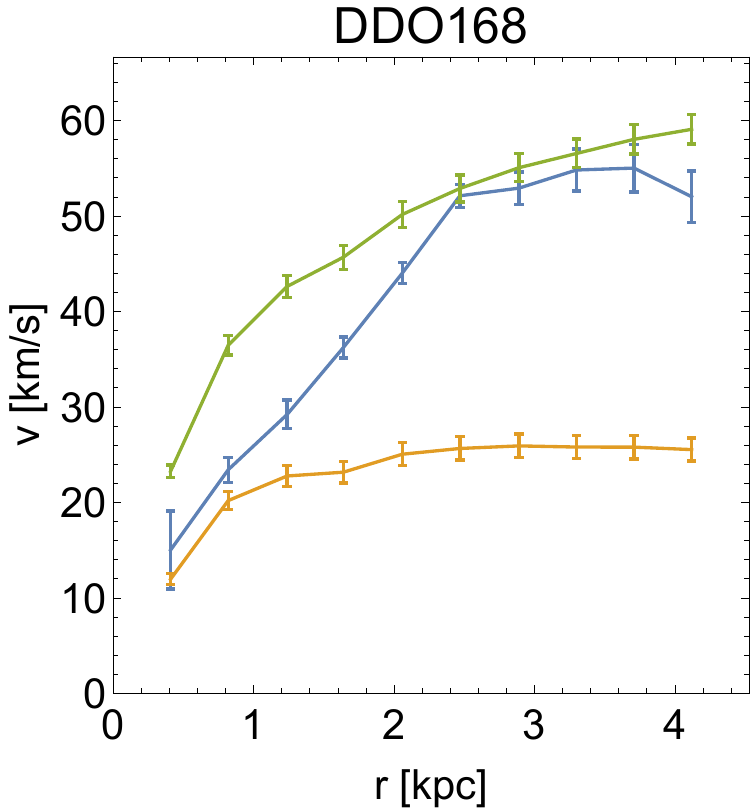}
	\includegraphics[width=0.2\textwidth]{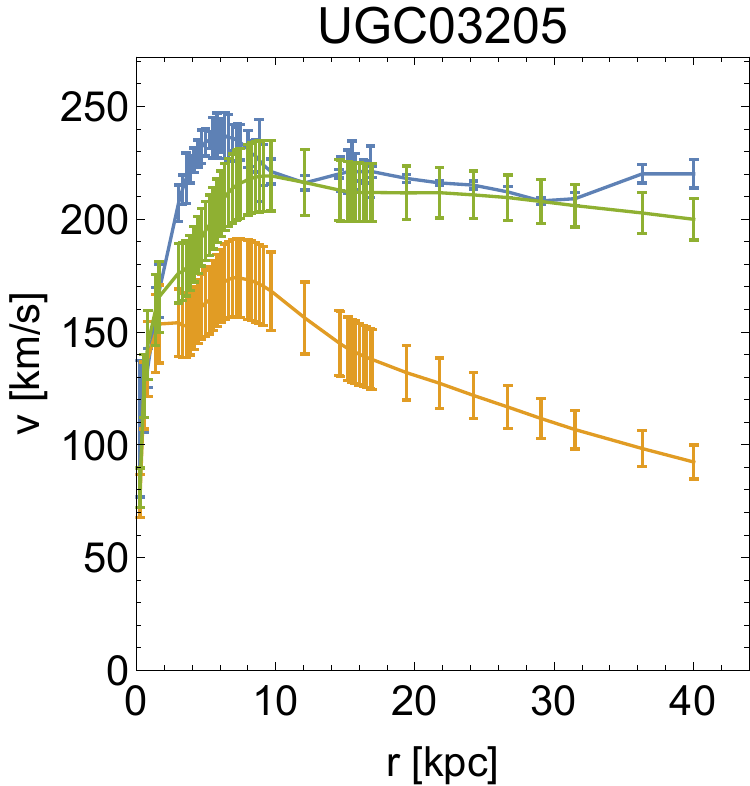}
	\includegraphics[width=0.2\textwidth]{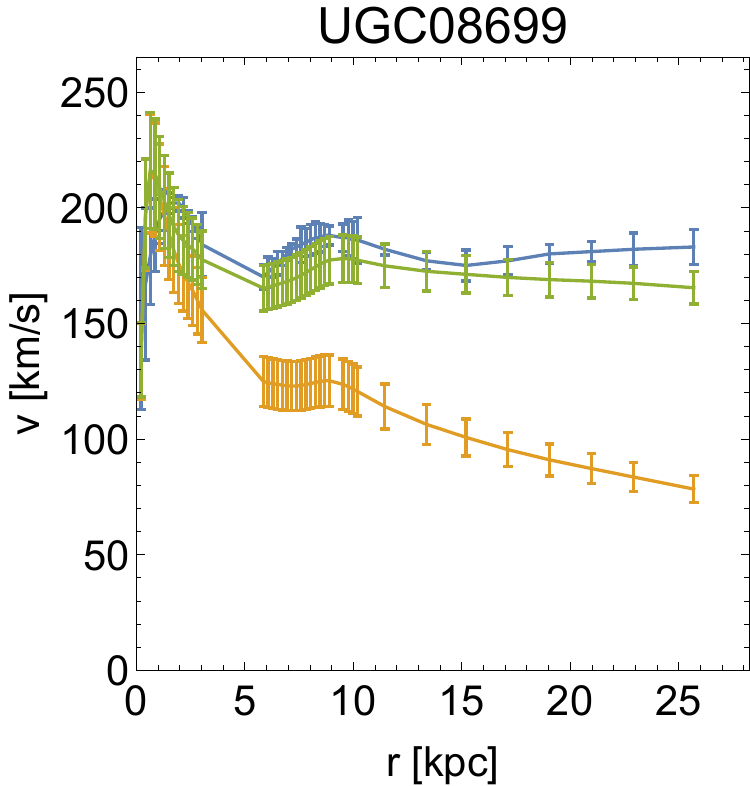}
	\includegraphics[width=0.2\textwidth]{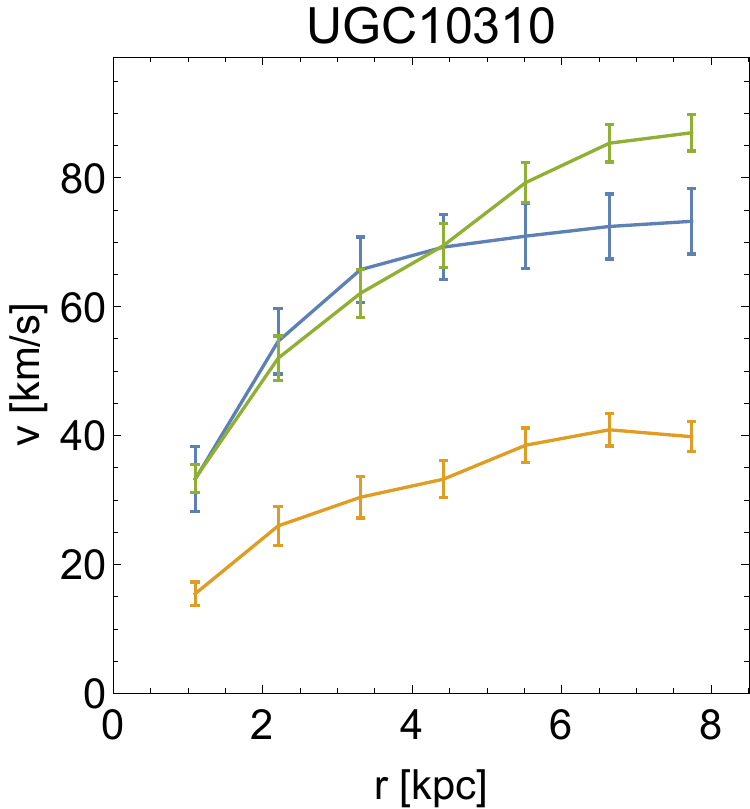}
	\includegraphics[width=0.2\textwidth]{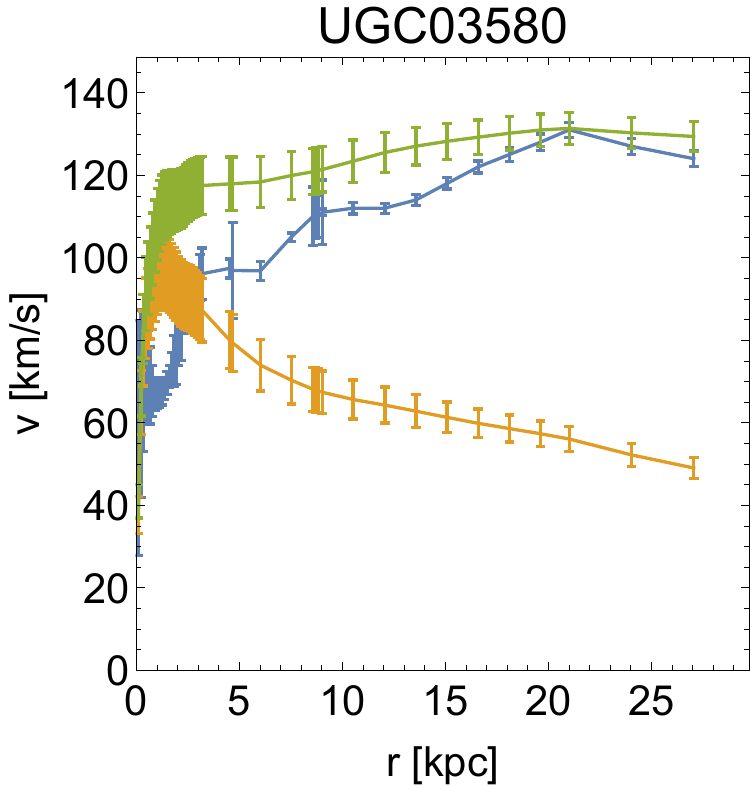}
	\includegraphics[width=0.2\textwidth]{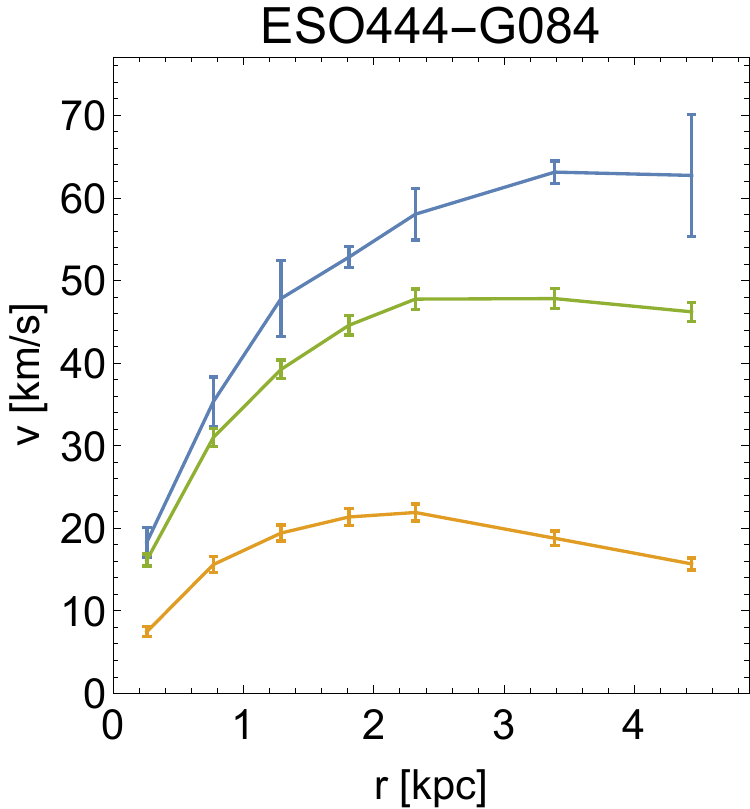}
	\includegraphics[width=0.2\textwidth]{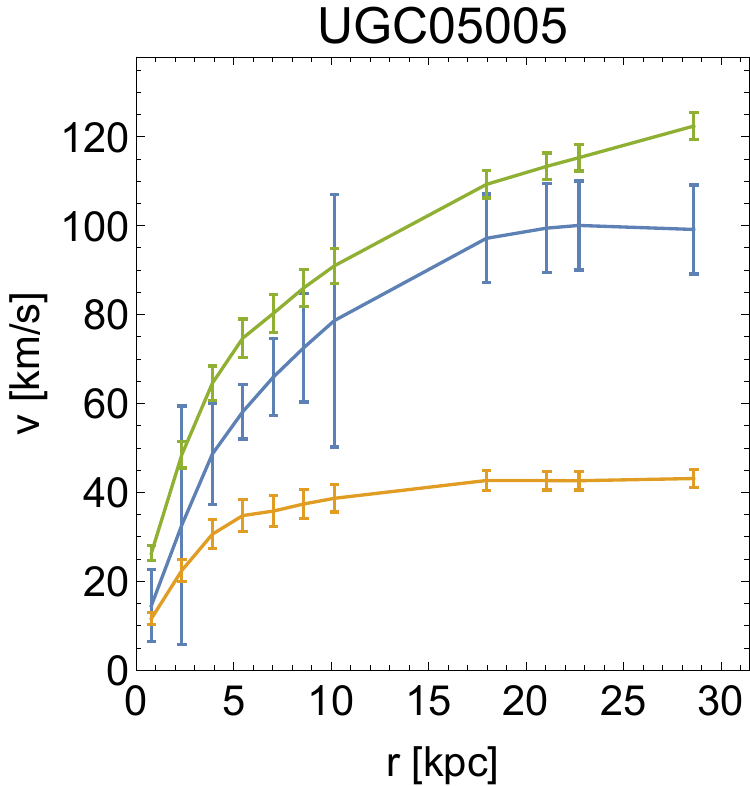}
	\includegraphics[width=0.2\textwidth]{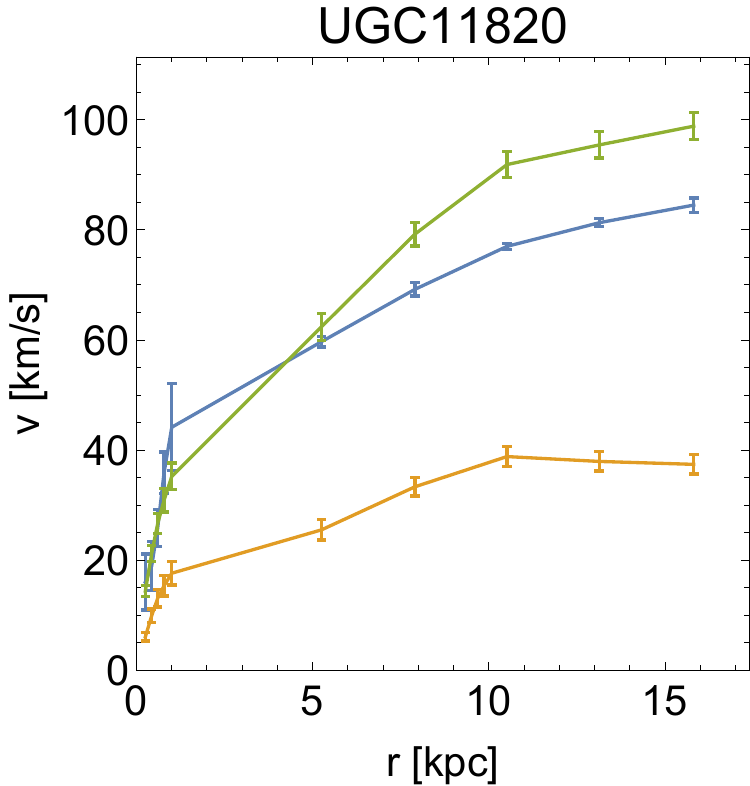}
	\includegraphics[width=0.2\textwidth]{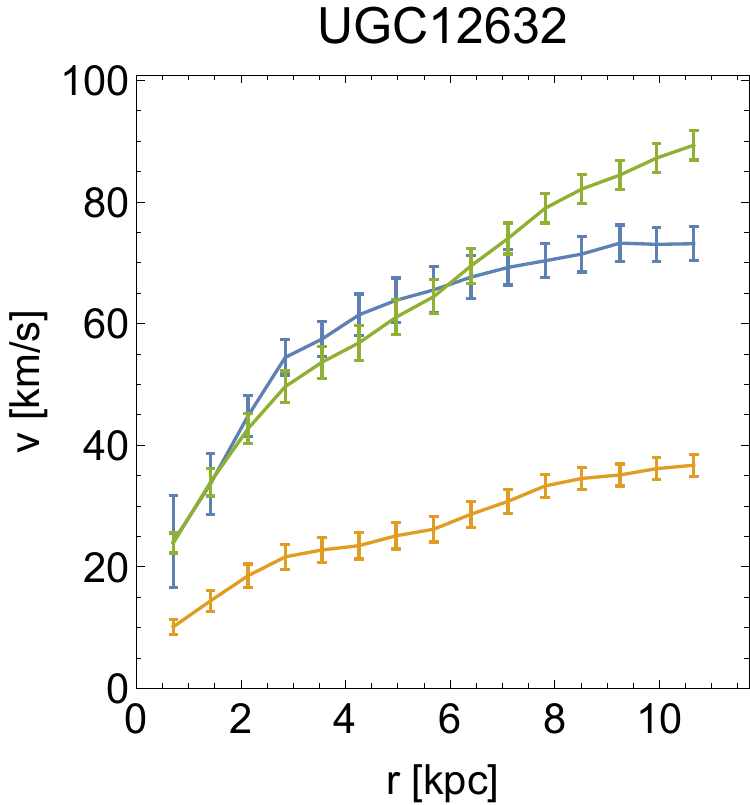}
	\includegraphics[width=0.2\textwidth]{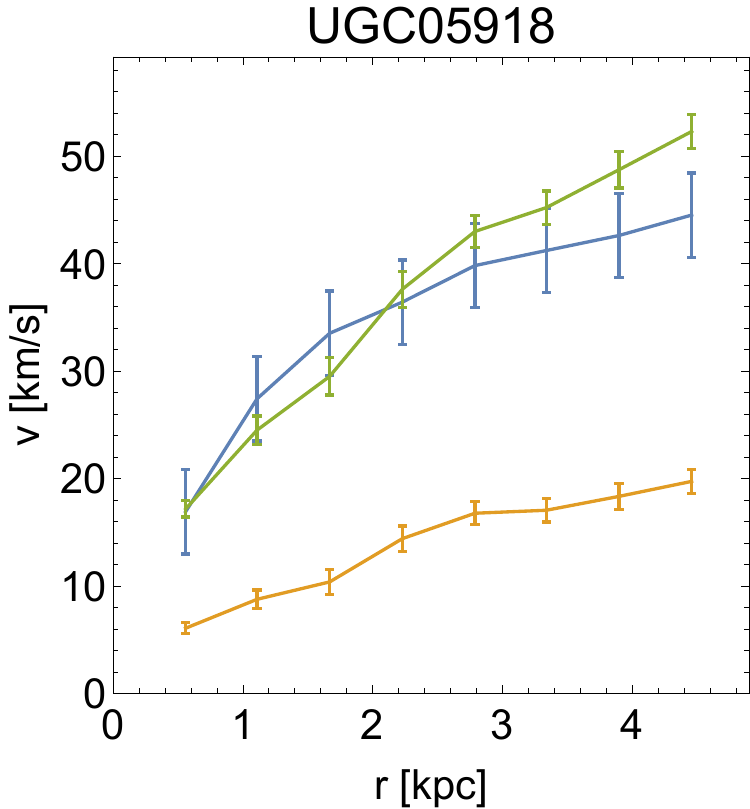}
	\includegraphics[width=0.2\textwidth]{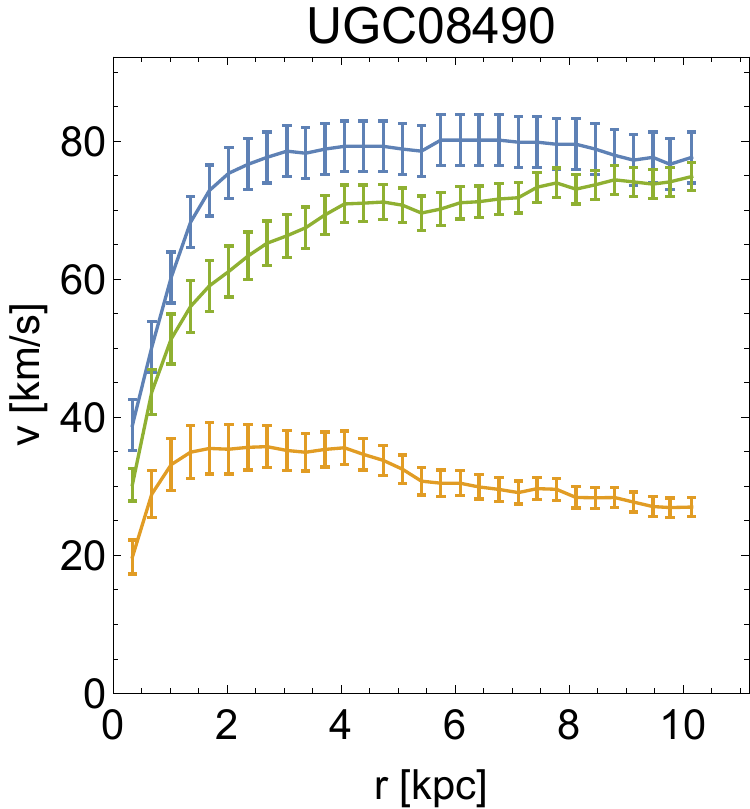}
	\includegraphics[width=0.2\textwidth]{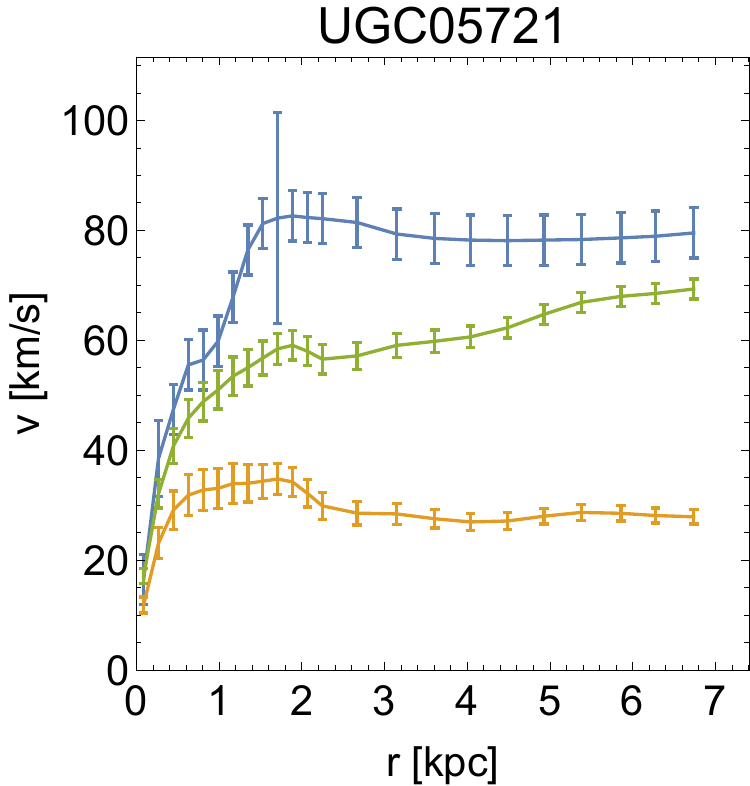}
	\includegraphics[width=0.2\textwidth]{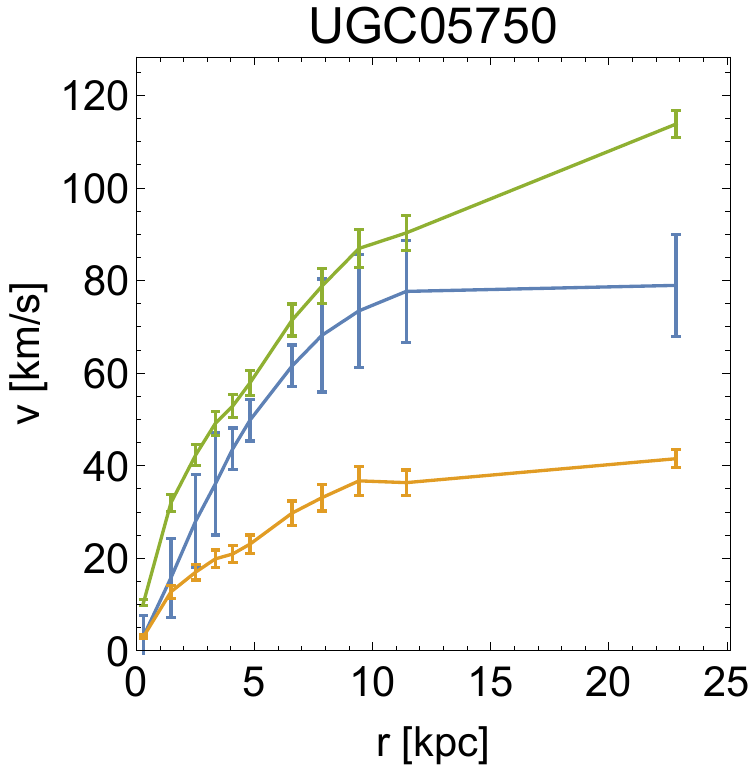}
	\caption{Rotation curve plots of galaxies corresponding to figure \ref{fig:p5} (same ordering as well). Blue denote $v_{obs}$, yellow denote $v_{bar}$ and green denote $v_{tot}^{RAR}$ based on observed $v_{bar}$.}
	\label{fig:p9}
\end{figure*}

\begin{figure*}
	\centering
	\includegraphics[width=0.2\textwidth]{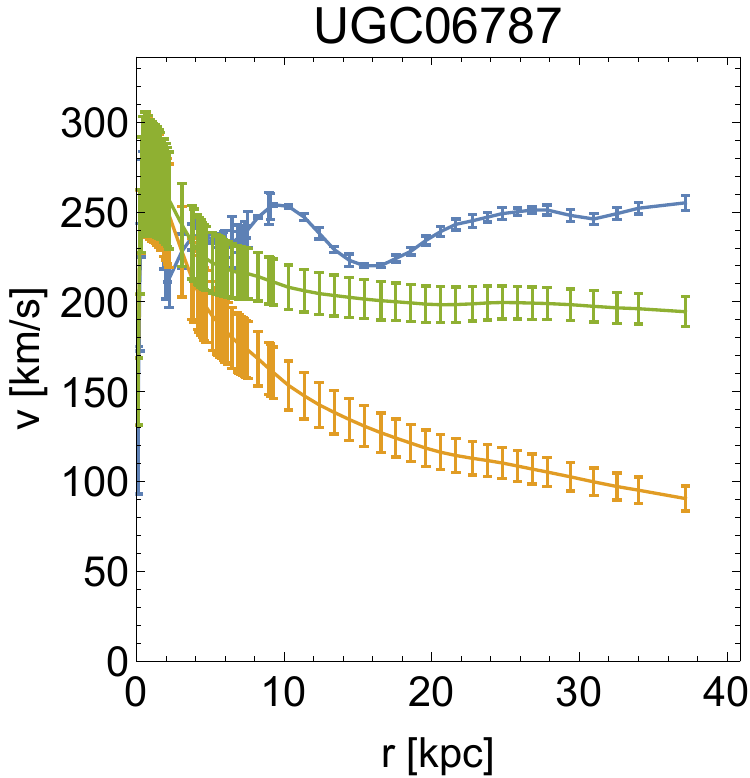}
	\includegraphics[width=0.2\textwidth]{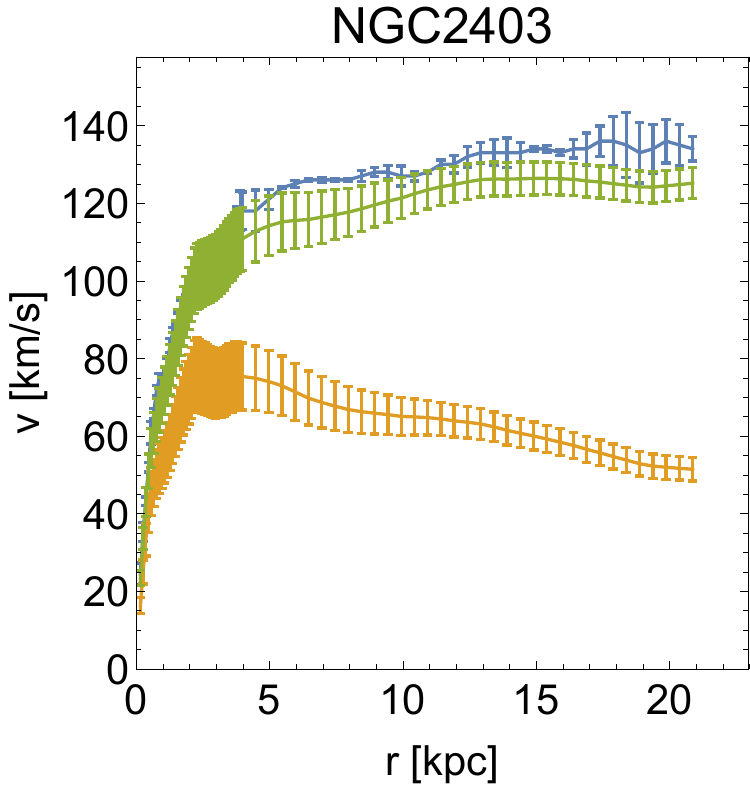}
	\includegraphics[width=0.2\textwidth]{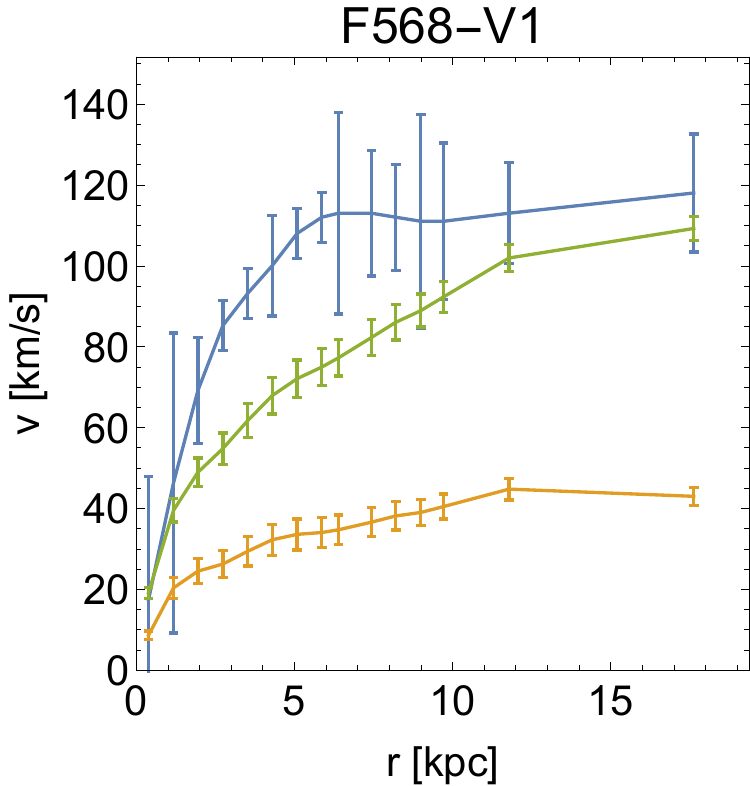}
	\includegraphics[width=0.2\textwidth]{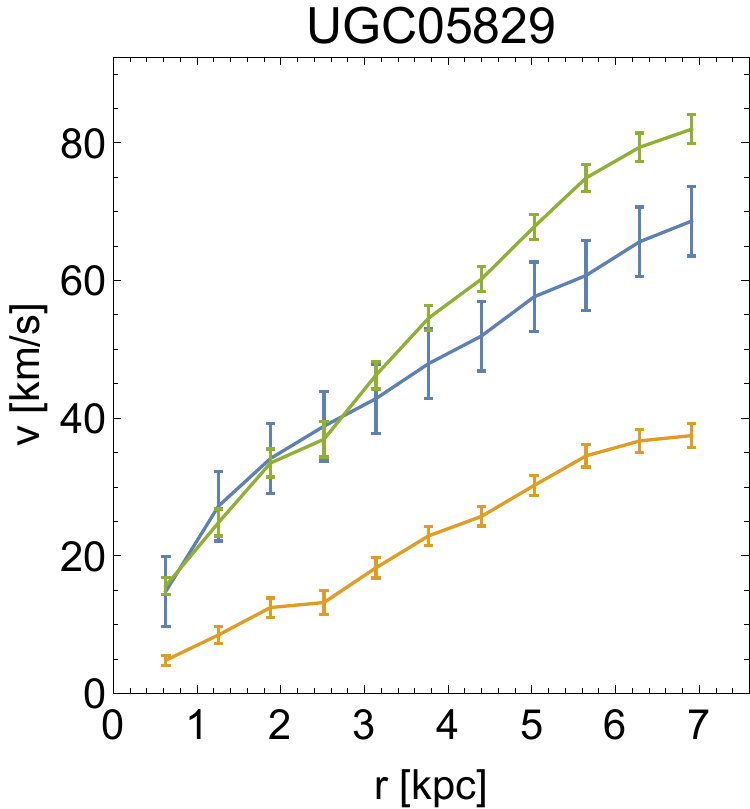}
	\includegraphics[width=0.2\textwidth]{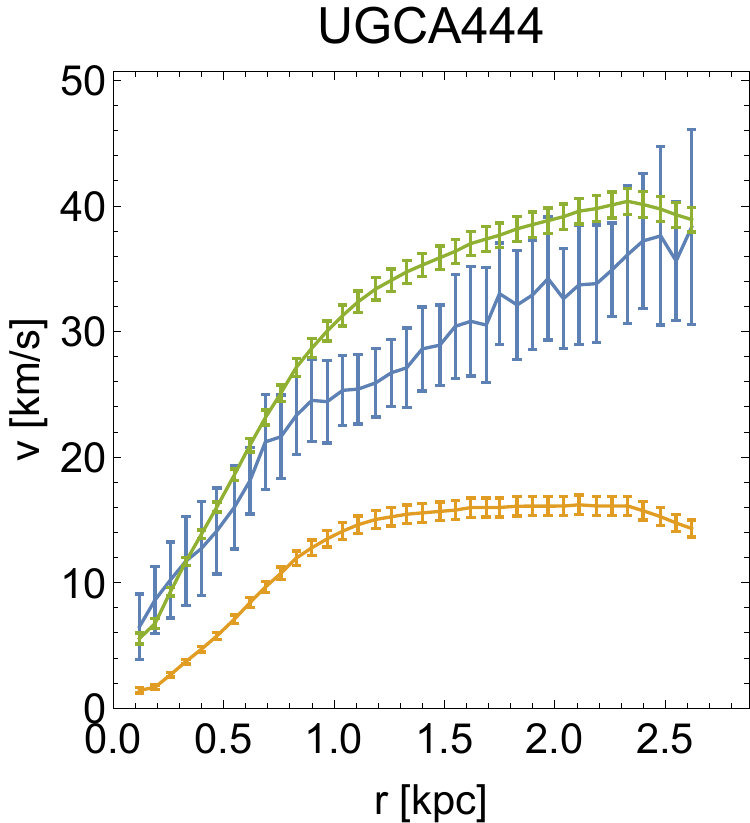}
	\includegraphics[width=0.2\textwidth]{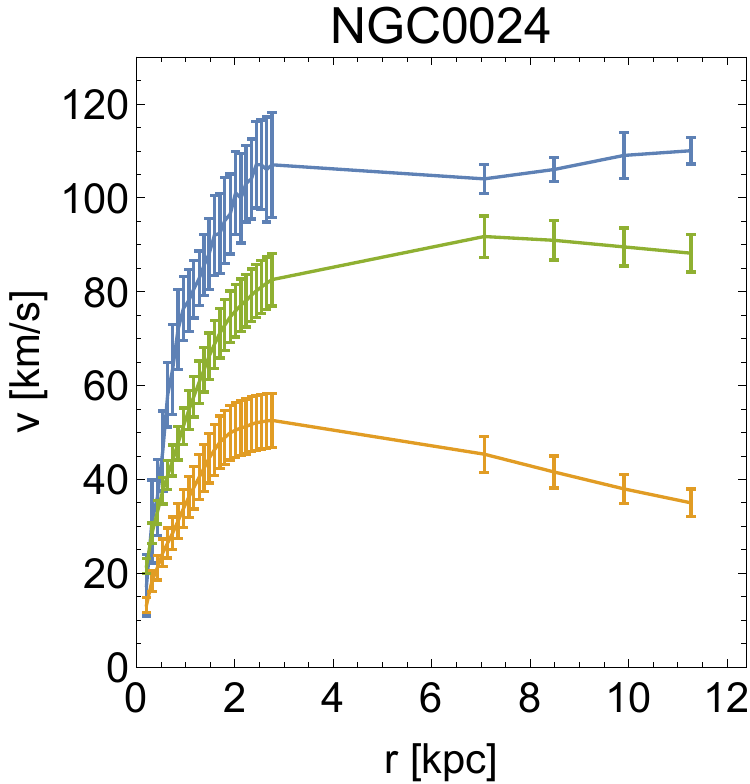}
	\includegraphics[width=0.2\textwidth]{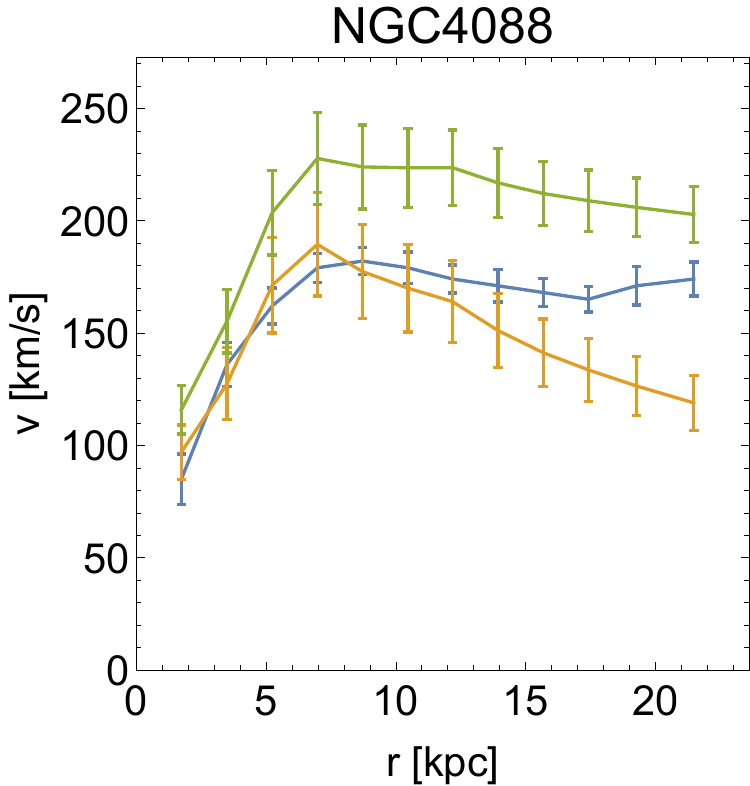}
	\includegraphics[width=0.2\textwidth]{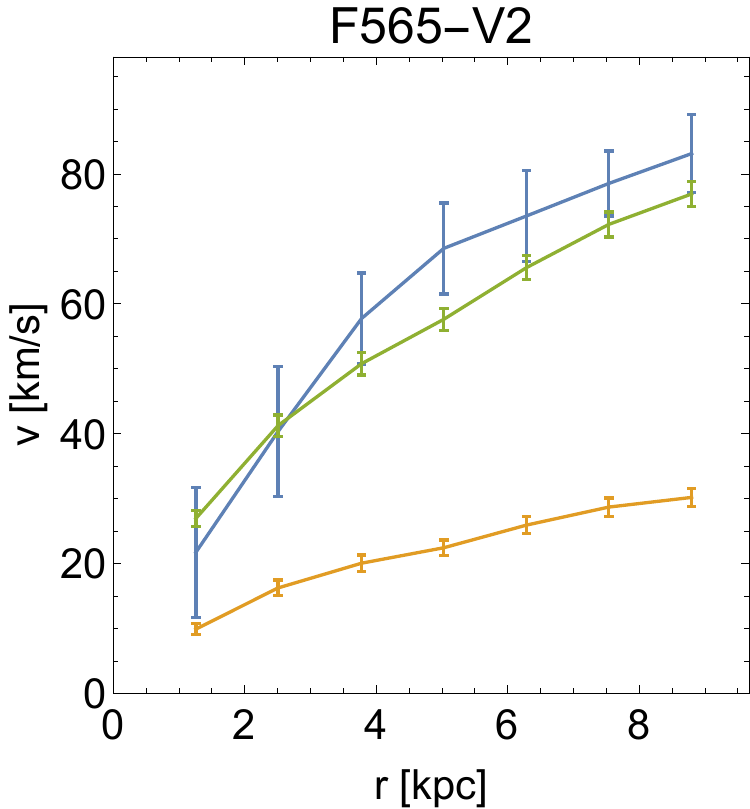}
	\includegraphics[width=0.2\textwidth]{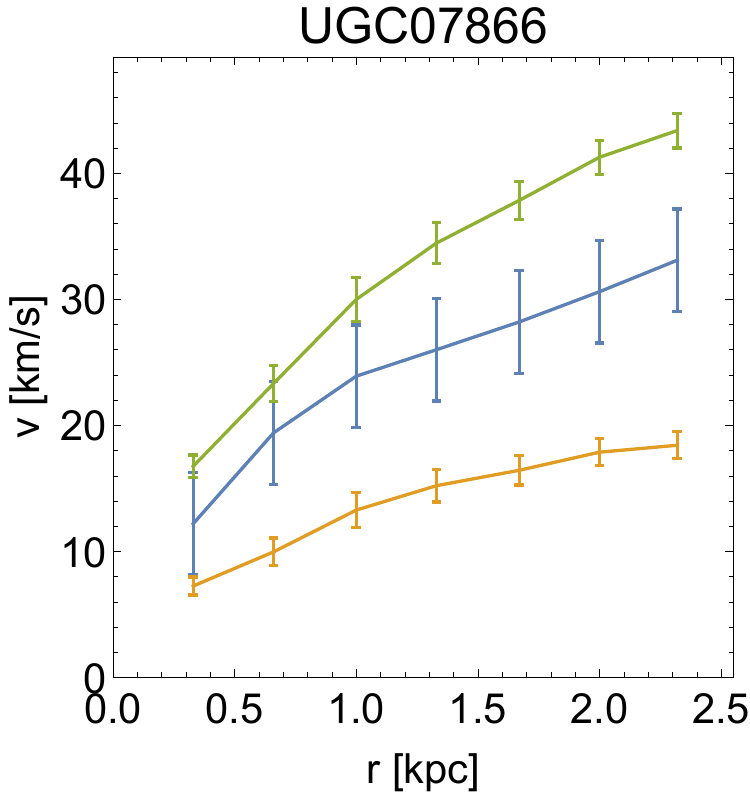}
	\includegraphics[width=0.2\textwidth]{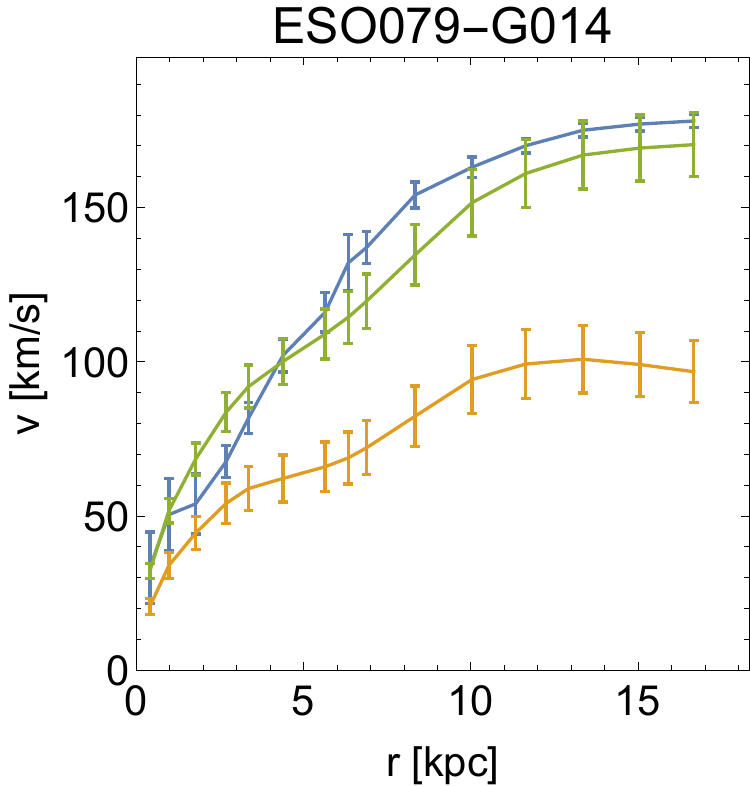}
	\includegraphics[width=0.2\textwidth]{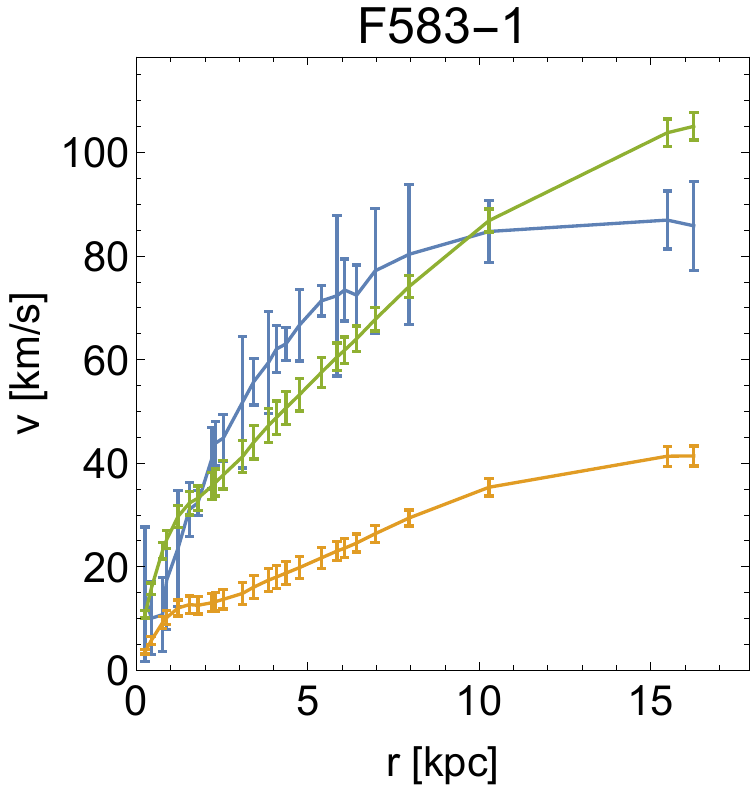}
	\includegraphics[width=0.2\textwidth]{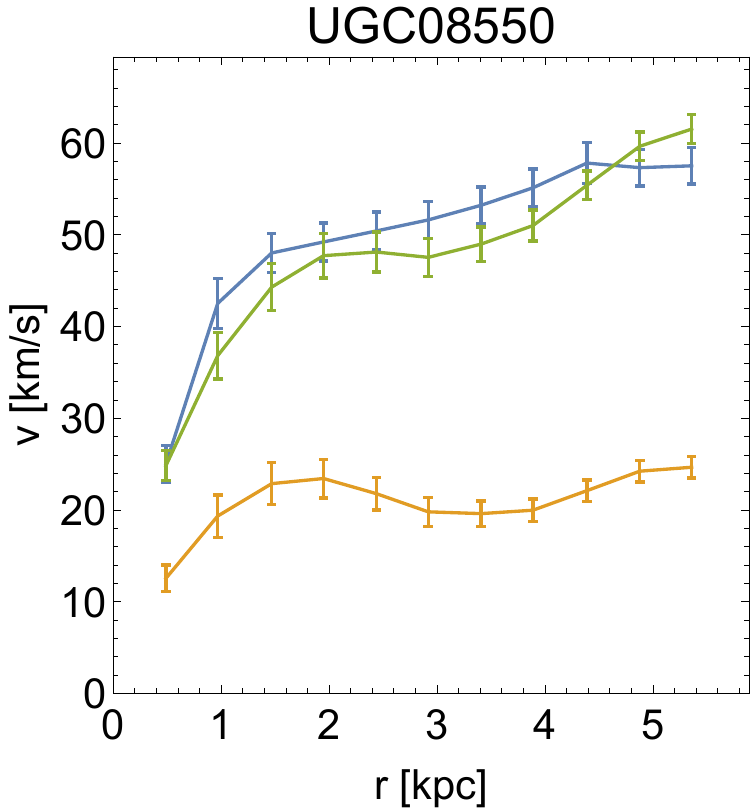}
	\includegraphics[width=0.2\textwidth]{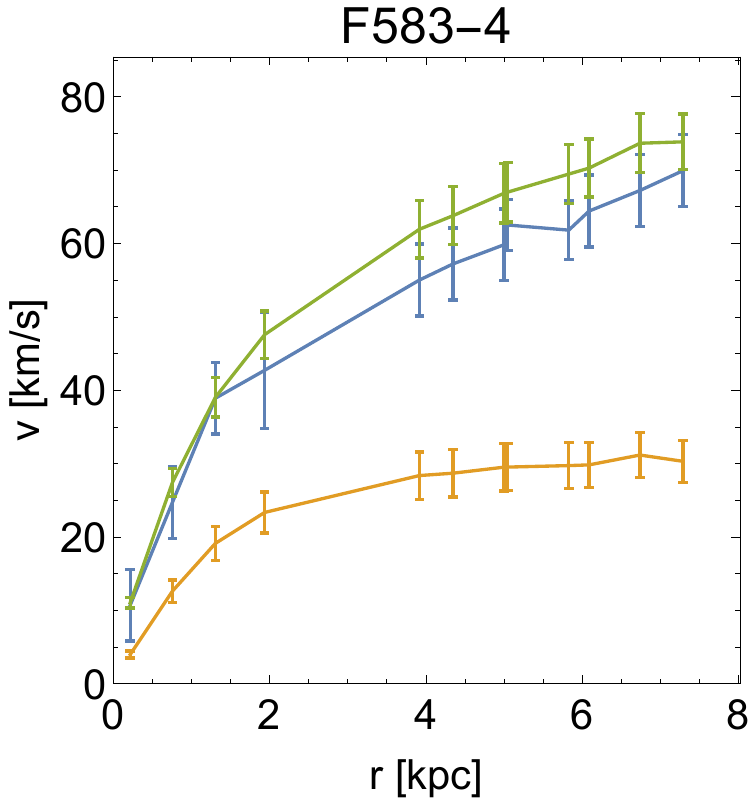}
	\caption{Rotation curve plots of galaxies corresponding to figure \ref{fig:p6} (same ordering as well). Blue denote $v_{obs}$, yellow denote $v_{bar}$ and green denote $v_{tot}^{RAR}$ based on observed $v_{bar}$.}
	\label{fig:p10}
\end{figure*}

In this appendix the $85$ galaxies used to test whether data at small and large radii coincide are examined. Figures corresponding to figure \ref{fig:p2} is shown for every galaxy in figures \ref{fig:p3}-\ref{fig:p6} with the squared residuals (which sum to the $\chi^2$) are listed in the plot labels. The square root of the residuals correspond to the magnitude of the statistical deviation from coinciding data at small and large radii in our test. In relation to the residuals shown on the figures we make three points: 1) the uncertainty of each point includes systematic uncertainties which propagate non-trivially to the uncertainty of the residual and 2) the statistical uncertainty is reduced when data are binned and 3) although there appears to be a `line' through points for a given galaxy, this does not necessarily mean a low value for $R_i^2$, which would require $\bar{g}^{obs}\simeq \bar{g}^{RAR}$. An example where this is not fulfilled is IC2574 (see figure \ref{fig:p5} bottom row, column 1); data at large radii (black) and small radii (brown) are concentrated in separate `clumps' in $g2$-space and so $\bar{g}^{obs}$ will have a relatively significant non-zero value. This value is not canceled by the expected value of MI since the slope of MI is much lower than the slope of the data. \newline
Rotation curve plots corresponding to figures \ref{fig:p3}-\ref{fig:p6} are shown in figures \ref{fig:p7}-\ref{fig:p10} alongside the prediction of the RAR. To be clear; the shown prediction of the RAR is based on data from the SPARC database (and fixed $g^\dagger=1.2\times 10^{-10}\frac{m}{s^2}$, $\tilde{\Upsilon}^{d}_i=0.5$ and $\tilde{\Upsilon}^{b}_i=0.7$) and hence we do not fit parameters (e.g. the distance, inclination angle, mass to light ratios or $g^\dagger$) as is e.g. done in \citet{Li:2018tdo}. In relation to the rotation curve plots we point out that due to the analytical relationship between the velocities and accelerations, it is not given that distinct features in $g2$-space is clearly visible in the rotation curve plots.

\clearpage

\bibliography{refs}

\end{document}